\documentclass[journal, final]{IEEEtran}

\usepackage{latexsym, amsfonts, amssymb, amsthm, setspace, verbatim,cite,mathrsfs,graphicx,bm,stfloats,url, tikz,xcolor, bm, enumerate, paralist, subfig}
\usepackage[cmex10]{amsmath}

\numberwithin{equation}{section}
\renewcommand{\theequation}{\arabic{section}.\arabic{equation}}
\newtheorem{thm}{Theorem}
\newtheorem{lem}{Lemma}
\newtheorem{fact}{Fact}
\newtheorem{applem}{Lemma}[section]

\newcommand{\be}{\begin{equation}}
\newcommand{\ee}{\end{equation}}
\newcommand{\ben}{\begin{equation*}}
\newcommand{\een}{\end{equation*}}
\newcommand{\mc}{\mathcal}
\newcommand{\mbf}{\mathbf}
\newcommand{\abs}[1]{\lvert#1\rvert}
\newcommand{\norm}[1]{\lVert#1\rVert}
\newcommand{\e}{\epsilon}
\newcommand{\expec}{\mathbb{E}}
\newcommand{\reals}{\mathbb{R}}
\newcommand{\tZ}{\tilde{Z}}
\newcommand{\tG}{\tilde{G}}
\newcommand{\hgam}{\hat{\gamma}}
\newcommand{\halph}{\hat{\alpha}}
\newcommand{\bpure}{b_{\rm pure}}
\newcommand{\hpure}{h_{\rm pure}}
\newcommand{\mscrs}{\mathscr{S}}
\newcommand{\sfc}{\mathsf{c}}
\newcommand{\sfd}{\mathsf{d}}
\newcommand{\un}{\underline}

\begin{document}

\title{Finite Sample Analysis of \\ Approximate Message Passing Algorithms}

\author{Cynthia~Rush,~\IEEEmembership{Member,~IEEE,}
        and~Ramji~Venkataramanan,~\IEEEmembership{Senior Member,~IEEE}
\thanks{This work was supported in part by a Marie Curie Career Integration Grant under Grant Agreement  Number 631489. This paper was presented in part at the 2016 IEEE International Symposium on Information Theory.}%
\thanks{C.~Rush is with the Department of Statistics, Columbia University,   New York, NY 10027, USA (e-mail: cynthia.rush@columbia.edu).}%
%
\thanks{R.~Venkataramanan is with Department of Engineering, University of Cambridge, Cambridge CB2 1PZ, UK (e-mail: rv285@cam.ac.uk).}
%
%
}

\maketitle

\begin{abstract}
Approximate message passing (AMP) refers to a class of efficient algorithms for statistical estimation in high-dimensional problems such as compressed sensing and low-rank matrix estimation.  This paper analyzes the performance of AMP in the regime where the problem dimension is large but finite.  For concreteness, we consider the setting of high-dimensional regression, where the goal is to estimate a high-dimensional vector $\beta_0$ from a noisy measurement  $y=A \beta_0 + w$. AMP is a low-complexity, scalable algorithm  for this problem. Under suitable assumptions on the measurement matrix $A$,  AMP has the attractive feature that its performance can be accurately characterized in the  large system limit by a simple scalar iteration called state evolution. Previous proofs of the validity of state evolution have all been asymptotic convergence results. In this paper, we derive a concentration inequality for AMP with i.i.d.\ Gaussian measurement matrices with finite size $n \times N$. The result shows that the probability of deviation from the state evolution prediction falls exponentially in $n$.     This provides theoretical support for empirical findings that have demonstrated excellent agreement of AMP performance with  state evolution predictions  for moderately large dimensions. The concentration inequality also indicates that the number of AMP iterations $t$ can grow no faster than order $\frac{\log n}{\log \log n}$  for the performance to be close to the state evolution predictions with high probability. The analysis can be extended to obtain similar non-asymptotic results for AMP in other settings such as low-rank matrix estimation.
\end{abstract}

\begin{IEEEkeywords}
Approximate message passing, compressed sensing, state evolution, non-asymptotic analysis, large deviations,  concentration inequalities.
\end{IEEEkeywords}

\section{Introduction} \label{sec:intro}

\IEEEPARstart{C}onsider the high-dimensional regression problem, where the goal is to estimate a vector 
 $\beta_0 \in \mathbb{R}^N$ from a noisy measurement $y \in \mathbb{R}^n$ given by
 \be  \label{eq:main_model}
 y= A\beta_0 + w.
 \ee
Here $A$ is a known $n \times N$ real-valued measurement matrix, and $w \in \mathbb{R}^n$ is the measurement noise.  The sampling ratio $\frac{n}{N} \in (0,\infty)$ is denoted by $\delta$.

Approximate Message Passing (AMP)  \cite{BayMont11, DonMalMont09, krz12, MontChap11, Rangan11, bayMontLASSO} is a class of low-complexity, scalable algorithms to solve the above problem, under suitable assumptions on $A$ and $\beta_0$. AMP algorithms are derived as Gaussian or quadratic approximations of loopy belief propagation algorithms (e.g., min-sum, sum-product) on the dense factor graph corresponding to \eqref{eq:main_model}.
 
Given the observed vector $y$,  AMP generates successive estimates of the unknown vector, denoted by  
$\beta^t \in \mathbb{R}^N$ for $t=1,2,\ldots$. Set $\beta^0=0$, the all-zeros vector. For $t=0,1,\ldots$, AMP computes
\begin{align}
z^t & = y - A\beta^t + \frac{z^{t-1}}{n} \sum_{i=1}^N \eta_{t-1}^{\prime}( [A^*z^{t-1} + \beta^{t-1}]_i), \label{eq:amp1}\\
\beta^{t+1} & = \eta_t( A^*z^t + \beta^t), \label{eq:amp2}
\end{align}
for an appropriately-chosen sequence of functions $\{\eta_t\}_{t \geq 0}: \mathbb{R} \to \mathbb{R}$.  In \eqref{eq:amp1} and \eqref{eq:amp2}, $A^*$ denotes the transpose of $A$, $\eta_t$ acts component-wise when applied to a vector, and $\eta^{\prime}_t$ denotes its (weak) derivative. Quantities with a negative index are set to zero throughout the paper.  For a demonstration of how the AMP updates \eqref{eq:amp1} and \eqref{eq:amp2} are derived from a min-sum-like message passing algorithm, we refer the reader to \cite{BayMont11}.

For a Gaussian measurement matrix $A$ with entries that are i.i.d.\ $\sim \mc{N}(0, 1/n)$, it was rigorously proven  \cite{BayMont11, JavMonState13}  that the performance of AMP can be characterized in the large system limit via a simple scalar iteration called \emph{state evolution}.  This result was extended to the class of matrices with i.i.d. sub-Gaussian entries in \cite{bayati2015}.  In particular,  these results imply that  performance measures such as the $L^2$-error $\frac{1}{N}\norm{\beta_0 - \beta^t}^2$ and the $L^1$-error $\frac{1}{N}\norm{\beta_0 - \beta^t}_1$ converge almost surely to constants that can be computed via the distribution of $\beta_0$.
 (The large system limit  is defined as $n,N \to \infty$ such that $\frac{n}{N} =\delta$, a constant.)
 
 AMP has also been applied to a variety of other high-dimensional estimation problems. Some examples are low-rank matrix estimation \cite{rangan2012iterative,Desh2014information,montanariRichard16, Desh16asymptotic,lesieur2015mmse,barbier2016mutual}, decoding of sparse superposition codes \cite{barbKrzISIT14, RushGV17, barbier2017approximate}, matrix factorization \cite{kabashima2016phase}, and estimation in generalized linear and bilinear models \cite{Rangan11, parkerSch2014A, parkerSch2014B}.
 
\emph{Main Contributions}: In this paper, we obtain a non-asymptotic result for the performance of the AMP iteration in \eqref{eq:amp1}--\eqref{eq:amp2}, when the measurement matrix $A$ has i.i.d.\ Gaussian entries $\sim \mc{N}(0, 1/n)$. We derive a concentration inequality (Theorem \ref{thm:main_amp_perf}) that implies that the probability of $\e$-deviation between various performance measures (such as  $\frac{1}{N}\norm{\beta_0 - \beta^t}^2$) and their limiting constant values fall exponentially in $n$. Our result provides theoretical support for empirical findings that have demonstrated excellent agreement of AMP performance with  state evolution predictions  for moderately large dimensions, e.g.,  $n$ of  the order of several hundreds \cite{DonMalMont09}. 
 
 In addition to refining earlier asymptotic results, the concentration inequality in Theorem \ref{thm:main_amp_perf} also clarifies the effect of the iteration number $t$ versus the problem dimension $n$. One implication is that the actual AMP performance is close to the state evolution prediction with high probability as long as $t$ is of order smaller than $\frac{\log n}{\log \log n}$. This is particularly relevant for settings where the number of  AMP iterations and the problem dimension are both large, e.g., solving the LASSO via AMP \cite{bayMontLASSO}.
 
We prove the concentration result in Theorem \ref{thm:main_amp_perf} by analyzing the following general recursion:
\be
\begin{split}
b^t & = A f_t(h^t, \beta_0) - \lambda_t g_{t-1}(b^{t-1}, w), \\ 
h^{t+1} & = A^* g_{t}(b^{t}, w) - \xi_t  f_t(h^t, \beta_0).
\end{split}
\label{eq:gen_rec0}
\ee
Here, for $t \geq 0$, the vectors $b^t \in \reals^n$, $h^{t+1} \in \reals^N$  describe the state of the algorithm, $f_t, g_t: \reals \to \reals$ are Lipschitz functions that are separable (act component-wise when applied to vectors), and $\lambda_t, \xi_t$ are scalars that can be computed from the state of the algorithm. The algorithm is initialized with 
$f_0(h^0=0, \beta_0)$. Further details on the recursion in \eqref{eq:gen_rec0}, including how the AMP in \eqref{eq:amp1}--\eqref{eq:amp2} can be obtained as a special case, are given in Section \ref{subsec:defs}.

For ease of exposition,  our analysis will focus  on the recursion \eqref{eq:gen_rec0} and the problem of high-dimensional regression.  However, it can be  extended to a number of related problems.  A  symmetric version of the  above recursion yields AMP algorithms for  problems such as  solving the TAP equations in statistical physics \cite{Bolthausen2014} and symmetric low-rank matrix estimation \cite{Desh2014information,Desh16asymptotic}. This recursion is defined in terms of a symmetric matrix $G \in \reals^{N \times N}$ with entries $\{G_{ij}\}_{i< j}$ i.i.d. $\sim \mc{N}(0,\frac{1}{N})$, and  $\{G_{ii}\}$ i.i.d. $\sim \mc{N}(0,\frac{2}{N})$ for $i \in [N]$. (In other words, $G$ can be generated as $(A + A^*)/2$, where $A \in \reals^{N \times N}$  has i.i.d. $\mc{N}(0, \frac{1}{N})$ entries.)
Then, for $t \geq 0$, let
\be
m^{t+1} = A \, p_t(m^t) - \mathsf{b}_t \,  p_{t-1} (m^{t-1}).
\label{eq:symm_rec}
\ee
Here, for $t \geq 0$, the state of the algorithm is represented by a single vector $m^t \in \reals^{N}$, the function $p_t: \reals \to \reals$ is Lipschitz and separable, and $\mathsf{b}_t$ is a constant computed from the state of the algorithm (see \cite[Sec. IV]{BayMont11} for details). The recursion  
\eqref{eq:symm_rec} is initialized with a deterministic vector $m^1 \in \reals^N$.

Our analysis of the recursion \eqref{eq:gen_rec0} can be easily extended to obtain an analogous non-asymptotic result  for the symmetric recursion in
\eqref{eq:symm_rec}. Therefore, for problems of estimating either symmetric or rectangular low-rank matrices in Gaussian noise, our analysis can be used to refine existing asymptotic  AMP guarantees (such as  those in \cite{rangan2012iterative,Desh2014information,montanariRichard16}), by providing a concentration result similar to that in Theorem  \ref{thm:main_amp_perf}. We also expect that the non-asymptotic analysis can be generalized to the case where the recursion in \eqref{eq:gen_rec0} generates matrices rather than vectors, i.e, $b^t \in \reals^{n \times q}$ and $h^{t+1} \in \reals^{N \times q}$ (where $q$ remains fixed as $n, N$ grow large; see \cite{JavMonState13} for details). Extending the analysis to this matrix recursion would yield non-asymptotic guarantees for the generalized AMP \cite{Rangan11} and  AMP for compressed sensing with spatially coupled measurement matrices \cite{DonSpatialC13}.

Since the publication of the conference version of this paper, the analysis described here has been used in a couple of recent papers: an error exponent for sparse regression codes with AMP decoding was obtained in \cite{RushV17error}, and a non-asymptotic result for AMP with non-separable denoisers was given in \cite{MaRB17analysis}.
 
\subsection{Assumptions} \label{sec:model_assumptions}

Before proceeding, we state  the assumptions on the model \eqref{eq:main_model} and the functions used to define the AMP.  In what follows, $K, \kappa > 0$ are generic positive constants whose values are not exactly specified but do not depend on $n$.  We use the notation $[n]$ to denote the set  $\{1, 2, \ldots, n\}$. 

\textbf{Measurement Matrix:} The entries of measurement matrix $A \in \mathbb{R}^{n \times N}$ are i.i.d.\ $\sim \mc{N}(0,1/n)$.

\textbf{Signal:} The entries of the signal $\beta_0 \in \mathbb{R}^N$ are i.i.d.\ according to a sub-Gaussian distribution $p_\beta$.  We recall that a zero-mean random variable $X$ is sub-Gaussian if there exist positive constants $K, \kappa$ such that $P(\abs{X - \mathbb{E}X} > \e) \leq K e^{-\kappa \e^2}$, $\forall \e >0$ \cite{BLMConc}. 

\textbf{Measurement Noise:} The entries of the measurement noise vector $w$ are i.i.d.\ according to some sub-Gaussian distribution $p_w$ with mean $0$ and $\mathbb{E}[w_i^2] = \sigma^2 < \infty$ for $i \in [n]$.  The sub-Gaussian assumption implies that, for $\e \in (0,1)$,
\be
P\left(\left \lvert \frac{1}{n} \norm{w}^2 - \sigma^2 \right \lvert \geq \e\right) \leq K e^{-\kappa n \e^2},
\label{eq:wassumption}
\ee
for some constants $K, \kappa >0$ \cite{BLMConc}.

 \textbf{The Functions $\eta_t$}: The denoising functions, $\eta_t: \mathbb{R} \rightarrow \mathbb{R}$, in \eqref{eq:amp2} are Lipschitz continuous for each $t \geq 0$, and are  therefore weakly differentiable. The weak derivative, denoted by $\eta'_t$, is assumed to be differentiable, except possibly at a finite number of points, with bounded derivative everywhere it exists. Allowing $\eta'_t$ to be non-differentiable at a finite number of points covers denoising functions like soft-thresholding which is used in applications such as the LASSO \cite{bayMontLASSO}. 

Functions defined with scalar inputs are assumed to act component-wise when applied to vectors.

The remainder of the paper is organized as follows. 
In Section \ref{sec:SE} we review state evolution, the formalism predicting the performance of AMP, and discuss how knowledge of the signal distribution $p_{\beta}$ and the noise distribution $p_w$ can help choose good denoising functions $\{\eta_t\}$. However, we emphasize that our result holds for the AMP with any choice of $\{\eta_t\}$ satisfying the above condition, even those that do not depend on $p_{\beta}$ and $p_w$. 
In Section \ref{subsec:stop_crit}, we  introduce a stopping criterion for termination of the AMP.
 In Section \ref{sec:mainresult}, we give our main result (Theorem \ref{thm:main_amp_perf}) which proves that the performance of AMP can be characterized accurately via state evolution for large but finite sample size $n$.   Section \ref{sec:amp_proof} gives the proof of Theorem \ref{thm:main_amp_perf}. The proof is based on two technical lemmas: Lemmas \ref{lem:hb_cond} and \ref{lem:main_lem}. The proof  of Lemma \ref{lem:main_lem} is long; we therefore give a brief summary of the main ideas in Section \ref{subsec:lem_comments} and then the full proof  in Section \ref{sec:main_lem_proof}.  In the appendices, we list a number of concentration inequalities that are used in the proof of Lemma \ref{lem:main_lem}. Some of these, such as the concentration inequality for the  sum of pseudo-Lipschitz functions of i.i.d.\ sub-Gaussian random variables (Lemma \ref{lem:PLsubgaussconc}), may be of independent interest. 

\section{State Evolution and the Choice of $\eta_t$} \label{sec:SE}

In this section, we briefly describe state evolution, the formalism that predicts the behavior of AMP in the large system limit.  We only review the main points followed by a few examples; a more detailed treatment can be found in \cite{BayMont11, MontChap11}.

Given $p_{\beta}$, let $\beta \in \mathbb{R} \sim p_{\beta}.$  Let $\sigma_0^2 = \mathbb{E}[\beta^2]/\delta > 0$, where $\delta=n/N$. Iteratively define the quantities $\{\tau_t^2\}_{t \geq 0}$ and $\{\sigma_t^2\}_{t \geq 1}$ as 
\begin{align}
\tau_t^2 = \sigma^2 + \sigma_t^2, \quad \quad \sigma_t^2 &= \frac{1}{\delta} \mathbb{E}\left [ \left(\eta_{t-1}(\beta + \tau_{t-1} Z) - \beta\right)^2\right ], \label{eq:sigmatdef_AMP}
\end{align}
where $\beta \sim p_{\beta}$ and $Z \sim \mc{N}(0,1)$ are independent random variables.

The AMP update \eqref{eq:amp2} is underpinned by the following key property of the vector $A^*z^t + \beta^t$: \emph{for large $n$, $A^*z^t + \beta^t$ is approximately distributed as $\beta_0 + \tau_t Z$, where $Z$ is an i.i.d.\ $\mc{N}(0,1)$ random vector independent of $\beta_0$.}  In light of this property, a natural way to generate $\beta^{t+1}$ from the ``effective observation" $A^*z^t + \beta^t =s$ is via the conditional expectation:
\be \beta^{t+1}(s) = \expec[\, \beta \mid \beta +\tau_t Z =s \,], \label{eq:opt_cond_expec} \ee
i.e., $\beta^{t+1}$ is the MMSE estimate of $\beta_0$ given the noisy observation $\beta_0 + \tau_t Z$. Thus if $p_\beta$ is known, the Bayes optimal choice for $\eta_t(s)$ is  the conditional expectation  in \eqref{eq:opt_cond_expec}.

 In the definition of the ``modified residual" $z^t$, the third term on the RHS of \eqref{eq:amp1} is crucial to ensure that the effective observation $A^*z^t + \beta^t$ has the above distributional  property. For intuition about the role of this `Onsager term', the reader is referred to \cite[Section I-C]{BayMont11}.

 We review two examples  to illustrate how full or partial  knowledge of $p_\beta$ can guide the choice of the denoising function $\eta_t$. In the first example, suppose we know that each element of $\beta_{0}$ is chosen uniformly at random from the set $\{+1, -1\}$.  Computing the conditional expectation in \eqref{eq:opt_cond_expec} with this $p_\beta$,  we obtain  $\eta_t(s) = \tanh(s/\tau_t^2)$ \cite{BayMont11}. The constants $\tau^2_{t}$ are determined iteratively from the state evolution equations \eqref{eq:sigmatdef_AMP}.
 
As a second example, consider the compressed sensing problem, where $\delta <1$, and $p_\beta$ is such that $P(\beta_0 =0) =1 - \xi$. The parameter $\xi  \in (0,1)$ determines the sparsity of $\beta_0$. For this problem, the authors in \cite{DonMalMont09,MontChap11} suggested the choice $\eta_t(s) = \eta(s;\theta_t)$,  where the soft-thresholding function $\eta$ is defined as 
\[ 
\eta(s; \theta) = \left\{
\begin{array}{ll}
(s-\theta),  & \text{ if } s > \theta, \\
0 & \text { if } -\theta \leq s \leq \theta, \\
(s-\theta),  &  \text{ if } s < -\theta.
\end{array}
\right.
\]
The threshold $\theta_t$ at step $t$ is set to $\theta_t = \alpha \tau_t$, where $\alpha$ is a tunable constant and 
$\tau_t$ is determined by \eqref{eq:sigmatdef_AMP}, making the threshold value proportional to the standard deviation of the noise in the effective observation.  However, computing $\tau_t$ using  \eqref{eq:sigmatdef_AMP} requires knowledge of  $p_\beta$. In the absence of such knowledge, we can estimate  $\tau_t^2$ by $\frac{1}{n}\norm{z^t}^2$: our concentration result (Lemma \ref{lem:main_lem}(e)) shows that this approximation is increasingly accurate as $n$ grows large. To fix $\alpha$, one could run the AMP with several different values of $\alpha$, and choose the one that gives the smallest value of $\frac{1}{n}\norm{z^t}^2$ for large $t$.

We note that in each of the above examples  $\eta_t$  is Lipschitz, and its derivative satisfies the assumption stated in Section \ref{sec:model_assumptions}.

\subsection{Stopping Criterion} \label{subsec:stop_crit}

To obtain a concentration result that clearly highlights the dependence on the iteration $t$ and the dimension $n$, we include a stopping criterion for the AMP algorithm. The intuition is that the AMP algorithm can be terminated once the expected squared error of the  estimates (as predicted by state evolution equations in \eqref{eq:sigmatdef_AMP})  is either very small or stops improving appreciably. 

For Bayes-optimal AMP where the denoising function $\eta_t(\cdot)$ is  the conditional expectation given in \eqref{eq:opt_cond_expec}, the stopping criterion is as follows. Terminate  the algorithm at the first iteration $t >0$ for which either
\be   \sigma_t^2 < \varepsilon_0,  \quad \text{ or  }  \quad  \frac{\sigma_t^2}{\sigma_{t-1}^2} > 1 - \varepsilon'_0,  \label{eq:Bayes_stop}\ee
where $\varepsilon_0 >0$ and $\varepsilon'_0 \in (0,1)$ are pre-specified constants. Recall from \eqref{eq:sigmatdef_AMP} that $\sigma^2_t$ is expected squared error in the estimate. Therefore, for suitably chosen values of  $\varepsilon_0, \varepsilon_0'$, the AMP will terminate when the expected squared error is either small enough, or has not significantly decreased from the previous iteration. 

For the  general case where $\eta_t( \cdot)$ is not the Bayes-optimal choice, the stopping criterion is: terminate the algorithm at the first iteration $t >0$ for which at least one of the following is true:
\be \sigma_t^2< \varepsilon_1,   \  \text{ or }  \  (\sigma^\perp_t)^2 < \varepsilon_2,  \  \text{ or }  \  (\tau^\perp_t)^2 < \varepsilon_3,  
\label{eq:stop_criteria} \ee
where $\varepsilon_1, \varepsilon_2, \varepsilon_3 >0$ are pre-specified constants, and $(\sigma^\perp_t)^2, (\tau^\perp_t)^2$ are defined in \eqref{eq:sigperp_defs}. The precise definitions of the scalars $(\sigma^\perp_t)^2, (\tau^\perp_t)^2$  are postponed to Sec. \ref{subsec:concvals} as a few other definitions are needed first. For now, it suffices to note that  $(\sigma^\perp_t)^2, (\tau^\perp_t)^2$  are measures of how close $\sigma_t^2$ and $\tau_t^2$ are to $\sigma_{t-1}^2$ and $\tau_{t-1}^2$, respectively. Indeed, for the Bayes-optimal case, we show in Sec \ref{subsec:bayes_opt} that \[  (\sigma_{t}^{\perp})^2 := \sigma_{t}^2\left(1 - \frac{\sigma_{t}^2}{\sigma_{t-1}^2}\right), \ \  (\tau^{\perp}_{t})^2 := \tau_{t}^2\left(1 - \frac{\tau_{t}^2}{\tau_{t-1}^2}\right). \]

Let $T^*>0$ be the first value of $t>0$ for which at least  one of the conditions  is met. Then the algorithm is run only for $0 \leq t < T^*$.  It follows that for $0 \leq t < T^*$,
\be
\sigma_t^2 > \varepsilon_1, \quad \tau_t^2 >\sigma^2+ \varepsilon_1,  \quad  (\sigma^{\perp}_t)^2 > \varepsilon_2, \quad   (\tau^{\perp}_t)^2 > \varepsilon_3.
\label{eq:sigmat_taut_bounds}
\ee
In the rest of the paper, we will use the stopping criterion to implicitly assume that $\sigma_t^2, \tau_t^2, (\sigma^{\perp}_t)^2, (\tau^{\perp}_t)^2$ are bounded below by positive constants. 

\section{Main Result} \label{sec:mainresult}
 
Our result, Theorem \ref{thm:main_amp_perf}, is a concentration inequality for \emph{pseudo-Lipschitz} (PL) loss functions. As defined in \cite{BayMont11}, a function 
$\phi: \mathbb{R}^m \to \mathbb{R}$ is pseudo-Lipschitz (of order $2$) if there exists a constant $L >0$ such that for all $x,y \in \mathbb{R}^m$, $\abs{\phi(x) - \phi(y)} \leq L ( 1 + \norm{x} + \norm{y} ) \norm{x-y},$ where $\norm{\cdot}$ denotes the Euclidean norm.

\begin{thm}\label{thm:main_amp_perf}
With the assumptions listed in Section \ref{sec:model_assumptions}, the following holds for any (order-$2$) pseudo-Lipschitz function $\phi: \mathbb{R}^2 \rightarrow \mathbb{R}$,  $\e \in (0,1)$ and  $0 \leq t < T^*$, where $T^*$ is the first iteration for which the stopping criterion in \eqref{eq:stop_criteria} is satisfied. 
\be
\begin{split}
&P\left(\left \lvert \frac{1}{N} \sum_{i=1}^N \phi(\beta^{t+1}_i, \beta_{0_i}) - \mathbb{E}[\phi(\eta_t (\beta + \tau_t Z), \beta)]\right \lvert \geq \e\right) \\
&\leq K_te^{-\kappa_t n \e^2}.
\label{eq:PL_conc_result}
\end{split}
\ee
In the expectation in \eqref{eq:PL_conc_result},  $\beta \sim p_{\beta}$ and $Z \sim \mc{N}(0,1)$ are independent, and  $\tau_t$ is given by \eqref{eq:sigmatdef_AMP}.  The  constants $K_t, \kappa_t$ are given by
$K_t =C^{2t} (t!)^{10},  \kappa_t = \frac{1}{c^{2t} (t!)^{22}}$, where $C,c  > 0$ are universal constants (not depending on $t$, $n$, or $\e$) that are not explicitly specified. 
\end{thm}

The probability in \eqref{eq:PL_conc_result} is with respect to the product measure on the space of the measurement matrix $A$, signal $\beta_0$, and the noise $w$.

\emph{Remarks}:

1. By considering the pseudo-Lipschitz function $\phi(a,b) = (a-b)^2$, Theorem \ref{thm:main_amp_perf} proves that state evolution tracks the mean square error of the AMP estimates with exponentially small probability of error in the sample size $n$.  Indeed, for all $t \geq 0$, 
\be
P\left(\left \lvert \frac{1}{N} \norm{\beta^{t+1} - \beta_{0}}^2 - \delta \sigma_{t+1}^2\right \lvert \geq \e\right) \leq K_t e^{-\kappa_t n \e^2}.
\label{eq:est_error1}
\ee
Similarly, taking $\phi(a,b) = \abs{a-b}$ the theorem implies that the normalized $L_1$-error $\frac{1}{N}\norm{\beta^{t+1}-\beta_0}_1$ is concentrated around $\mathbb{E}\abs{\eta_t (\beta + \tau_t Z) - \beta}$.

2. Asymptotic convergence results of the kind given in \cite{BayMont11,bayMontLASSO} are implied by Theorem \ref{thm:main_amp_perf}.  Indeed, from Theorem \ref{thm:main_amp_perf}, the sum
\ben
\begin{split}
\sum_{N=1}^{\infty} P\Big(\Big \lvert \frac{1}{N} \sum_{i=1}^N \phi(\beta^{t+1}_i, \beta_{0_i}) - 
\mathbb{E}[\phi(\eta_t (\beta + \tau_t Z),\beta)]\Big \lvert \geq \e\Big)
\end{split}
\een
is finite for any fixed $t \geq 0$. Therefore the Borel-Cantelli lemma implies that for any fixed $t \geq 0$:
\[\lim_{N \to \infty}  \frac{1}{N} \sum_{i=1}^N \phi(\beta^{t+1}_i, \beta_{0_i}) \stackrel{a.s.}{=} \mathbb{E}[\phi(\eta_t (\beta + \tau_t Z), \beta)]. \]

3. Theorem \ref{thm:main_amp_perf} also refines the asymptotic convergence result  by specifying how large $t$ can be (compared to the dimension $n$) for the state evolution predictions to be meaningful. Indeed, if we require the bound in \eqref{eq:PL_conc_result} to go to zero with growing $n$,  we need $\kappa_t n \e^2 \to \infty$ as $n\to \infty$. Using the expression for $\kappa_t$ from the theorem then yields $t = o\left( \frac{\log n}{\log \log n} \right)$.  

 Thus, when the AMP is run for a growing number of iterations, the state evolution predictions are guaranteed to be valid until iteration $t$ if the problem dimension grows faster than exponentially in $t$. Though the constants $K_t, \kappa_t$ in the bound have not been optimized, we believe that the dependence of  these constants on $t!$ is inevitable in any induction-based proof of the result. An open question is whether this relationship between $t$ and $n$ is fundamental, or a different analysis of the AMP can yield constants which allow $t$ to grow faster with $n$.

4.  As mentioned in the introduction, we expect that non-asymptotic results similar to Theorem \ref{thm:main_amp_perf} can be  obtained for other estimation problems (with Gaussian matrices) for which rigorous  asymptotic results have been proven for  AMP. Examples of such problems include  low-rank matrix estimation \cite{rangan2012iterative,Desh2014information,montanariRichard16}, robust high-dimensional M-estimation \cite{DonMEst15}, AMP with spatially coupled matrices \cite{DonSpatialC13},  and generalized AMP \cite{JavMonState13, KamilovRFU14}. 

 As our proof technique depends heavily  on  $A$ being i.i.d. Gaussian, extending Theorem \ref{thm:main_amp_perf} to AMP  with sub-Gaussian matrices \cite{bayati2015} and to variants of AMP with structured measurement matrices (e.g., \cite{maLiOAMP,Takeuchi17, RanganVAMP16}) is non-trivial, and an interesting direction for future work.


\section{Proof of Theorem \ref{thm:main_amp_perf}} \label{sec:amp_proof}

We first lay down the notation that will be used in the proof, then  state  two technical lemmas (Lemmas \ref{lem:hb_cond} and \ref{lem:main_lem}) and use them to prove Theorem \ref{thm:main_amp_perf}. 

\subsection{Notation and Definitions} \label{subsec:defs}
For consistency and ease of comparison, we use notation similar to \cite{BayMont11}.  To prove the technical lemmas, we use the general recursion in \eqref{eq:gen_rec0}, which we write in a slightly different form below.  Given $w \in \mathbb{R}^n$, $\beta_0 \in \mathbb{R}^N$, define the  column vectors $h^{t+1}, q^{t+1} \in \mathbb{R}^N$ and $b^t, m^t \in \mathbb{R}^n$ for $t \geq 0$ recursively as follows, starting with initial condition $q^0 \in \mathbb{R}^N$:
\begin{equation}
\begin{split}
b^t := A q^t - \lambda_t m^{t-1},\qquad & m^t  := g_t(b^t, w), \\ 
h^{t+1} := A^*m^t - \xi_t q^t, \qquad &  q^{t}  := f_t(h^t, \beta_0).
\end{split}
\label{eq:hqbm_def}
\end{equation}
where the scalars $\xi_t$ and $\lambda_t$ are defined as 
\be \xi_t := \frac{1}{n} \sum_{i=1}^n g_t'(b^t_i, w_i), \quad \lambda_t := \frac{1}{\delta N} \sum_{i=1}^N f_t'(h^t_i, \beta_{0_i}).  \label{eq:xi_lamb_def} \ee 
In \eqref{eq:xi_lamb_def}, the derivatives of $g_t: \mathbb{R}^2 \rightarrow \mathbb{R}$ and $f_t: \mathbb{R}^2 \rightarrow \mathbb{R}$ are with respect to the first argument. The functions $f_t, g_t$ are assumed to be Lipschitz continuous for $t \geq 0$,  hence the weak derivatives  $g_t'$ and $f_t'$ exist.  Further, $g'_t$ and $f'_t$ are  each assumed to be differentiable, except possibly at a finite number of points, with bounded derivative everywhere it exists.

Let $\sigma_0^2 := \mathbb{E}[ f^2_0(0,\beta)] > 0$ with $\beta \sim p_{\beta}$.  We let $q^0 = f_0(0, \beta_0)$ and assume that there exist constants $K, \kappa > 0$ such that
\be
P\left(\left\lvert \frac{1}{n} \norm{q^0}^2 - \sigma_0^2\right \lvert \geq \e\right) \leq K e^{-\kappa n \e^2}.
\label{eq:qassumption}
\ee
 Define the state evolution scalars $\{\tau_t^2\}_{t \geq 0}$ and $\{\sigma_t^2\}_{t \geq 1}$ for the general recursion as follows.
\be
\tau_t^2 := \mathbb{E}\left[ (g_{t}(\sigma_{t} Z, W))^2\right],   \quad \sigma_t^2 := \frac{1}{\delta} \mathbb{E}\left[ (f_{t}(\tau_{t-1} Z, \beta))^2\right], \label{eq:sigmatdef}
\ee
where $\beta \sim p_{\beta}, W \sim p_w$, and $Z \sim \mc{N}(0,1)$ are independent random variables.  We assume that both 
$\sigma_0^2$ and $\tau_0^2$ are strictly positive.

The AMP algorithm is a special case of the general recursion  in \eqref{eq:hqbm_def} and \eqref{eq:xi_lamb_def}.  Indeed, the AMP can be recovered by defining the following vectors  recursively for $t\geq 0$, starting with $\beta^0=0$ and $z^0=y$. 
\begin{equation}
\begin{split}
h^{t+1} = \beta_0 - (A^*z^t + \beta^t), \qquad &  q^t  =\beta^t - \beta_0, \\
b^t = w-z^t,\qquad & m^t  =-z^t.
\end{split}
\label{eq:hqbm_def_AMP}
\end{equation}
It can be verified that these vectors satisfy \eqref{eq:hqbm_def} and \eqref{eq:xi_lamb_def} with
\be
f_t(a, \beta_0) = \eta_{t-1}(\beta_0 - a) - \beta_0, \quad \text{ and } \quad g_t(a, w) = a-w.
\label{eq:fg_def}
\ee
Using this choice of $f_t, g_t$ in  \eqref{eq:sigmatdef} yields the expressions for $\sigma_t^2, \tau_t^2$  given in \eqref{eq:sigmatdef_AMP}. Using \eqref{eq:fg_def}  in \eqref{eq:xi_lamb_def}, we also see that for AMP,
\be
\lambda_t =-\frac{1}{\delta N} \sum_{i=1}^N \eta'_{t-1}([A^*\beta^{t-1} + z^{t-1}]_i), \qquad \xi_t=1.
\label{eq:lambda_xi_def_AMP}
\ee

Recall that $\beta_0 \in \mathbb{R}^N$ is the vector we would like to recover and $w \in \mathbb{R}^n$  is the measurement noise.   The vector $h^{t+1}$ is the noise in the effective observation $A^*z^t + \beta^t$, while $q^t$ is the error in the estimate $\beta^t$.  The proof will show that $h^t$ and $m^t$ are approximately i.i.d.\ $\mc{N}(0, \tau_t^2)$, while $q^t$  is approximately i.i.d.\ with zero mean and variance $\sigma_t^2$.  

For the analysis, we work with the general recursion given by \eqref{eq:hqbm_def} and \eqref{eq:xi_lamb_def}. Notice from \eqref{eq:hqbm_def} that  for all $t$,
\begin{equation}
b^{t} + \lambda_t m^{t-1} = A q^t, \quad \quad h^{t+1} + \xi_t q^t = A^* m^t.
\label{eq:bmq}
\end{equation}
Thus we have the matrix equations $X_t = A^* M_t$ and $Y_t =AQ_t,$ where
\be
\label{eq:XYMQt}
\begin{split}
X_t  &  := [h^1 + \xi_0 q^0 \mid h^2 + \xi_1 q^1 \mid \ldots \mid h^t + \xi_{t-1} q^{t-1}],  \\  
Y_t  &  := [b^0 \mid b^1 + \lambda_1 m^0 \mid \ldots \mid b^{t-1} + \lambda_{t-1} m^{t-2}], \\
M_t & := [m^0 \mid \ldots \mid m^{t-1} ], \\
Q_t  &:=  [q^0 \mid \ldots \mid q^{t-1}].
\end{split}
\ee
The notation $[c_1 \mid c_2 \mid \ldots \mid c_k]$ is used to denote a matrix with columns $c_1, \ldots, c_k$.
Note that $M_0$ and $Q_0$ are the all-zero vector.  Additionally define the matrices 
\be
\begin{split}
 H_t := [h^1 | \ldots | h^{t}],  &\qquad  \Xi_{t}:= \text{diag}(\xi_0, \ldots, \xi_{t-1}), \\
 B_t := [b^0 | \ldots | b^{t-1}],  &\qquad \Lambda_t := \text{diag}(\lambda_0, \ldots, \lambda_{t-1}). 
 \end{split}
 \ee 
Note that $B_0$, $H_0$, $\Lambda_0$, and $\Xi_{0}$ are all-zero vectors.  Using the above we see that $Y_t = B_t + [0 | M_{t-1}] \Lambda_t $ and $X_t = H_t + Q_{t}\Xi_t .$

We use the notation $m^t_{\|}$ and $q^t_{\|}$ to denote the projection of $m^t$ and $q^t$ onto the column space of $M_t$ and $Q_t$, respectively. Let
 \be 
 \alpha^t := (\alpha^t_0, \ldots, \alpha^t_{t-1})^*, \qquad  \gamma^t :=  (\gamma^t_0, \ldots, \gamma^t_{t-1})^* 
 \label{eq:vec_alph_gam_conc}
 \ee 
 be the coefficient vectors of these projections, i.e.,
 \be
 m^t_{\| } := \sum_{i=0}^{t-1} \alpha^t_i m^i, \qquad  q^t_{\|} := \sum_{i=0}^{t-1} \gamma^t_i q^i.
 \label{eq:mtqt_par}
 \ee
 The projections of $m^t$ and $q^t$ onto the orthogonal complements of $M_t$ and $Q_t$, respectively,  are denoted by
 \be
 m^t_{\perp} := m^t - m^t_{\|}, \qquad  q^t_{\perp} := q^t - q^t_{\|}.
  \label{eq:mtqt_perp}
 \ee
 Lemma \ref{lem:main_lem}  shows that for large $n$, the entries of $\alpha^t$ and ${\gamma}^t$ are concentrated around constants. We now specify these constants and provide some intuition about their values in the special case where the denoising function in the AMP recursion is the Bayes-optimal choice, as in \eqref{eq:opt_cond_expec}. 


\subsection{Concentrating Values} \label{subsec:concvals}
Let $\{\breve{Z}_t \}_{t \geq 0}$ and $\{ \tilde{Z}_t \}_{t \geq 0}$ each be sequences of of zero-mean jointly Gaussian random variables whose covariance is defined recursively as follows. For $r,t \geq 0$,
 \be 
\expec[\breve{Z}_r \breve{Z}_t] = \frac{\tilde{E}_{r,t}}{\sigma_r \sigma_t}  , \qquad \expec[\tilde{Z}_r \tilde{Z}_t] = \frac{\breve{E}_{r,t}}{\tau_r \tau_t}, 
\label{eq:tildeZcov}
\ee
where 
\be
\begin{split}
\tilde{E}_{r,t} &:= \frac{1}{\delta}\mathbb{E}[f_{r}(\tau_{r-1} \tilde{Z}_{r-1}, \beta) f_{t}(\tau_{t-1} \tilde{Z}_{t-1}, \beta)], \\
 \breve{E}_{r,t} &:= \mathbb{E}[g_{r}(\sigma_{r} \breve{Z}_{r}, W) g_{t}(\sigma_{t} \breve{Z}_{t}, W)],
\end{split}
\label{eq:Edef}
\ee
where $\beta \sim p_{\beta}$ and $W \sim p_w$ are independent random variables.  In the above, we take $f_0(\cdot, \beta) := f_0(0,\beta)$, the initial condition.  Note that $\tilde{E}_{t,t} = \sigma_t^2$ and $\breve{E}_{t,t} = \tau_t^2$, thus $\expec[\tilde{Z}^2_t] = \expec[\breve{Z}^2_t] =1$.

Define matrices $\tilde{C}^t, \breve{C}^t \in \mathbb{R}^{t \times t}$ for $t \geq 1$ such that
\be
\tilde{C}^{t}_{i+1,j+1} = \tilde{E}_{i,j}, \ \text{ and } \ \breve{C}^{t}_{i+1,j+1} = \breve{E}_{i,j}, \ 0\leq i,j \leq t-1. 
\label{eq:Ct_def}
\ee
With these definitions, the concentrating values for $\gamma^t$ and $\alpha^t$ (if $\tilde{C}^{t}$ and $\breve{C}^{t}$ are invertible) are
\be
\hat{\gamma}^{t} := (\tilde{C}^t)^{-1}\tilde{E}_t,  \quad \text{ and } \quad \hat{\alpha}^{t} := (\breve{C}^t)^{-1}\breve{E}_t,
\label{eq:hatalph_hatgam_def}
\ee
 with
\be \tilde{E}_t:=(\tilde{E}_{0,t} \ldots, \tilde{E}_{t-1,t})^*, \ \text{ and } \ \breve{E}_t:=(\breve{E}_{0,t} \ldots, \breve{E}_{t-1,t})^*. 
\label{eq:Et_def}
\ee
Let $(\sigma^{\perp}_0)^2 := \sigma_0^2$ and $(\tau^{\perp}_0)^2 := \tau_0^2$, and for $t > 0$ define 
\be
\begin{split}
& (\sigma_{t}^{\perp})^2 := \sigma_{t}^2 - (\hat{\gamma}^{t})^* \tilde{E}_{t}= \tilde{E}_{t,t} -   \tilde{E}^*_{t} (\tilde{C}^{t})^{-1}  \tilde{E}_{t}, \\
& (\tau^{\perp}_{t})^2 := \tau_{t}^2 - (\hat{\alpha}^{t})^* \breve{E}_t =  \breve{E}_{t,t} -   \breve{E}^*_{t} (\breve{C}^{t})^{-1}  \breve{E}_{t}.
\label{eq:sigperp_defs}
\end{split}
\ee
Finally, we define the concentrating values for $\lambda_{t}$ and $\xi_{t}$ as
\be
\hat{\lambda}_{t} := \frac{1}{\delta} \mathbb{E}[f'_{t}(\tau_{t-1} \tilde{Z}_{t-1}, \beta)], \  \text{ and } \  \hat{\xi}_{t} = \mathbb{E}[g'_{t}(\sigma_{t} \breve{Z}_{t}, W)].
\label{eq:hatlambda_hatxi}
\ee
Since $\{f_t\}_{t \geq 0}$ and $\{g_t\}_{t \geq 0}$ are assumed to be Lipschitz continuous, the derivatives $\{f'_t\}$ and $\{g'_t\}$ are bounded for $t \geq 0$.  Therefore 
$\lambda_t, \xi_t$ defined in \eqref{eq:xi_lamb_def} and $\hat{\lambda}_t, \hat{\xi}_t$ defined in \eqref{eq:hatlambda_hatxi} are also bounded. For the AMP recursion, it follows from  \eqref{eq:fg_def} that 
\be 
\hat{\lambda}_{t} = -\frac{1}{\delta} \mathbb{E}[ \eta'_{t-1}(\beta - \tau_{t-1} \tilde{Z}_{t-1})], \ \text{ and } \  \hat{\xi}_{t} =1.
\label{eq:hat_values_AMP}
\ee

\begin{lem} \label{lem:Ct_invert}
If $(\sigma_k^{\perp})^2$ and $(\tau_k^{\perp})^2$ are bounded below by some positive constants (say $\tilde{c}$ and $\breve{c}$, respectively) for $1 \leq k \leq t$, then the matrices $\tilde{C}^k$ and $\breve{C}^k$ defined in \eqref{eq:Ct_def} are invertible for $1\leq k \leq t$. \end{lem}
\begin{IEEEproof}
We prove the result using induction. Note that  $\tilde{C}^1= \sigma_0^2$ and $\breve{C}^1 = \tau_0^2$ are both strictly positive by assumption and hence invertible.  Assume  that for some $k <t$, $\tilde{C}^k$ and $\breve{C}^k$  are invertible.   The matrix $\tilde{C}^{k+1}$ can be written as
\ben
\tilde{C}^{k+1} = \begin{bmatrix}
    \mathsf{M}_1  &  \mathsf{M}_2 \\
    \mathsf{M}_3 &  \mathsf{M}_4
\end{bmatrix},
\een
where $\mathsf{M}_1 = \tilde{C}^{k} \in \mathbb{R}^{k \times k}$, $\mathsf{M}_4 = \tilde{E}_{k,k} = \sigma_{k}^2$, and $\mathsf{M}_2 = \mathsf{M}^*_3 = \tilde{E}_{k} \in \mathbb{R}^{k \times 1}$ defined in \eqref{eq:Et_def}.   By the block inversion formula, $\tilde{C}^{k+1}$ is invertible if $\mathsf{M}_1$ and the Schur complement $\mathsf{M}_4 - \mathsf{M}_3\mathsf{M}_1^{-1}\mathsf{M}_2$ are both invertible. By the induction hypothesis $\mathsf{M}_1 = \tilde{C}^{k}$ is invertible, and 
\be \mathsf{M}_4 - \mathsf{M}_3\mathsf{M}_1^{-1}\mathsf{M}_2 =   \tilde{E}_{k,k} -   \tilde{E}^*_{k} (\tilde{C}^{k})^{-1}  \tilde{E}_{k} = (\sigma_{k}^{\perp})^2  \geq \tilde{c} > 0.
\label{eq:sigt_perp_pos}
\ee
Hence  $\tilde{C}^{t+1}$ is invertible.  Showing that $\breve{C}^{t+1}$ is invertible is very similar.
\end{IEEEproof}
We note that the stopping criterion ensures that $\tilde{C}^t$ and $\breve{C}^t$ are invertible for all $t$ that are relevant to Theorem \ref{thm:main_amp_perf}.

\subsection{Bayes-optimal AMP} \label{subsec:bayes_opt}

The concentrating constants in \eqref{eq:tildeZcov}--\eqref{eq:sigperp_defs} have simple representations in the special case where the denoising function  $\eta_t(\cdot)$ is chosen to be Bayes-optimal, i.e.,  the conditional expectation of $\beta$ given the noisy observation $\beta+ \tau_t Z$, as in \eqref{eq:opt_cond_expec}.  In this case:
 \begin{enumerate}
\item  It can be shown that $\tilde{E}_{r,t}$ in \eqref{eq:Edef} equals $\sigma_{t}^2$ for $0\leq r \leq t$. This is done in two steps. First  verify that the following Markov property holds for the jointly Gaussian  $\tilde{Z}_r, \tilde{Z}_t$ with covariance given by \eqref{eq:tildeZcov}:
\[ \expec[\beta \mid \beta + \tau_t \tilde{Z}_t, \  \beta +\tau_r \tilde{Z}_r ] = \expec[\beta \mid \beta + \tau_t \tilde{Z}_t], \quad 0 \leq r \leq t.  \]
We then use the above in the definition of $\tilde{E}_{r,t}$ (with $f_t$ given by \eqref{eq:fg_def}), and  apply the orthogonality principle to show that $\tilde{E}_{r,t} = \sigma_t^2$ for $r \leq t$.

\item Using $\tilde{E}_{r,t} = \sigma_{t}^2$ in \eqref{eq:tildeZcov} and \eqref{eq:Edef}, we obtain $\breve{E}_{r,t} =  \sigma^2 + \sigma_t^2= \tau_t^2$.

\item 
From the orthogonality principle, it also follows that for $0 \leq r \leq t$,
\ben
\mathbb{E}[\norm{\beta^{t}}^2] = \mathbb{E}[\beta^*\beta^{t}], \ \  \text{ and } \ \  \mathbb{E}[\norm{\beta^{r}}^2] = \mathbb{E}[(\beta^{r})^*\beta^{t}],
\een
where $\beta^{t}=\expec[\beta \mid \beta + \tau_{t-1} \tilde{Z}_{t-1}]$.
\item With $\tilde{E}_{r,t} = \sigma_{t}^2$ and $\breve{E}_{r,t}=\tau_t^2$  for $r \leq t$, the quantities in \eqref{eq:hatalph_hatgam_def}--\eqref{eq:sigperp_defs} simplify to the following for $t >0$:
\be
\begin{split}
& \hat{\gamma}^{t}  = [0,\ldots,0, {\sigma^2_t}/{\sigma^2_{t-1}} ], \quad  \hat{\alpha}^{t}  = [0,\ldots,0, {\tau^2_t}/{\tau^2_{t-1}} ], \\
& (\sigma_{t}^{\perp})^2 := \sigma_{t}^2\left(1 - \frac{\sigma_{t}^2}{\sigma_{t-1}^2}\right), \ \  (\tau^{\perp}_{t})^2 := \tau_{t}^2\left(1 - \frac{\tau_{t}^2}{\tau_{t-1}^2}\right),
\end{split}
\label{eq:simpl_exps}
\ee
where $\hat{\gamma}^{t}, \hat{\alpha}^{t} \in \mathbb{R}^t$.  
\end{enumerate} 

For the AMP, $m^t = -z^t$ is the modified residual  in iteration $t$, and $q^t = \beta^t - \beta$ is the error in the estimate $\beta^t$. Also recall that $\gamma^t$ and $\alpha^t$ are the coefficients of the projection of $m^t$ and $q^t$ onto $\{ m^0, \ldots, m^{t-1}\}$ and $\{ q^0, \ldots, q^{t-1}\}$, respectively.  The fact that only the last entry of $\hat{\gamma}^t$  is non-zero in the Bayes-optimal case indicates that residual $z^{t}$ can be well approximated as a linear combination of $z^{t-1}$ and a vector that is independent of $\{z^0, \ldots, z^{t-1}\}$; a similar interpretation holds for the error $q^t = \beta^t - \beta$.  


\subsection{Conditional Distribution Lemma} \label{sec:cond_dist_lemma}

We next characterize the conditional distribution of the vectors $h^{t+1}$ and $b^t$ given the matrices in \eqref{eq:XYMQt} as well as $\beta_0, w$.  Lemmas \ref{lem:hb_cond}  and  \ref{lem:bhpure} show that the conditional distributions of  $h^{t+1}$ and $b^t$ can each be expressed in terms of a standard normal vector and a deviation vector. Lemma  \ref{lem:main_lem} shows that the norms of the deviation vectors are small with high probability, and provides concentration inequalities for various inner products and functions involving
$\{ h^{t+1}, q^t, b ^t, m^t \}$.

We use the following notation in the lemmas. Given two random vectors $X, Y$ and a sigma-algebra $\mscrs$, $X |_\mscrs \stackrel{d}{=} Y$ denotes that the conditional distribution of $X$  given $\mscrs$ equals the distribution of $Y$.   The $t \times t$ identity matrix is denoted by $\mathsf{I}_{t}$. We suppress the subscript on the matrix if the dimensions are clear from context.  For a matrix $A$ with full column rank, $\mathsf{P}^{\parallel}_{A} := A(A^*A)^{-1}A^*$ denotes  the orthogonal projection matrix onto the column space of $A$, and $\mathsf{P}^\perp_{A}:=\mathsf{I}- \mathsf{P}^{\parallel}_{A}$. If $A$ does not have full column rank, $(A^*A)^{-1}$ is interpreted as the pseudoinverse.

Define $\mathscr{S}_{t_1, t_2}$ to be the sigma-algebra generated by
\[ b^0, ..., b^{t_1 -1}, m^0, ..., m^{t_1 - 1}, h^1, ..., h^{t_2}, q^0, ..., q^{t_2},\text{ and }  \beta_0, w. \]
A key ingredient in the proof is the distribution of $A$ conditioned on the sigma algebra $\mscrs_{t_1,t}$ where $t_1$ is either $t+1$ or $t$ from which we are able to specify the conditional distributions of $b^t$ and $h^{t+1}$ given $\mathscr{S}_{t, t}$ and $\mathscr{S}_{t+1, t}$, respectively.  Observing that conditioning on $\mscrs_{t_1,t}$ is equivalent to conditioning on the linear constraints\footnote{While conditioning on the linear constraints, we emphasize that only $A$ is treated as random.}
\[ A Q_{t_1} = Y_{t_1}, \ A^*M_t=X_t, \]
the following lemma from \cite{BayMont11} specifies the conditional distribution of $A |_{\mscrs_{t_1,t}}$. 
 
\begin{lem}\cite[Lemma $10$, Lemma $12$]{BayMont11}
The conditional distributions of the vectors in \eqref{eq:bmq} satisfy the following, provided $n >t$ and $M_t, Q_t$ have full column rank.
\begin{align*}
& A^* m^t |_{\mscrs_{t+1,t}} \stackrel{d}{=} X_t (M_t^* M_t)^{-1} M_t^* m^t_{\parallel}  \\
&\hspace{0.7in} +   Q_{t+1}(Q^*_{t+1} Q_{t+1})^{-1} Y^*_{t+1} m_\perp^t  + \mathsf{P}^\perp_{Q_{t+1}} \tilde{A}^* m^t_{\perp}, \\
& A q^t |_{\mscrs_{t,t}} \stackrel{d}{=} Y_t (Q_t^* Q_t)^{-1} Q_t^* q^t_{\parallel} + M_{t}(M^*_{t} M_{t})^{-1} X^*_{t} q_\perp^t\\
& \hspace{0.7in} + \mathsf{P}^\perp_{M_t} \hat{A} q^t_{\perp},
\end{align*}
where $m^t_{\parallel}, m_\perp^t, q^t_{\|}, q_\perp^t$ are defined in \eqref{eq:mtqt_par} and \eqref{eq:mtqt_perp}.  Here $\tilde{A}, \hat{A} \stackrel{d}{=}A$ are random matrices independent of $\mscrs_{t+1,t}$ and $\mscrs_{t,t}$.
\label{lem:Et1t}
\end{lem}


\begin{lem}[Conditional Distribution Lemma]
For the vectors $h^{t+1}$ and $b^t$ defined in \eqref{eq:hqbm_def}, the following hold for $t \geq 1$, provided $n >t$ and $M_t, Q_t$ have full column rank.
\begin{equation}
\begin{split}
&b^{0} \lvert_{\mscrs_{0, 0}} \stackrel{d}{=} \sigma_0 Z'_0 + \Delta_{0,0}, \\
 &b^{t} \lvert_{\mscrs_{t, t}}\stackrel{d}{=} \sum_{r=0}^{t-1} \hat{\gamma}^{t}_r b^r + \sigma^{\perp}_t Z'_t + \Delta_{t,t}, \label{eq:Ba_dist} 
 \end{split}
 \end{equation}
 \begin{equation}
 \begin{split}
 h^{1} \lvert_{\mscrs_{1, 0}} &  \stackrel{d}{=} \tau_0 Z_0 + \Delta_{1,0},  \\
  h^{t+1} \lvert_{\mscrs_{t+1, t}} &  \stackrel{d}{=} \sum_{r=0}^{t-1} \hat{\alpha}^t_r h^{r+1} + \tau^{\perp}_t Z_t + \Delta_{t+1,t}, \label{eq:Ha_dist}
 \end{split}
\end{equation}
where $Z_0, Z_t \in \mathbb{R}^N$ and $Z'_0, Z'_t \in \mathbb{R}^n$ are i.i.d.\ standard Gaussian random vectors that are independent of the corresponding conditioning sigma-algebras. The terms $\hat{\gamma}^{t}_i$ and $\hat{\alpha}^t_{i}$ for $i \in [t-1]$ are defined in \eqref{eq:hatalph_hatgam_def} and the terms $(\tau_{t}^{\perp})^2$ and $(\sigma_{t}^{\perp})^2$ in \eqref{eq:sigperp_defs}.  The deviation terms are 
\begin{align}
\Delta_{0,0} &= \Big(\frac{\norm{q^0}}{\sqrt{n}} - \sigma_0\Big)Z'_0, \label{eq:D00} \\
\Delta_{1,0} &= \Big[ \Big(\frac{\norm{m^0}}{\sqrt{n}}  - \tau_0\Big)\mathsf{I}_N -\frac{\norm{m^0}}{\sqrt{n}} \mathsf{P}^{\parallel}_{q^0}\Big] Z_0 \nonumber \\
& + q^0 \Big(\frac{\norm{q^0}^2}{n}\Big)^{-1} \Big(\frac{(b^0)^*m^0}{n} - \xi_0 \frac{\norm{q^0}^2}{n}\Big), \label{eq:D10}
\end{align}
and for $t >0$, defining $\mathbf{Q}_{t} := Q_{t}^* Q_{t}$ and $\mathbf{M}_{t} := M_{t}^* M_{t}$,
\begin{align}
&\Delta_{t,t} =  \sum_{r=0}^{t-1} (\gamma^t_r - \hat{\gamma}^{t}_r) b^r + \Big[  \Big(\frac{\norm{q^t_{\perp}}}{\sqrt{n}} - \sigma_{t}^{\perp}\Big) \mathsf{I}_n  - \frac{\norm{q^t_{\perp}} }{\sqrt{n}} \mathsf{P}^{\parallel}_{M_t}\Big]Z'_t  \nonumber \\
& + M_t\Big(\frac{\mathbf{M}_{t}}{n}\Big)^{-1} \Big(\frac{H_t^* q^t_{\perp}}{n} - \frac{M_t}{n}^*\Big[\lambda_t m^{t-1} - \sum_{i=1}^{t-1} \lambda_{i} \gamma^t_{i} m^{i-1}\Big]\Big),\label{eq:Dtt} \\
&\Delta_{t+1,t} =  \sum_{r=0}^{t-1} (\alpha^t_r - \hat{\alpha}^t_r) h^{r+1} \nonumber \\
&+ \Big[\Big(\frac{\norm{m^t_{\perp}}}{\sqrt{n}} - \tau_{t}^{\perp}\Big)  \mathsf{I}_N  -\frac{\norm{m^t_{\perp}}}{\sqrt{n}} \mathsf{P}^{\parallel}_{Q_{t+1}}\Big]Z_t \nonumber \\
& + Q_{t+1} \Big(\frac{\mathbf{Q}_{t+1}}{n}\Big)^{-1} \Big(\frac{B^*_{t+1} m^t_{\perp}}{n} - \frac{Q_{t+1}^*}{n}\Big[\xi_t q^t - \sum_{i=0}^{t-1} \xi_i \alpha^t_i q^i\Big]\Big).\label{eq:Dt1t}  
\end{align} 
\label{lem:hb_cond}
\end{lem}

\begin{IEEEproof}
We begin by demonstrating  \eqref{eq:Ba_dist}.  By \eqref{eq:hqbm_def} it follows that
\ben
b^{0}\lvert_{\mscrs_{0,0}} = A q^0 \overset{d}{=} (\norm{q^0}/\sqrt{n}) Z'_0,
\een
where $Z'_0 \in \mathbb{R}^n$ is an i.i.d.\ standard Gaussian random vector, independent of $\mscrs_{0,0}$.

For the case $t \geq 1$, we use Lemma \ref{lem:Et1t} to write
\be
\begin{split}
&b^t  \lvert_{\mscrs_{t, t}}  = (A q^t - \lambda_t m^{t-1}) \lvert_{\mscrs_{t, t}} \\
&\overset{d}{=} Y_t\mathbf{Q}_{t}^{-1} Q_t^* q^t_{\parallel} + M_t \mathbf{M}_{t}^{-1} X_t^*  q_{\perp}^t + \mathsf{P}^{\perp}_{M_{t}} \tilde{A} q^t_{\perp} - \lambda_t m^{t-1}\\
& = B_t \mathbf{Q}_t^{-1} Q_t^* q^t_{\parallel} + [ 0 | M_{t-1}] \Lambda_t \mathbf{Q}_{t}^{-1} Q_t^* q^t_{\parallel} + M_t \mathbf{M}_{t}^{-1} H_t^*  q_{\perp}^t \\
&\qquad + \mathsf{P}^{\perp}_{M_{t}} \tilde{A} q^t_{\perp}- \lambda_t m^{t-1}. \nonumber
\end{split}
\label{eq:lemma13a}
\ee
The last equality above is obtained using $Y_t = B_t + [0| M_{t-1}] \Lambda_t$, and $X_t=H_t + \Xi_{t} Q_t$.  Noticing that $B_t \mathbf{Q}_{t}^{-1} Q_t^* q^t_{\parallel} = \sum_{i=0}^{t-1} \gamma^t_i b^i$ and $\mathsf{P}^{\perp}_{M_{t}} \tilde{A} q^t_{\perp} = (\mathsf{I}  - \mathsf{P}^{\parallel}_{M_t})\tilde{A}q^t_{\perp} \overset{d}{=}  (\mathsf{I}  - \mathsf{P}^{\parallel}_{M_t}) \frac{\norm{q^t_{\perp}}}{\sqrt{n}} Z'_t$ where $Z'_t \in \mathbb{R}^n$ is an i.i.d.\ standard Gaussian random vector,  it follows that
\be
\begin{split}
b^t |_{\mscrs_{t,t}} 
&\overset{d}{=}  (\mathsf{I}  - \mathsf{P}^{\parallel}_{M_t}) \frac{\norm{q^t_{\perp}}}{\sqrt{n}} Z'_t + \sum_{i=0}^{t-1} \gamma^t_i b^i \\
&+ [0 | M_{t-1}] \Lambda_t \mathbf{Q}_{t}^{-1} Q^*_t q^t_{\parallel} + M_t \mathbf{M}_{t}^{-1} H_t^* q^t_{\perp} - \lambda_t m^{t-1}.
\end{split}
\label{eq:btdef}
\ee
All the quantities in the RHS of \eqref{eq:btdef} except $Z'_{t}$ are in the conditioning sigma-field.  We can rewrite \eqref{eq:btdef} with the following pair of values:
\begin{align*}
b^{t}\lvert_{\mscrs_{t, t}} &\overset{d}{=} \sum_{r=0}^{t-1} \hat{\gamma}^t_{r} b^{r} + \sigma_{t}^{\perp} Z'_t + \Delta_{t,t}, \\
\Delta_{t,t} &= \sum_{r=0}^{t-1} (\gamma^t_r - \hat{\gamma}^t_{r}) b^r  +   \Big[ \Big(\frac{\norm{q^t_{\perp}}}{\sqrt{n}} - \sigma_{t}^{\perp}\Big)\mathsf{I}  - \frac{\norm{q^t_{\perp}} }{\sqrt{n}} \mathsf{P}^{\parallel}_{M_t}\Big]Z'_t   \\
&+  [0 | M_{t-1}] \Lambda_t \mathbf{Q}_{t}^{-1} Q^*_t q^t_{\parallel}  +  M_t \mathbf{M}_{t}^{-1} H_t^* q^t_{\perp}  -  \lambda_t m^{t-1}. 
\end{align*}
The above definition of $\Delta_{t,t}$ equals that given in \eqref{eq:Dtt} since
\begin{align*}
& [0 | M_{t-1}] \Lambda_t \mathbf{Q}_{t}^{-1} Q^*_t q^t_{\parallel}  -\lambda_t m^{t-1}  = \sum_{i=1}^{t-1} \lambda_{i} \gamma^t_{i} m^{i-1}  -\lambda_t m^{t-1}  \\
&=  - M_t \mathbf{M}_{t}^{-1} M_t^*\Big(\lambda_t m^{t-1} - \sum_{i=1}^{t-1} \lambda_{i} \gamma^t_{i} m^{i-1}\Big).
\end{align*}
This completes the proof of \eqref{eq:Ba_dist}.  Result \eqref{eq:Ha_dist} can be shown similarly.  
\end{IEEEproof}


The conditional distribution representation in Lemma \ref{lem:hb_cond} implies that for each $t \geq 0$,  $h^{t+1}$ is the sum of an i.i.d.\ $\mc{N}(0, \tau_t^2)$ random vector plus a deviation term.  Similarly $b^t$ is the sum of an i.i.d.\ $\mc{N}(0, \sigma_t^2)$ random vector and a deviation term.  This is made precise in the following lemma. 

\begin{lem}
For $t  \geq 0$, let $Z'_t \in \reals^n$,  $Z_t \in  \reals^N$  be independent standard normal random vectors.
 Let $\bpure^0=  \sigma_0 Z'_0 $, $\hpure ^1=  \tau_0 Z_0$, and recursively define for $t \geq 1$:
 \begin{align}
 \bpure^t &= \sum_{r=0}^{t-1} \hgam^t_r \bpure^r + \sigma^\perp_t Z'_t,  \quad 
\hpure^{t+1}  = \sum_{r=0}^{t-1} \halph^t_{r} \hpure^{r+1} + \tau^\perp_t Z_t.
\label{eq:pure_bh_def}
 \end{align}
Then for  $t \geq 0$,  the following statements hold.
\begin{enumerate}
\item 
For $j \in [N]$ and $k \in [n]$,
\be
\begin{split}
 ({\bpure^0}_{j}, \ldots, {\bpure^t}_{j} ) \, &\stackrel{d}{=} \,  (\sigma_0 \breve{Z}_0, \ldots, \sigma_t \breve{Z}_t), \\
 ({\hpure^1}_{k}, \ldots, {\hpure^{t+1}}_{k} ) \, &\stackrel{d}{=} \,  (\tau_0 \tilde{Z}_0, \ldots, \tau_t \tilde{Z}_t),
 \label{eq:pure_bh_dist}
\end{split}
\ee
where $\{\breve{Z}_t \}_{t \geq 0}$ and $\{ \tilde{Z}_t \}_{t \geq 0}$ are the jointly Gaussian random variables defined in Sec. \ref{subsec:concvals}.
\item For $t  \geq 0$,
\begin{align}
\bpure^t = \sum_{i=0}^t Z'_i \, \sigma_i^{\perp} \sfc^t_i, \quad \hpure^t = \sum_{i=0}^t Z_i  \, \tau_i^{\perp}  \sfd^t_i,
\label{eq:bhZZ}
\end{align}
where the constants $\{ \sfc^t_i \}_{0 \leq i \leq t}$ and $\{ \sfd^t_i \}_{0 \leq i \leq t}$ are recursively defined as follows, starting with $\sfc^0_0=1$ and 
$\sfd^0_0=1$. For $t >0$, 
\begin{align}
& \sfc^t_t =1, \quad \sfc^t_i= \sum_{r=i}^{t-1} \sfc^r_i \hgam^{t}_r,  \ \text { for } 0 \leq i \leq (t-1), \\ 
&  \sfd^t_t  =1, \quad \sfd^t_i= \sum_{r=i}^{t-1} \sfd^r_i \halph^{t}_r,  \ \text { for } 0 \leq i \leq (t-1).  
\label{eq:sfcd_def}
\end{align}

\item The conditional distributions in Lemma \ref{lem:hb_cond} can be expressed as 
\be
\begin{split}
 b^{t} \lvert_{\mscrs_{t, t}} &\stackrel{d}{=}  \bpure^t + \sum_{r=0}^{t}  \sfc^t_r \, \Delta_{r,r}, \\ 
h^{t+1} \lvert_{\mscrs_{t+1, t}} &\stackrel{d}{=}  \hpure^{t+1} + \sum_{r=0}^{t}  \sfd^t_r \, \Delta_{r+1,r}.
\label{eq:btht1_pure_exp}
\end{split}
\ee
\end{enumerate}
\label{lem:bhpure}
\end{lem}

\begin{IEEEproof}
We prove \eqref{eq:pure_bh_dist} by induction.  We prove the $\bpure^t$ result; the proof for $\hpure^t$ is  very similar. The base case of $t=0$ holds by the definition of  $\bpure^0$. Assume towards induction that \eqref{eq:pure_bh_dist} holds for  $(\bpure^0, \ldots, \bpure^{t-1})$.  Then using \eqref{eq:pure_bh_def}, $\bpure^t$ has the same distribution as 
$\sum_{r=0}^{t-1} \hat{\gamma}^{t}_{r} \sigma_r \breve{Z}_r +\sigma^{\perp}_t Z$ where $Z \in \mathbb{R}^n$ is a standard Gaussian random vector  independent of $ \breve{Z}_0, \ldots,  \breve{Z}_{t-1}$. 
We now show that $\sum_{r=0}^{t-1} \hat{\gamma}^{t}_{r} \sigma_r \breve{Z}_r +\sigma^{\perp}_t Z \stackrel{d}{=} \sigma_t \breve{Z}_t$ by demonstrating that: \\
(i) var$ (\sum_{r=0}^{t-1} \hat{\gamma}^{t}_{r} \sigma_r \breve{Z}_r +\sigma^{\perp}_t Z) = \sigma_t^2$; and \\
(ii) $\expec[\sigma_k \breve{Z}_k ( \sum_{r=0}^{t-1} \hat{\gamma}^{t}_{r} \sigma_r \breve{Z}_r +\sigma^{\perp}_t Z) ] = \sigma_k \sigma_t \expec[\breve{Z}_k \breve{Z}_t] = \tilde{E}_{k,t}$, for $0 \leq k \leq (t-1)$.  \\
The variance is
\ben 
\expec( \sum_{r=0}^{t-1} \hat{\gamma}^{t}_{r} \sigma_r \breve{Z}_r +\sigma^{\perp}_t Z )^2  
= \sum_{r=0}^{t-1}\sum_{k=0}^{t-1}  \hat{\gamma}^{t}_{r}  \hat{\gamma}^{t}_{k} \tilde{E}_{k,r} + (\sigma^{\perp}_t )^2
= \sigma_t^2,
\een
 where the last equality follows from rewriting the double sum as follows using the definitions in Section \ref{subsec:defs}:
\be
\begin{split}
& \sum_{r,k} \hat{\gamma}^{t}_{r}  \hat{\gamma}^{t}_{k} \tilde{E}_{k,r}  =  (\hat{\gamma}^{t})^* \tilde{C}^t \hat{\gamma}^{t}  = [\tilde{E}_t^* (\tilde{C}^t)^{-1}]  \tilde{C}^t [ (\tilde{C}^t)^{-1} \tilde{E}_t] \\
 &=  \tilde{E}_t^* (\tilde{C}^t)^{-1} \tilde{E}_t = \tilde{E}_{t,t} -  (\sigma^{\perp}_t)^2. 
\label{eq:gamr_gamk}
\end{split}
\ee
 Next, for any $0 \leq k \leq t-1$, we have
 \ben
 \begin{split}
 \expec[\sigma_k \breve{Z}_k ( \,  \sum_{r=0}^{t-1} \hat{\gamma}^{t}_{r} \sigma_r \breve{Z}_r +\sigma^{\perp}_t Z \, ) ] 
&\stackrel{(a)}{=} \sum_{r=0}^{t-1}  \tilde{E}_{k,r}  \hat{\gamma}^{t}_{r}  \\
&\stackrel{(b)}{=}  [\tilde{C} \hat{\gamma}^{t}]_{k+1} \stackrel{(c)}{=}  \tilde{E}_{k,t}.
 \end{split}
 \een
In the above, step $(a)$ follows from \eqref{eq:tildeZcov}; step $(b)$ by recognizing from \eqref{eq:Ct_def} that the required sum is the inner product of $\hat{\gamma}^{t}$ with row  $(k+1)$ of $\tilde{C}^t$; step $(c)$ from the definition of  $\hat{\gamma}^{t}$ in \eqref{eq:hatalph_hatgam_def}. This proves \eqref{eq:pure_bh_dist}.

Next we show the expression for $\bpure^t$ in \eqref{eq:bhZZ} using induction; the proof for $\hpure^t$ is similar. The base case of $t=0$ holds by definition because $\sigma_1^{\perp} = \sigma_1$. Using the induction hypothesis that \eqref{eq:bhZZ} holds for $\bpure^0, \ldots, \bpure^{t-1}$,  the defintion \eqref{eq:pure_bh_def} can be written as 
\begin{align*}
\bpure^t & = \sum_{r=0}^{t-1} \hgam^t_r \Big( \sum_{i=0}^r  Z'_i \sigma_i^{\perp} \sfc^r_i \Big)  + \sigma_t^{\perp} Z'_t \\
& = \sum_{i=0}^{t-1} Z'_i \sigma_i^{\perp} \Big( \sum_{r=i}^{t-1} \hgam^t_r \sfc^r_i \Big)  + \sigma_t^{\perp} Z'_t   =  \sum_{i=0}^{t} Z'_i \sigma_i^{\perp} \sfc^t_i,
\end{align*}
where the last inequality follows from the definition of $c^t_i$ for $0 \leq i \leq t$ in \eqref{eq:sfcd_def}. This proves \eqref{eq:bhZZ}.

The expressions for the conditional distribution of $b^t$ and $h^{t+1}$ in \eqref{eq:btht1_pure_exp}
can be similarly  obtained from \eqref{eq:Ba_dist}  and \eqref{eq:Ha_dist} using an induction argument.
\end{IEEEproof}


\subsection{Main Concentration Lemma}

For $t \geq 0$, let
\be
\begin{split}
K_t =C^{2t} (t!)^{10}, & \quad \kappa_t = \frac{1}{c^{2t} (t!)^{22}},  \\
K'_t = C (t+1)^5K_t,  &\quad  \kappa'_t  = \frac{\kappa_t}{  c(t+1)^{11}},
\end{split}
\label{eq:Kkappa_def}
\ee 
where $C, c > 0$ are universal constants (not depending on $t$, $n$, or $\e$).  To keep the notation compact,  we  use $K, \kappa, \kappa'$ to denote generic positive universal constants whose values may change through  the  lemma statement and the proof.
\begin{lem}
The following statements hold for $1 \leq t < T^*$ and $\e \in (0,1)$.

\begin{enumerate}[(a)]

\item 
\begin{align}
P\Big(\frac{1}{N}\norm{\Delta_{{t+1,t}}}^2 \geq \epsilon \Big) \leq K t^2 K'_{t-1} e^{-{ \kappa \kappa'_{t-1} n \epsilon}/{ t^4}}, \label{eq:Ha} \\
P\Big(\frac{1}{n}\norm{\Delta_{{t,t}}}^2 \geq \epsilon\Big) \leq  K t^2 K_{t-1} e^{-{\kappa \kappa_{t-1} n \epsilon}/{ t^4}}. \label{eq:Ba} 
\end{align}

\item \emph{i)} Let $X_n \overset{\mathbf{. .}}{=} c$ be shorthand for $P(\abs{X_n -c} \geq \e) \leq K t^3 K'_{t-1} e^{- \kappa \kappa'_{t-1} n \epsilon^2/t^7 }$.  Then for pseudo-Lipschitz functions $\phi_h: \mathbb{R}^{t+2} \rightarrow \mathbb{R}$
 \be
\begin{split}
&\frac{1}{N}\sum_{i=1}^N \phi_h(h^1_i, \ldots, h^{t+1}_i, \beta_{0_i})  \overset{\mathbf{. .}}{=} \expec\, \phi_h(\tau_0 \tilde{Z}_0, \ldots, \tau_t \tilde{Z}_t, \beta).
 \end{split}
  \label{eq:Hb1}
\ee
The random variables $\tilde{Z}_{0}, \ldots, \tilde{Z}_t$ are jointly Gaussian with zero mean and covariance given by \eqref{eq:tildeZcov}, and are  independent of $\beta \sim p_{\beta}$.

\emph{ii)} Let $\psi_h: \mathbb{R}^2 \rightarrow \mathbb{R}$ be a bounded function that is differentiable in the first argument except possibly at a finite number of points, with bounded derivative where it exists.  Then,
\be
\begin{split}
&P\Big( \Big \lvert\frac{1}{N} \sum_{i=1}^N \psi_h(h^{t+1}_i, \beta_{0_i}) -  \mathbb{E}\, \psi_h( \tau_{t} \tilde{Z}_{t}, \beta) \Big \lvert \geq \e \Big) \\
 &\leq K t^2 K'_{t-1} e^{{-\kappa \kappa'_{t-1} n \epsilon^2}/{t^4}}.  \label{eq:Hb2}
\end{split}
 \ee
 As above, $\tilde{Z}_t \sim \mc{N}(0,1)$ and $\beta \sim p_\beta$ are independent.  

\emph{iii)} Let $X_n \doteq c$ be shorthand for $P(\abs{X_n -c} \geq \e) \leq K t^3 K_{t-1} e^{- \kappa \kappa_{t-1} n \epsilon^2/t^7 }$. Then for pseudo-Lipschitz functions $\phi_b: \mathbb{R}^{t+2} \rightarrow \mathbb{R}$
\be
\begin{split}
& \frac{1}{n}\sum_{i=1}^n \phi_b(b^0_i, \ldots, b^{t}_i, w_{i}) \doteq
 \expec \, \phi_b(\sigma_0 \breve{Z}_0, \ldots, \sigma_t \breve{Z}_t, W ).
 \end{split} 
\label{eq:Bb1}
\ee
The random variables $\breve{Z}_{0}, \ldots, \breve{Z}_t$ are jointly Gaussian with zero mean and covariance given by \eqref{eq:tildeZcov}, and are independent of $W \sim p_{w}$.

\emph{iv)} Let $\psi_b: \mathbb{R} \rightarrow \mathbb{R}$ be a bounded function that is differentiable in the first argument except possibly at a finite number of points, with bounded derivative where it exists. Then,
\be
\begin{split}
&P\Big( \Big \lvert \frac{1}{n} \sum_{i=1}^n \psi_b(b^{t}_i, w_{i}) -  \mathbb{E}\, \psi_b(\sigma_{t} \breve{Z}_{t}, W) \Big \lvert \geq \e \Big) \\
&\leq  K t^2 K_{t-1} e^{{ -\kappa \kappa_{t-1} n \epsilon^2}/{t^4}}.
\label{eq:Bb2}
\end{split}
\ee
 As above, $\breve{Z}_t \sim \mc{N}(0,1)$ and $W \sim p_w$ are independent. 


\item
\begin{align}
\frac{1}{n}(h^{t+1})^* q^0 & \overset{\mathbf{. .}}{=} 0, \quad \frac{1}{n}(h^{t+1})^* \beta_0 \overset{\mathbf{. .}}{=} 0, \label{eq:Hc} \\
\frac{1}{n}(b^t)^* w &\doteq 0.  \label{eq:Bc}
\end{align}


\item For all $0 \leq r \leq t$, 
\begin{align}
\frac{1}{N}(h^{r+1})^* h^{t+1} &\overset{\mathbf{. .}}{=} \breve{E}_{r,t}, \label{eq:Hd} \\
\frac{1}{n}(b^r)^*b^t &\doteq \tilde{E}_{r,t}. \label{eq:Bd}
\end{align}

\item For all $0 \leq r \leq t$,
\begin{align}
\frac{1}{n}(q^{0})^* q^{t+1} & \overset{\mathbf{. .}}{=} \tilde{E}_{0,t+1}, \quad \frac{1}{n}(q^{r+1})^*q^{t+1} \overset{\mathbf{. .}}{=} \tilde{E}_{r+1,t+1},  \label{eq:He}  \\
\frac{1}{n}(m^r)^* m^t &\doteq \breve{E}_{r,t}. \label{eq:Be}
\end{align}

\item For all $0 \leq r \leq t$,  
\begin{equation}
\begin{split}
&\lambda_t \overset{\mathbf{. .}}{=} \hat{\lambda}_{t},  \quad  \frac{1}{n}(h^{t+1})^*q^{r+1}  \overset{\mathbf{. .}}{=} \hat{\lambda}_{r+1} \breve{E}_{r,t}, \\
& \frac{1}{n}(h^{r+1})^*q^{t+1} \overset{\mathbf{. .}}{=}  \hat{\lambda}_{t+1} \breve{E}_{r,t}, 
\end{split} 
\label{eq:Hf} 
\end{equation}
\begin{align}
\xi_t \doteq \hat{\xi}_{t},  \quad \frac{1}{n}(b^r)^*m^t\doteq \hat{\xi}_{t}\tilde{E}_{r,t} ,  \quad \frac{1}{n}(b^t)^*m^r \doteq \hat{\xi}_{r} \tilde{E}_{r,t}.  \label{eq:Bf}
\end{align}


\item Let $\mathbf{Q}_{t+1} := \frac{1}{n} Q_{t+1}^* Q_{t+1}$ and $\mathbf{M}_{t} := \frac{1}{n} M_{t}^* M_{t}$. Then,
\begin{align}
& P\left(\mathbf{Q}_{t+1} \text{ is  singular}\right) \leq t K_{t-1} e^{-\kappa_{t-1} \kappa n},  \label{eq:Qsing}\\
& P\left(\mathbf{M}_t \text{  is singular}\right) \leq t K_{t-1} e^{-\kappa_{t-1} \kappa n}. \label{eq:Msing}
\end{align}
When the inverses of $\mathbf{Q}_{t+1}, \mathbf{M}_t$ exist, for $1\leq i, j \leq t+1$,
\begin{equation}
\begin{split}
& P\Big( \Big \lvert[\mathbf{Q}_{t+1}^{-1}  -  (\tilde{C}^{t+1})^{-1}]_{i,j} \Big \lvert \geq \e \Big) \leq K  K'_{t-1} e^{-\kappa \kappa'_{t-1}n\e^2 }, \\
&  P\Big(\lvert \gamma^{t+1}_{i-1} - \hat{\gamma}^{t+1}_{i-1}  \lvert \geq \e \Big) \leq K t^4 K'_{t-1}  e^{{-\kappa \kappa'_{t-1}n\e^2}/{ t^9}}.
\end{split}  
\label{eq:Hg}
\end{equation} 
For $1 \leq  i, j  \leq t,$
\begin{equation}
\begin{split}
& P\Big( \Big \lvert [\mathbf{M}_t^{-1} - (\breve{C}^t)^{-1} ]_{i,j} \Big \lvert \geq \e\Big) \leq K  K_{t-1} e^{-\kappa \kappa_{t-1}n\e^2 },  \\
&  P\Big( \lvert\alpha^{t}_{i-1} - \hat{\alpha}^{t}_{i-1} \lvert \geq \e \Big) \leq K t^4 K_{t-1}  e^{{-\kappa \kappa_{t-1}n\e^2}/{ t^9}}.
\end{split}
\label{eq:Bg}   
\end{equation}
where $\hat{\gamma}^{t+1}$ and $\hat{\alpha}^{t}$ are defined in \eqref{eq:hatalph_hatgam_def}.
\item With $\sigma_{t+1}^{\perp}, \tau_{t}^{\perp}$ defined in \eqref{eq:sigperp_defs},
\begin{align} 
&P\Big( \Big \lvert\frac{1}{n}\norm{q^{t+1}_{\perp}}^2 - (\sigma_{t+1}^{\perp})^2 \Big \lvert \geq \e \Big) \leq K t^5 K'_{t-1}  e^{{-\kappa \kappa'_{t-1}n\e^2}/{t^{11}}}, \label{eq:Hh} \\
&P\Big( \Big \lvert\frac{1}{n}\norm{m^t_{\perp}}^2 - (\tau_{t}^{\perp})^2 \Big \lvert \geq \e \Big) \leq K t^5 K_{t-1} e^{{-\kappa \kappa_{t-1}n\e^2}/{ t^{11}}}. \label{eq:Bh}
\end{align}

\end{enumerate}
\label{lem:main_lem}
\end{lem}


\subsection{Remarks on Lemma \ref{lem:main_lem}} \label{subsec:lem_comments}

The proof of Theorem  \ref{thm:main_amp_perf} below only requires the concentration result in part $(b)$.(i) of Lemma \ref{lem:main_lem}, but the proof of part $(b)$.(i) hinges on the other parts of the lemma. The proof of Lemma \ref{lem:main_lem}, given in Section \ref{sec:main_lem_proof}, uses induction starting at time $t=0$,  sequentially proving the concentration results in parts $(a)-(h)$. The proof is long, but is based on a sequence of a few key steps which we summarize here.

The main result that needs to be proved (part $(b)$.(i), \eqref{eq:Hb1}) is that within the normalized sum of the pseudo-Lipschitz function $\phi_h$,  the  inputs $h^1, \ldots,  h^{t+1}$  can be effectively replaced by $\tau_0 \tilde{Z}_0, \ldots, \tau_t \tilde{Z}_t$, respectively. To prove this, we use the representation for $h^{t+1}$ given by Lemma  \ref{lem:hb_cond}, and show that the deviation term given by \eqref{lem:hb_cond} can be effectively dropped. In order to show that the deviation term can be dropped, we need to prove the concentration results in parts $(c)$ -- $(h)$ of Lemma  \ref{lem:main_lem}. Parts $(b)$.(ii), $(b)$.(iii), and $(b)$.(iv) of the lemma are used to establish the results in parts $(c)$ -- $(h)$.

\emph{The concentration constants $\kappa_t, K_t$}: The concentration results in Lemma \ref{lem:main_lem} and Theorem  \ref{thm:main_amp_perf} for  AMP iteration $t \geq 1$ are of the form $K_t e^{-\kappa_t n \e^2}$, where $\kappa_t, K_t$ are given in \eqref{eq:Kkappa_def}. Due to the inductive nature of the proof, the concentration results for step $t$ depend on those corresponding to all the previous steps --- this determines how $\kappa_t, K_t$ scale with $t$.

The $t!$ terms in  $\kappa_t, K_t$ can be understood as follows. Suppose that we want prove a concentration result for a quantity that can be expressed as a sum of $t$ terms with step indices $1,\ldots, t$. (A typical example  is $\Delta_{t+1,t}$ in \eqref{lem:hb_cond}.) For such a term, the deviation from the deterministic  concentrating value  is less than $\e$ if the deviation in each of the terms in the sum is less than $\e/t$. The induction hypothesis (for steps $1, \ldots, t$) is then used to bound the $\e/t$-deviation probability for each term in the sum. This introduces factors of $1/t$ and $t$ multiplying the exponent and pre-factor, respectively, in each step $t$ (see Lemma \ref{sums}), which results in the $t!$ terms in $K_t$ and $\kappa_t$.

The $(C_2)^t$ and $(c_2)^t$ terms in  $\kappa_t, K_t$  arise due to quantities  that can be expressed as the \emph{product} of  two terms, for each of which we have a concentration result available (due to the induction hypothesis). This can be used to bound the  $\e$-deviation probability of the product, but with a smaller exponent  and  a larger prefactor (see Lemma \ref{products}). Since this occurs in each step of the induction, the constants $K_t,\kappa_t$ have terms of the form $(C_2)^t, (c_2)^t$, respectively.

\emph{Comparison with earlier work}:  Lemmas \ref{lem:hb_cond} and \ref{lem:main_lem} are similar to the main technical lemma in \cite[Lemma $1$]{BayMont11}, in that they both analyze the behavior of similar functions and inner products arising in the AMP.  The key difference is that  Lemma  \ref{lem:main_lem}  replaces the asymptotic convergence statements in \cite{BayMont11} with  concentration inequalities. Other  differences from \cite[Lemma 1]{BayMont11} include:
\begin{itemize} 
\item[--] Lemma \ref{lem:main_lem} gives explicit values for the deterministic limits in parts $(c)$--$(h)$, which are needed in other parts of our proof.  

\item[--] Lemma \ref{lem:hb_cond} characterizes the the conditional distribution of the vectors $h^{t+1}$ and $b^t$ as the sum of an ideal distribution and a deviation term.    \cite[Lemma $1$(a)]{BayMont11}  is  a similar  distributional characterization of $h^{t+1}$ and $b^t$, however it does not use the ideal distribution.  We found that working with the ideal distribution throughout Lemma \ref{lem:main_lem} simplified our proof.

\end{itemize}

\subsection{Proof of Theorem \ref{thm:main_amp_perf}} \label{subsec:proof_thm1}

Applying Part $(b)$.(i) of Lemma \ref{lem:main_lem} to a pseudo-Lipschitz  function of the form $\phi_h(h^{t+1}, \beta_0)$, for $0 \leq t \leq T^*$ we have
\be
P\Big(\Big\lvert \frac{1}{N} \sum_{i=1}^N \phi_h(h^{t+1}_i, \beta_{0_i}) - \mathbb{E}[\phi_h(\tau_t Z, \beta)] \Big \lvert \geq \e \Big) \leq K_t e^{-\kappa_t n \e^2},
\label{eq:Prphi_h}
\ee
where the random variables $Z\sim{N}(0,1)$ and $\beta \sim p_{\beta}$ are independent.  (Though Lemma \ref{lem:main_lem} is stated for $1 \leq t \leq T^*$,  one can see that \eqref{eq:Prphi_h} holds for $t=0$ by considering the pseudo-Lipschitz (PL) function $\phi_h(h^{1}, \beta_0)$.) Now let $\phi_h(h^{t+1}_i, \beta_{0_i}) := \phi(\eta_t(\beta_{0_i} - h^{t+1}_i), \beta_{0_i}),$ where $\phi$ is the PL function in the statement of the theorem. The function $\phi_h(h^{t+1}_i, \beta_{0_i})$ is PL since $\phi$ is PL and $\eta_t$ is Lipschitz. We therefore obtain
\be
\begin{split}
&P\Big(\Big \lvert  \frac{1}{N} \sum_{i=1}^N \phi(\eta_t(\beta_{0_i} - h^{t+1}_i), \beta_{0_i})   \\
& \hspace{0.5in }- \mathbb{E}[ \phi(\eta_t(\beta - \tau_t Z), \beta)] \big  \lvert \geq \e \Big)   \leq K_t e^{-\kappa_t n \e^2}. \nonumber
\end{split}
\ee
The proof is completed by noting from \eqref{eq:amp2} and \eqref{eq:hqbm_def_AMP} that 
$\beta^{t+1} = \eta_t(A^* z^t + \beta^t) = \eta_t(\beta_0 - h^{t+1})$.
\hfill \IEEEQED


\section{Proof of Lemma \ref{lem:main_lem}} \label{sec:main_lem_proof}

\subsection{Mathematical Preliminaries} \label{subsec:mathpre}
Some of the results below can be found in \cite[Section III.G]{BayMont11}, but we summarize them here for completeness.

\begin{fact}
 Let  $u \in \mathbb{R}^N$ and $v \in \mathbb{R}^n$ be deterministic vectors, and let $\tilde{A} \in \mathbb{R}^{n \times N}$ be a matrix with independent  $\mc{N}(0, 1/n)$ entries. Then:
 
(a) \begin{equation*}
\tilde{A} u \overset{d}{=} \frac{1}{\sqrt{n}} \norm{u} Z_u  \quad \text{ and }  \quad \tilde{A}^*v \overset{d}{=} \frac{1}{\sqrt{n}} \norm{v} Z_v,
\end{equation*}
where $Z_u \in \mathbb{R}^n$ and $Z_v \in \mathbb{R}^N$ are i.i.d.\ standard Gaussian random vectors. 

(b) Let $\mc{W}$  be a $d$-dimensional subspace of $\mathbb{R}^n$ for $d \leq n$. Let $(w_1, ..., w_d)$ be an orthogonal basis of $\mc{W}$ with 
$\norm{w_\ell}^2 = n$ for
$\ell \in [d]$, and let  $\mathsf{P}^{\parallel}_\mc{W}$ denote  the orthogonal projection operator onto $\mc{W}$.  Then for $D = [w_1\mid \ldots \mid w_d]$, we have 
$\mathsf{P}^{\parallel}_{\mc{W}} \tilde{A} u \overset{d}{=}   \frac{1}{\sqrt{n}} \norm{u} \mathsf{P}^{\parallel}_{\mc{W}} Z_u\overset{d}{=}  \frac{1}{\sqrt{n}} \norm{u} Dx$ where $x \in \mathbb{R}^d$ is a random vector with i.i.d.\ $\mc{N}(0, 1/n)$ entries. 
\label{fact:gauss_p0}
\end{fact}

\begin{fact}[Stein's lemma]
For zero-mean jointly Gaussian random variables $Z_1, Z_2$, and any function $f:\mathbb{R} \to \mathbb{R}$ for which $\expec[Z_1 f(Z_2)]$ and $\expec[f'(Z_2)]$  both exist, we have $\expec[Z_1 f(Z_2)] = \expec[Z_1Z_2] \expec[f'(Z_2)]$.
\label{fact:stein}
\end{fact}

\begin{fact}
Let $v_1, \ldots, v_t$ be a sequence of vectors in $\mathbb{R}^n$ such that for $i \in [t]$,
$\frac{1}{n} \norm{v_i - \mathsf{P}^{\parallel}_{i-1}(v_i)}^2 \geq c$,
where $c$ is a positive constant that does not depend on $n$, and $\mathsf{P}^{\parallel}_{i-1}$ is the orthogonal projection onto the span of $v_1, \ldots, v_{i-1}$. Then the matrix $C \in \mathbb{R}^{t \times t}$ with $C_{ij} = v^*_i v_j / n$ has minimum eigenvalue $\lambda_{\min} \geq c'_t$, where $c'_t$ is a  positive constant (not depending on $n$).
\label{fact:eig_proj}
\end{fact}

\begin{fact}
Let $g:\mathbb{R} \to \mathbb{R}$ be a bounded function. For all $s, \Delta \in \mathbb{R}$ such that $g$ is differentiable in the closed interval between $s$ and $s+\Delta$, there exists a constant $c > 0$ such that
$\left \lvert g(s + \Delta) - g(s) \right \lvert \leq c \abs{\Delta}$.
\label{fact:lip_deriv}
\end{fact}

We also use several concentration results listed in Appendices \ref{app:conc_lemma} and  \ref{app:gauss_lemmas}, with proofs provided for the results that are non-standard. Some of these may be of independent interest, e.g., concentration of sums of a pseudo-Lipschitz function of sub-Gaussians (Lemma \ref{lem:PLsubgaussconc}).

The proof of  Lemma \ref{lem:main_lem}. proceeds by induction on $t$.  We label as $\mathcal{H}_{t+1}$ the results \eqref{eq:Ha}, \eqref{eq:Hb1}, \eqref{eq:Hb2}, \eqref{eq:Hc}, \eqref{eq:Hd}, \eqref{eq:He}, \eqref{eq:Hf}, \eqref{eq:Qsing}, \eqref{eq:Hg}, \eqref{eq:Hh} and similarly as $\mathcal{B}_t$ the results \eqref{eq:Ba}, \eqref{eq:Bb1}, \eqref{eq:Bb2}, \eqref{eq:Bc}, \eqref{eq:Bd}, \eqref{eq:Be}, \eqref{eq:Bf}, \eqref{eq:Msing}, \eqref{eq:Bg}, \eqref{eq:Bh}.  The proof consists of showing four steps:

\begin{enumerate}

\item $\mathcal{B}_0$ holds.

\item $\mathcal{H}_1$ holds.

\item If $\mathcal{B}_r, \mathcal{H}_s$ holds for all $r < t $ and $s \leq t $, then $\mathcal{B}_t$ holds.

\item if $\mathcal{B}_r, \mathcal{H}_s$ holds for all $r \leq t $ and $s \leq t$, then $\mathcal{H}_{t+1}$ holds.
\end{enumerate}

For the proofs of parts $(b)$.(ii) and $(b)$.(iv),  for brevity we assume that the functions $\psi_h$ and $\psi_b$ are differentiable everywhere. The case where they are not differentiable at a finite number of points involves additional technical details; see Appendix \ref{supA}.

\subsection{Step 1: Showing $\mc{B}_0$ holds}

We wish to show results (a)-(h) in \eqref{eq:Ba}, \eqref{eq:Bb1}, \eqref{eq:Bb2}, \eqref{eq:Bc}, \eqref{eq:Bd}, \eqref{eq:Be}, \eqref{eq:Bf}, \eqref{eq:Msing}, \eqref{eq:Bg}, \eqref{eq:Bh}.

\textbf{(a)} We have
\begin{align*}
&P \Big(\frac{1}{n}\norm{\Delta_{{0,0}}}^2 \geq \epsilon\Big) \\
&\overset{(a)}{\leq} P \Big(\Big \lvert \frac{1}{\sqrt{n}} \norm{q^0} - \sigma^{\perp}_0\Big \lvert  \geq \sqrt{\frac{\epsilon}{2}} \Big) + P\Big(\Big\lvert \frac{1}{\sqrt{n}}\norm{Z'_0} - 1\Big \lvert \geq \sqrt{\frac{\epsilon}{2}} \Big) \\
&\overset{(b)}{\leq} Ke^{-\kappa \varepsilon_2 n \e/4} + 2e^{-n \e/8}.
\end{align*}
Step (a) is obtained using the definition of $\Delta_{0,0}$ in  \eqref{eq:D00}, and then applying Lemma \ref{products}. For step (b), we use \eqref{eq:qassumption},  Lemma \ref{sqroots}, and Lemma \ref{subexp}.

\textbf{(b).(iii)}
For $t=0$, the LHS of \eqref{eq:Bb1} can be bounded as
\begin{equation}
\begin{split}
& P\Big(\Big \lvert \frac{1}{n} \sum_{i=1}^n \phi_b(b^0_i, w_{i}) - \mathbb{E}[\phi_b(\sigma_0 \breve{Z}_0, W)] \Big \lvert \geq \epsilon \Big) \\
&\overset{(a)}{=} P\Big(\Big \lvert \frac{1}{n} \sum_{i=1}^n \phi_b(\sigma_0 Z'_{0_i} + [\Delta_{0,0}]_i, w_{i})  \\ 
& \hspace{1in} - \mathbb{E}[\phi_b(\sigma_0 \breve{Z}_0, W)] \Big\lvert \geq \epsilon \Big)  \\
&\overset{(b)}{\leq} P\Big(\Big \lvert \frac{1}{n} \sum_{i=1}^n \phi_b(\sigma_0 Z'_{0_i}, w_{i})  \\ 
& \hspace{1in}- \mathbb{E}[\phi_b(\sigma_0 \breve{Z}_0, W)] \Big \lvert \geq \frac{\epsilon}{2}\Big) \\
& \ + P\Big(\Big \lvert \frac{1}{n} \sum_{i=1}^n \Big[\phi_b(\sigma_0 Z'_{0_i} + [\Delta_{0,0}]_i, w_{i})  \\ 
& \hspace{1in} - \phi_b(\sigma_0 Z'_{0_i}, w_{i})\Big] \Big \lvert \geq \frac{\epsilon}{2} \Big).
\end{split}
\label{eq:phib_0}
\end{equation}
Step (a) uses the conditional distribution of $b^0$ given in \eqref{eq:Ba_dist}, and step (b) follows from Lemma \ref{sums}.  
Label the terms on the RHS of \eqref{eq:phib_0} as $T_1$ and $T_2$.   Term $T_1$ can be upper bounded by $K e^{-\kappa n \e^2}$ using Lemma \ref{lem:PLsubgaussconc}.  We now show a similar upper bound for term $T_2$.
\begin{align}
&T_2 \nonumber \\
&\overset{(a)}{\leq} P\Big(\frac{L}{n} \sum_{i=1}^n  (1 + 2\abs{\sigma_0 Z'_{0_i}} + \abs{\Delta_{{0,0}_i}} + 2\abs{w_i}) \abs{{\Delta_{0,0}}_i} \geq \frac{\epsilon}{2} \Big) \nonumber \\
& \overset{(b)}{\leq} P\Big( \frac{\norm{\Delta_{0,0}}}{\sqrt{n}} \, \norm{ \frac{1}{\sqrt{n}} + \frac{\abs{\Delta_{0,0}}}{\sqrt{n}} + 2\sigma_0 \frac{\abs{Z'_{0}}}{\sqrt{n}} + 2 \frac{\abs{w}}{\sqrt{n}} } \geq \frac{\epsilon}{2 L} \Big)  \nonumber \\
&\overset{(c)}{\leq} P\Big( \frac{\norm{\Delta_{0,0}}}{\sqrt{n}} \Big(1 + \frac{\norm{\Delta_{0,0}}}{\sqrt{n}} + 2\sigma_0 \frac{\norm{Z'_{0}}}{\sqrt{n}} + 2 \frac{\norm{w}}{\sqrt{n}}\Big)\geq \frac{\epsilon}{4 L} \Big),   \label{eq:B1func1eq1}
\end{align}
where inequality (a) holds because $\phi_b$ is pseudo-Lipschitz with constant $L >0$. Inequality (b) follows from Cauchy-Schwarz (with $\mathbf{1}$ denoting the all-ones vector). Inequality $(c)$ is obtained by applying Lemma \ref{lem:squaredsums}.
From \eqref{eq:B1func1eq1},  we have
\be
\begin{split}
T_2 & \leq  P\Big( \frac{\norm{w}}{\sqrt{n}} \geq \sigma + 1 \Big)  + P\Big( \frac{\norm{Z'_{0}}}{\sqrt{n}} \geq 2 \Big)  \\
& \qquad + P\Big( \frac{\norm{\Delta_{0,0}}}{\sqrt{n}} \geq \frac{\e  \min\{1, (4 L)^{-1}\}}{4 + 4 \sigma_0 + 2 \sigma}\Big)  \\ 
 & \overset{(a)}{\leq} K e^{-\kappa n} + e^{-n} + K e^{-\kappa n \e^2},
\end{split}
\ee
where to obtain $(a)$, we use assumption \eqref{eq:wassumption}, Lemma \ref{subexp}, and  $\mc{B}_0 (a)$ proved above.

\textbf{(b).(iv)} For $t=0$, the probability in \eqref{eq:Bb2} can be bounded as 
\be
\begin{split}
&P\Big(\Big \lvert \frac{1}{n} \sum_{i=1}^n \psi_b(b^{0}_i, w_i) - \mathbb{E}[\psi_b( \sigma_{0} \breve{Z}_{0}, W)] \Big \lvert \geq \e \Big)  \\
&\overset{(a)}{=} P\Big(\Big \lvert \frac{1}{n} \sum_{i=1}^n \psi_b(\sigma_0 Z'_{0_i} + [\Delta_{0,0}]_i, w_i)  \\
& \hspace{1in} - \mathbb{E}[\psi_b( \sigma_{0} \breve{Z}_{0}, W)] \Big \lvert \geq \e \Big)  \\ 
&\overset{(b)}{\leq}  P \Big(\Big \lvert \frac{1}{n} \sum_{i=1}^n [\psi_b(\sigma_0 Z'_{0_i} + [\Delta_{0,0}]_i, w_i) \\
& \quad - \psi_b(\sigma_0 Z'_{0_i}, w_i)] \Big \lvert \geq \frac{\e}{2} \Big) \label{eq:Bj12}  \\
& \quad + P \Big(\Big \lvert \frac{1}{n} \sum_{i=1}^n \psi_b(\sigma_0 Z'_{0_i}, w_i) - \mathbb{E}[ \psi_b( \sigma_{0} \breve{Z}_{0}, W)] \Big\lvert \geq \frac{\e}{2} \Big).
\end{split}
\ee
Step $(a)$ uses the conditional distribution of $b^0$ given in \eqref{eq:Ba_dist}, and step (b) follows from Lemma \ref{sums}. Label the two terms on the RHS of \eqref{eq:Bj12} as $T_1$ and $T_2$, respectively.  We now show that each term is  bounded by $Ke^{-\kappa n \e^2}$.   Since $\abs{\psi_b}$ is bounded  (say it takes values in an interval of length $B$), the term $T_2$ can be bounded using Hoeffding's inequality (Lemma \ref{lem:hoeff_lem}) by $2 e^{-n{\e^2}/(2B^2)}$.

Next, consider $T_1$. Let $\Pi_0$ be the event under consideration, so that $T_1 = P(\Pi_0)$, and define an event $\mc{F}$ as follows. 
\be
\mc{F} := \Big\{ \Big \lvert \frac{1}{\sqrt{n}}\norm{q^0} - \sigma_0 \Big\lvert \geq \e_0  \Big\}, \label{eq:BFdef}
\ee
where  $\e_0 > 0$ will be specified later. With this definition, 
\begin{align}
T_1 =P(\Pi_0) & \leq P(\mc{F}) + P(\Pi_0 | \mc{F}^c) \nonumber \\
& \leq K e^{-\kappa n \e_0^2} + P(\Pi_0 | \mc{F}^c). \label{eq:Bprobcondition}
\end{align}
The final inequality in \eqref{eq:Bprobcondition} follows from the concentration of $\norm{q^0}$ in \eqref{eq:qassumption}. To bound the last term $P(\Pi_0| \mc{F}^c)$, we write it as
\begin{equation}
\begin{split}
P(\Pi_0 | \mc{F}^c) = \mathbb{E}[ \mathsf{I}\{\Pi_0\} | \mc{F}^c]  &
 = \mathbb{E}[\mathbb{E}[ \mathsf{I}\{\Pi_0\} | \mc{F}^c, \mscrs_{0, 0}] \mid \mc{F}^c]  \\ 
& = \mathbb{E}[P(\Pi_0 | \mc{F}^c, \mscrs_{0, 0}) \mid \mc{F}^c], 
\end{split} 
\label{eq:Bprobcondition2}
\end{equation}
where $ \mathsf{I}\{\cdot \}$ denotes the indicator function, and $P\left(\Pi_0 | \mc{F}^c, \mscrs_{0, 0}\right)$ equals
\be
\begin{split} 
 & P \Big( \Big \lvert \frac{1}{n}  \sum_{i=1}^n \Big[\psi_b\Big(\frac{\norm{q^0}}{\sqrt{n}}  Z'_{0_i}, w_i\Big)  \\
&  \hspace{1in} - \psi_b(\sigma_0 Z'_{0_i}, w_i) \Big]  \Big \lvert \geq \frac{\e}{2}  \Big \lvert \mc{F}^c, \mscrs_{0, 0}  \Big). 
 \end{split}
\label{eq:Bprob1}
\ee
 To obtain \eqref{eq:Bprob1}, we use the fact that $\sigma_0 Z'_{0_i} + [\Delta_{0,0}]_i = \frac{1}{\sqrt{n}} \norm{q^0} Z'_{0_i}$ which follows from the definition of $\Delta_{0,0}$ in Lemma \ref{lem:hb_cond}. Recall from Section \ref{sec:cond_dist_lemma}  that $\mscrs_{0, 0}$ is the sigma-algebra generated by $\{ w, \beta_0, q^0 \}$; so in \eqref{eq:Bprob1}, only $Z'_{0}$ is random ---  all other terms are in $\mscrs_{0, 0}$.    We now derive a bound for the upper tail of the probability in \eqref{eq:Bprob1}; the lower tail bound is similarly obtained.  From here on, we suppress the conditioning on $\mc{F}^c, \mscrs_{0, 0}$  for brevity.
 
 Define the shorthand  $\textsf{diff}(Z'_{0_i}) :=\psi_b(\frac{1}{\sqrt{n}} \norm{q^0} Z'_{0_i}, w_i) -  \psi_b(\sigma_0 Z'_{0_i}, w_i)$. Since $\psi_b$ is bounded, so is 
 $\textsf{diff}(Z'_{0_i})$. Let $\abs{\psi_b} \leq B/2$, so that $\abs{\textsf{diff}(Z'_{0_i})} \leq B$ for all $i$. Then the  upper tail of the probability in \eqref{eq:Bprob1} can be written as 
 \be
  \label{eq:diffi_bound}
 P\Big( \frac{1}{n} \sum_{i=1}^n \textsf{diff}(Z'_{0_i}) - \expec[\textsf{diff}(Z'_{0_i})] \geq \frac{\e}{2} - \frac{1}{n}\sum_{i=1}^n   \expec[\textsf{diff}(Z'_{0_i})] \Big).
 \ee
We now show that $\abs{\expec[\textsf{diff}(Z'_{0_i})]} \leq  \frac{1}{4} \e$ for all $i \in [n]$.   Denoting the standard normal density by $\phi$, we have
 \ben
 \begin{split}
 \abs{\expec[\textsf{diff}(Z'_{0_i})] }& \leq  \int_{\mathbb{R} } \phi(z) \,  \abs{\textsf{diff}(z)} dz \\
  &\overset{(a)}{\leq} \int_{\mathbb{R} } \phi(z) \, C \Big \lvert z\Big(\frac{\norm{q^0}}{\sqrt{n}} - \sigma_0\Big) \Big \lvert dz \stackrel{(b)}{\leq} 2C\e_0.
 \end{split}
 \een
 The above is bounded by $\frac{1}{4}\e$ if we choose $\e_0 \leq \e/8C$.  In the chain above, $(a)$ follows by Fact \ref{fact:lip_deriv} for a suitable constant $C>0$ as $\psi_b$ is bounded and assumed to be differentiable. Step $(b)$ follows since $\abs{ \frac{1}{\sqrt{n}}\norm{q^0} - \sigma_0} \leq  \e_0$ under $\mc{F}^c$. 
 
 The probability in \eqref{eq:diffi_bound} can then be bounded using Hoeffding's inequality (Lemma \ref{lem:hoeff_lem}):
 \[ 
  P\Big( \frac{1}{n} \sum_{i=1}^n \textsf{diff}(Z'_{0_i}) - \expec[\textsf{diff}(Z'_{0_i})] \geq \frac{\e}{4} \,  \Big{\lvert} \, \mc{F}^c, \mscrs_{0, 0} \Big) \leq e^{-\frac{n\e^2}{(8B^2)}}.
 \]
 Substituting in \eqref{eq:Bprob1} and using a similar bound for the lower tail, we have shown via \eqref{eq:Bprobcondition2} that $P (\Pi_0 \mid  \mc{F}^c) \leq  2 e^{-n\e^2/(8B^2)}$. 
Using this in \eqref{eq:Bprobcondition} with $\e_0 \leq \e/8C$ proves that the first term in \eqref{eq:Bj12} is  bounded by $K e^{-n \kappa \e^2}$. 
 
 \textbf{(c)} The function $\phi_b(b^0_i, w_{i}) := b^0_i w_i  \in PL(2)$ by Lemma \ref{lem:Lprods}.  By $\mathcal{B}_0 (b).\text{(iii)}$,
\begin{align*}
P\Big(\Big \lvert \frac{1}{n}(b^0)^*w - \mathbb{E}[\sigma_0 \breve{Z}_0 W] \Big \lvert \geq \epsilon \Big) &\leq K e^{-\kappa n\e^2}.
\end{align*}
This result follows since $\mathbb{E}[\sigma_0 \breve{Z}_0 W] = 0$ by the independence of $W$ and $\hat{Z}_0$.


\textbf{(d)} The function $\phi_b(b^0_i, w_{i}) := (b^0_i)^2 \in PL(2)$ by Lemma \ref{lem:Lprods}.  By $\mathcal{B}_0 (b).\text{(iii)}$,
\begin{align*}
P\Big(\Big \lvert \frac{1}{n}\norm{b^0}^2 - \mathbb{E}[(\sigma_0 \breve{Z}_0)^2]\Big \lvert \geq \epsilon \Big) &\leq K e^{-\kappa  n\e^2}.
\end{align*}
This result follows since $\mathbb{E}[(\sigma_0 \hat{Z}_0)^2] = \sigma_0^2$.


\textbf{(e)} Since $g_0$ is Lipschitz, the function $\phi_b(b^0_i, w_{i}) := (g_0(b^0_i, w_{i}))^2 \in PL(2)$ by Lemma \ref{lem:Lprods}.  By $\mathcal{B}_0 (b).\text{(iii)}$,
\begin{align*}
P\Big(\Big \lvert \frac{1}{n}\norm{m^0}^2 - \mathbb{E}[(g_0(\sigma_0 \breve{Z}_0, W))^2] \Big \lvert \geq \epsilon \Big) &\leq K e^{-\kappa n\e^2}.
\end{align*}
This result follows since $\mathbb{E}[(g_0(\sigma_0 \breve{Z}_0, W))^2] = \tau_0^2$ by \eqref{eq:sigmatdef}.


\textbf{(f)} The concentration of $\xi_0$ around $\hat{\xi}_0$ follows from $\mathcal{B}_0 (b).$(iv) applied to the function $\psi_b(b^0_i,w_i):=g_0'(b^0_i,w_i)$.
Next, the function $\phi_b(b^0_i, w_{i}) := b^0_i \, g_0(b^0_i, w_{i}) \in PL(2)$ by Lemma \ref{lem:Lprods}.  Then by $\mathcal{B}_0 (b).\text{(iii)}$,
\begin{align*}
P\Big(\Big \lvert \frac{1}{n}(b^0)^* m^0 - \mathbb{E}[\sigma_0 \breve{Z}_0 g_0(\sigma_0 \breve{Z}_0, W)]\Big \lvert \geq \epsilon\Big) &\leq K e^{-\kappa n\e^2}.
\end{align*}
This result follows since $\mathbb{E}[\sigma_0 \breve{Z}_0 g_0(\sigma_0 \breve{Z}_0, W)] = \sigma_0^2 \mathbb{E}[ g'_0(\sigma_0 \breve{Z}_0, W)] = \hat{\xi}_0 \tilde{E}_{0,0}$ by Stein's Lemma given in Fact \ref{fact:stein}.


\textbf{(g)} Nothing to prove.


\textbf{(h)} The result is equivalent to $\mathcal{B}_0 (e)$ since $\norm{m^0_{\perp}} = \norm{m^0}$ and $(\tau_{0}^{\perp})^2 = \tau_0^2$.

\subsection{Step 2: Showing $\mc{H}_1$ holds}

We wish to show results (a)--(h) in \eqref{eq:Ha}, \eqref{eq:Hb1}, \eqref{eq:Hb2}, \eqref{eq:Hc}, \eqref{eq:Hd}, \eqref{eq:He}, \eqref{eq:Hf}, \eqref{eq:Qsing}, \eqref{eq:Hg}, \eqref{eq:Hh}.


\textbf{(a)} From the definition of $\Delta_{1,0}$ in \eqref{eq:D10} of Lemma \ref{lem:hb_cond}, we have 
\be
\begin{split}
\Delta_{1,0} & \stackrel{d}{=} Z_0\Big(\frac{\norm{m^0}}{\sqrt{n}} - \tau^{\perp}_0\Big) -\frac{\norm{m^0}\tilde{q}^0  \bar{Z}_0}{\sqrt{n}}  \\
& \qquad + q^0 \Big(\frac{n}{\norm{q^0}^2}\Big)\Big(\frac{(b^0)^*m^0}{n} - \frac{\xi_0 \norm{q^0}^2}{n}\Big). \label{eq:newDelta10}
\end{split}
\ee
where $\tilde{q}^0=q^0/\norm{q^0}$,
and $\bar{Z}_0 \in \mathbb{R}$ is a standard Gaussian random variable. The equality in \eqref{eq:newDelta10} is obtained using Fact \ref{fact:gauss_p0} to write
$ \mathsf{P}^{\parallel}_{q^0} Z_0 \overset{d}{=}  \tilde{q}^0 \bar{Z}_0$.
Then, from \eqref{eq:newDelta10} we have
\begin{equation}
\begin{split}
P\Big(\frac{1}{N}\norm{\Delta_{{1,0}}}^2 \geq \epsilon\Big)& \overset{(a)}{\leq} P \Big(\Big \lvert \frac{\norm{m^0}}{\sqrt{n}} - \tau_0 \Big\lvert \frac{\norm{Z_{0}}}{\sqrt{N}} \geq \sqrt{\frac{\epsilon}{9}} \Big) \\
 &+ P \Big(\frac{\norm{m^0}\abs{\bar{Z}_0}}{\sqrt{nN}} \geq \sqrt{\frac{\epsilon}{9}}\Big)  \\
 & + P\Big(\Big\lvert \frac{(b^0)^*m^0}{\sqrt{n} \norm{q^0}} -  \frac{\xi_0 \norm{q^0}}{\sqrt{n}}\Big \lvert  \geq \sqrt{\frac{\epsilon}{9 \delta}} \Big). 
 \end{split} 
 \label{eq:del10_split}
\end{equation}
Step (a) follows from Lemma \ref{lem:squaredsums} applied to  $\Delta_{1,0}$ in \eqref{eq:newDelta10} and Lemma \ref{sums}. Label the terms on the RHS of \eqref{eq:del10_split} as $T_1-T_3$.  To complete the proof, we show that each term is  bounded by $Ke^{-\kappa n \e}$ for generic positive constants $K, \kappa$ that do not depend on $n,\e$.  

Indeed, $T_1 \leq K e^{-\kappa n \e}$ using Lemma \ref{products}, Lemma \ref{sqroots}, result $\mathcal{B}_0 (e)$, and Lemma \ref{subexp}.  Similarly, $T_2 \leq K e^{-\kappa  n \e}$ using Lemma \ref{products}, Lemma \ref{sqroots}, result $\mathcal{B}_0 (e)$, and Lemma \ref{lem:normalconc}.  Finally,
\begin{align*}
T_3 &\overset{(a)}{\leq} P \Big(\Big \lvert \frac{(b^0)^* m^0}{n} \cdot \frac{\sqrt{n}}{\norm{q^0}} - \hat{\xi}_0 \sigma_0 \Big \lvert \geq \frac{1}{2}\sqrt{\frac{\epsilon}{9\delta}} \Big) \\
&\quad + P\Big(\Big \lvert\xi_0 \frac{\norm{q^0}}{\sqrt{n}} - \hat{\xi}_0 \sigma_0 \Big \lvert \geq \frac{1}{2} \sqrt{\frac{\epsilon}{9 \delta}} \Big) \\
&\overset{(b)}{\leq}  2K e^{\frac{-\kappa n \e}{ \delta\max(1, \hat{\xi}_0^2 \sigma_0^4, \sigma_0^{-2})}} + 2K e^{\frac{-\kappa  n \e}{\delta \max(1, \hat{\xi}_0^2, \sigma_0^{2})}}.
\end{align*}
Step (a) follows from Lemma \ref{sums}, and step (b) from Lemma \ref{products}, $\mathcal{B}_0 (f)$, the concentration of $\norm{q^0}$ given in \eqref{eq:qassumption}, and Lemma \ref{inverses}. 

\textbf{(b)(i)} The proof of \eqref{eq:Hb1} is similar to analogous $\mc{B}_0 (b)$(iii) result \eqref{eq:Bb1}.

\textbf{(b)(ii)}
First,
\begin{align}
&P \Big(\Big  \lvert \frac{1}{N} \sum_{i=1}^N \psi_h(h^{1}_i, \beta_{0_i}) - \mathbb{E}[\psi_h( \tau_{0} \tilde{Z}_{0}, \beta)] \Big \lvert \geq \e \Big) \nonumber \\
&\overset{(a)}{=}  P\Big(\Big  \lvert \frac{1}{N} \sum_{i=1}^N \psi_h(\tau_0 Z_{0_i} + [\Delta_{1,0}]_i, \beta_{0_i})  \nonumber \\
& \hspace{1in} -  \mathbb{E}[\psi_h(\tau_{0} \tilde{Z}_{0}, \beta)]\Big \lvert \geq \e \Big)\nonumber  \\ 
&\overset{(b)}{\leq}  P \Big(\Big  \lvert \frac{1}{N} \sum_{i=1}^N \left[\psi_h(\tau_0 Z_{0_i}+ [\Delta_{1,0}]_i, \beta_{0_i}) \right.  \nonumber \\
& \hspace{1in}  \left. - \psi_h(\tau_0 Z_{0_i}, \beta_{0_i})\right] \Big \lvert \geq \frac{\e}{2} \Big) \nonumber \\
& \quad + P \Big(\Big  \lvert \frac{1}{N} \sum_{i=1}^N \psi_h(\tau_0 Z_{0_i}, \beta_{0_i}) -  \mathbb{E}[\psi_h(\tau_{0} \tilde{Z}_{0}, \beta)] \Big\lvert \geq \frac{\e}{2}\Big). \label{eq:Hj12}
\end{align}
Step $(a)$ follows from the conditional distribution of $h^1$ stated in \eqref{eq:Ha_dist} and step $(b)$ from Lemma \ref{sums}.  Label the two terms on the RHS as $T_1$ and $T_2$.  Term $T_2$ is upper bounded by $K e^{-\kappa n \e^2}$ by Hoeffding's inequality (Lemma \ref{lem:hoeff_lem}).   To complete the proof, we show that $T_1$ has the same bound.

Consider the first term in \eqref{eq:Hj12}.  From the  definition of $\Delta_{1,0}$ in Lemma \ref{lem:hb_cond},
\begin{align}
\tau_0 Z_{0_i}  + [\Delta_{1,0}]_i = \frac{1}{\sqrt{n}}\norm{m^0} [(\mathsf{I} -  \mathsf{P}^{\parallel}_{q^0}) Z_{0}]_i + u_i,  \label{eq:vdef}
\end{align}
where
\[u_{i} := q^0_i \Big( \frac{(b^0)^*m^0}{\norm{q^0}^2} - \xi_0 \Big)\].
For $\e_0 > 0$ to be specified later, define event $\mc{F}$ as
\be
\mc{F} := \Big\{\Big \lvert \frac{\norm{m^0}}{\sqrt{n}} - \tau_0 \Big \lvert \geq \e_0 \Big\}  \cup \Big\{ \Big \lvert \frac{(b^0)^*m^0}{n} - \frac{\xi_0  \norm{q^0}^2 }{n} \Big \lvert \geq \e_0 \Big\}. \label{eq:Fdef}
\ee
Denoting the event we are considering in $T_1$ by $\Pi_1$,  so that $T_1 = P(\Pi_1)$, we write
\be \label{eq:PPi0} T_1= P(\Pi_1)  \leq P(\mc{F}) + P(\Pi_1 \mid \mc{F}^c) \leq K e^{-\kappa n \e_0^2} + P(\Pi_1 \mid \mc{F}^c)
 \ee
where the last inequality is by $\mc{B}_0 (e), \mc{B}_0 (f)$ and the concentration assumption \eqref{eq:qassumption} on $q^0$. Writing  $P(\Pi_1| \mc{F}^c)= 
\expec[P(\Pi_1 | \mc{F}^c, \mscrs_{1, 0}) \mid \mc{F}^c]$, we now bound $P(\Pi_1 | \mc{F}^c, \mscrs_{1, 0})$.  In what follows, we drop the explicit conditioning on $\mc{F}^c$ and  
$\mscrs_{1, 0}$ for brevity. Then using Lemma \ref{sums}, $P(\Pi_1 | \mc{F}^c, \mscrs_{1, 0})$ can be written as 
\begin{equation}
\begin{split}
& P \Big(\Big \lvert \frac{1}{N}  \sum_{i=1}^N \Big[\psi_h\Big(\frac{\norm{m^0}}{\sqrt{n}} [(\mathsf{I} -  \mathsf{P}^{\parallel}_{q^0}) Z_{0}]_i + u_i, \beta_{0_i}\Big)  \\
& \hspace{0.7in} - \psi_h(\tau_0 Z_{0_i}, \beta_{0_i})\Big] \Big\lvert \geq \frac{\e}{2} \Big) \\ 
& \leq P \Big(\Big\lvert \frac{1}{N} \sum_{i=1}^N \psi_h\Big(\frac{\norm{m^0}}{\sqrt{n}} [(\mathsf{I} -  \mathsf{P}^{\parallel}_{q^0}) Z_{0}]_i + u_i, \beta_{0_i}\Big)  \\
& \hspace{0.7in} - \psi_h\Big(\frac{\norm{m^0}}{\sqrt{n}} Z_{0_i} + u_i, \beta_{0_i}\Big) \Big \lvert \geq \frac{\e}{4} \Big)  \\
& \quad  + P \Big(\Big \lvert \frac{1}{N} \sum_{i=1}^N \psi_h\Big(\frac{\norm{m^0}}{\sqrt{n}} Z_{0_i} + u_i, \beta_{0_i}\Big) 
 \\ 
& \hspace{0.7in}  -\psi_h(\tau_0 Z_{0_i}, \beta_{0_i}) \Big \lvert \geq \frac{\e}{4} \Big). \label{eq:prob2}
\end{split}
\end{equation}
%
Note that in \eqref{eq:prob2}, only $Z_{0}$ is random  as the other terms are all in $\mscrs_{1, 0}$.   Label the two terms on the RHS of \eqref{eq:prob2} as $T_{1,a}$ and $T_{1,b}$.  To complete the proof we show that both are bounded by $K e^{-\kappa n \e^2}$.

First consider $T_{1,a}$.
\ben
\begin{split}
& T_{1,a} \overset{(a)}{\leq} P\Big(\frac{C}{N} \sum_{i=1}^N\Big \lvert \frac{\norm{m^0}}{\sqrt{n}} [\mathsf{P}^{\parallel}_{q^0} Z_{0}]_i \Big \lvert \geq \frac{\e}{4} \Big)\\
& \overset{(b)}{\leq} P\Big(\frac{C}{N} \sum_{i=1}^N \abs{\tau_0 + \e_0}\Big \lvert [\mathsf{P}^{\parallel}_{q^0} Z_{0}]_i \Big \lvert \geq \frac{\e}{4} \Big) \\
&  \overset{(c)}{\leq} P\Big(\frac{C}{N} \sum_{i=1}^N \frac{\abs{q^0_i}}{\norm{q^0}} \abs{Z} \geq \frac{\e}{4 \abs{\tau_0 + \e_0}} \Big) \\
&\overset{(d)}{\leq} P\Big(\frac{\abs{Z}}{\sqrt{N}} \geq \frac{\e}{4 C \abs{\tau_0 + \e_0}} \Big) \overset{(e)}{\leq} e^{-\kappa N \e^2}. 
\end{split}
\een
Step $(a)$ holds by Fact \ref{fact:lip_deriv} for a suitable constant $C>0$.  Step $(b)$ follows because we are conditioning on  $\mc{F}^c$ defined in \eqref{eq:Fdef}. Step $(c)$ is obtained by writing out the expression for the vector $\mathsf{P}^{\parallel}_{q^0} Z_{0}$:
\ben
 \mathsf{P}^{\parallel}_{q^0} Z_{0}   = \frac{q^0}{\norm{q^0}} \sum_{j=1}^N \frac{q^0_j}{\norm{q^0}} Z_{0_j} \overset{d}{=} \frac{q^0}{\norm{q^0}}Z,
\een
where $Z \in \mathbb{R}$ is standard Gaussian (Fact \ref{fact:gauss_p0}). Step $(d)$ follows from Cauchy-Schwarz and step $(e)$ by Lemma \ref{lem:normalconc}.

Considering $T_{1,b}$, the second term of \eqref{eq:prob2},  and noting  that all quantities except $Z_0$ are in $\mscrs_{1, 0}$, define the shorthand  $\textsf{diff}(Z_{0_i}) :=\psi_h(\frac{1}{\sqrt{n}} \norm{m^0} Z_{0_i} +u_i, \beta_{0_i} ) -  \psi_h(\tau_0 Z_{0_i}, \beta_{0,i})$. Then the  upper tail of $T_{1,b}$  can be written as 
 \be
 P\Big( \frac{1}{N} \sum_{i=1}^N \textsf{diff}(Z_{0_i}) - \expec[\textsf{diff}(Z_{0_i})]  \geq \frac{\e}{4} - \frac{1}{N}\sum_{i=1}^N \expec[\textsf{diff}(Z_{0_i})]  \Big).
 \label{eq:diffi_boundH1}
 \ee
Since $\psi_h$ is bounded, so is  $\textsf{diff}(Z_{0_i})$.  Using the conditioning on $\mc{F}^c$ and steps similar to those in $\mc{B}_0(b)$(iv), we can show that $\frac{1}{N} \sum_{i=1}^N  \expec[\textsf{diff}(Z_{0_i})] \leq \frac{1}{8} \e$ for $\e_0 \leq C \tau_0 \e$, where $C > 0$ can be explicitly computed. For such $\e_0$, using Hoeffding's inequality the probability in 
\eqref{eq:diffi_boundH1} can be bounded by $e^{-n\e^2/(128B^2)}$ when $\psi_h$ takes values within an interval of length $B$. A similar bound holds for the lower tail of $T_{1,b}$. Thus we have now bounded both terms of \eqref{eq:prob2} by $Ke^{-n \kappa \e^2}$.  The result follows by substituting the value of $\e_0$ (chosen as described above) in \eqref{eq:PPi0}. 

\textbf{(c),(d),(e),(f)} These results can be proved by appealing to $\mc{H}_1(b)$ in a  manner similar to $\mc{B}_0 (c)(d)(e)(f)$.

\textbf{(g)}  
From the definitions in Section \ref{subsec:defs} and defining $\mathbf{Q}_{1} := \frac{1}{n}\norm{q^0}^2$, we have $\gamma^1_0 = \mathbf{Q}_{1}^{-1}\frac{1}{n} (q^0)^*q^1$ and 
$\hat{\gamma}_0^1= {\tilde{E}_{0,1}}/{\tilde{E}_{0,0}}  = \tilde{E}_{0,1} \sigma_0^{-2}$. Therefore, 
\begin{equation}
\begin{split}
P( \lvert \gamma^1_0 - \hat{\gamma}^1_0  \lvert \geq \epsilon )
&\overset{(a)}{\leq} P(\lvert  \mathbf{Q}_{1}^{-1} - \sigma_0^{-2} \lvert \geq \tilde{\e} )\\
&\qquad  + P\Big(\Big \lvert  \frac{1}{n}(q^0)^*q^1 -\tilde{E}_{0,1}  \Big \lvert \geq \tilde{\e} \Big) 
\end{split}
\label{eq:gam0_conc}
\end{equation}
where $(a)$ follows from Lemma \ref{products} with $\tilde{\e}:= \min\{ \sqrt{\e/3}, \ \e/(3 \tilde{E}_{0,1}), \ \e \sigma_0^2/3 \}$.   We now show that each of the two terms in \eqref{eq:gam0_conc} is bounded by $Ke^{-\kappa n \tilde{\e}^2}$.  Since $\sigma^2_0 >0$, by Lemma \ref{inverses} and \eqref{eq:qassumption}, we have 
$P(\lvert \mathbf{Q}_{1}^{-1}- \sigma_0^{-2} \lvert \geq \tilde{\e} ) \leq 2 K e^{-\kappa n \tilde{\e}^2 \sigma_0^{2} \min(1, \sigma_0^{2})}$.  The concentration bound for $ \frac{1}{n}(q^0)^*q^1$ follows from $\mc{H}_1(e)$. 

\textbf{(h)} From the definitions in Section \ref{subsec:defs},  we have  $\norm{q^1_{\perp}}^2 = \norm{q^1}^2 -\norm{q^1_{\parallel}}^2 =   \norm{q^1}^2- (\gamma_0^1)^2 \norm{q^0}^2$, and $(\sigma_{1}^{\perp})^2= \sigma_1^2 - (\hat{\gamma}^1_0)^2 \sigma_0^2$. We therefore have
\begin{align*}
& P\Big(\Big \lvert \frac{1}{n}\norm{q^1_{\perp}}^2 - (\sigma_{1}^{\perp})^2 \Big \lvert \geq \epsilon\Big) \\
& \overset{(a)}{\leq} P\Big(\Big \lvert \frac{\norm{q^1}^2}{n} - \sigma_1^2 \Big \lvert \geq \frac{\epsilon}{2} \Big)  \\
& \quad + P\Big(\Big \lvert (\gamma_0^1)^2 \frac{\norm{q^0}^2}{n} - (\hat{\gamma}^1_0)^2 \sigma_0^2 \Big \lvert \geq \frac{\epsilon}{2}\Big) \\
& \overset{(b)}{\leq} K \exp\{-\kappa n \e^2\} + K \exp\Big\{\frac{- \kappa n \e^2}{4(9) \max(1, (\hat{\gamma}^1_0)^4, \sigma_0^4)}\Big\} 
\end{align*}
In the chain above, $(a)$ uses Lemma \ref{sums} and $(b)$ is obtained using $\mc{H}_1(e)$ for bounding the first term and by applying Lemma \ref{products} to the second term along with the concentration of $\norm{q^0}$ in \eqref{eq:qassumption}, $\mc{H}_1(g)$, and Lemma \ref{powers} (for concentration of the square).

\subsection{Step 3: Showing $\mc{B}_t$ holds}

We prove the statements in $\mc{B}_t$ assuming that $\mc{B}_{0}, \ldots, \mc{B}_{t-1}$, and $\mc{H}_1, \ldots, \mc{H}_t$ hold due to the induction hypothesis.  The induction hypothesis implies that for $0 \leq r \leq (t-1)$, the  deviation probabilities  $P(\frac{1}{n} \| \Delta_{r,r} \|^2 \geq \epsilon)$ in  \eqref{eq:Ba} and 
$P(\frac{1}{n} \| \Delta_{r+1,r} \|^2 \geq \epsilon)$ in \eqref{eq:Ha}  are each bounded by $K_r e^{-\kappa_{r} n \e}$. Similarly,  the LHS in each of \eqref{eq:Hb1} -- \eqref{eq:Bh} is bounded by  $K_r e^{-\kappa_{r} n \e^2}$.

We begin with a lemma that is required to prove $\mc{B}_t(a)$. The lemma as well as other parts of $\mc{B}_t$ assume the invertibility of $\mbf{M}_1, \ldots, \mbf{M}_t$, but for the sake of brevity,  we do not explicitly specify the conditioning.

\begin{lem}
\label{lem:Mv_conc}
Let $v := \frac{1}{n}H_t^* q^t_{\perp} - \frac{1}{n}M_t^*[\lambda_t m^{t-1} - \sum_{i=1}^{t-1} \lambda_{i} \gamma^t_{i} m^{i-1}]$ and $\mathbf{M}_t := \frac{1}{n}M_{t}^* M_{t}$.  If \, $\mbf{M}_1, \ldots, \mathbf{M}_t$ are invertible, we have for $j \in [t]$,
\ben
P( \lvert [\mathbf{M}_t^{-1} v]_{j}  \lvert \geq \e) \leq 
K t^2 K_{t-1} \exp\{ -n \kappa \kappa_{t-1} \e^2/t^2 \}.
\een
\end{lem}
\begin{IEEEproof}
We can represent $\mathbf{M}_{t}$ as
\ben
\mathbf{M}_{t} = \frac{1}{n}\Big[ \begin{array}{cc}
n\mathbf{M}_{t-1} & M_{t-1}^* m^{t-1} \\
 (M_{t-1}^* m^{t-1})^* & \norm{m^{t-1}}^2 \end{array} \Big],
 \een
 Then, if $\mathbf{M}_{t-1}$ is invertible, by the block inversion formula we have
\be
 \mathbf{M}_{t}^{-1} = \left[ \begin{array}{cc}
\mathbf{M}_{t-1}^{-1} + \frac{n \alpha^{t-1}(\alpha^{t-1})^*}{\norm{m^{t-1}_{\perp}}^{2}} & -\frac{n\alpha^{t-1}}{\norm{m^{t-1}_{\perp}}^{2}}  \\
-\frac{n (\alpha^{t-1})^* }{\norm{m^{t-1}_{\perp}}^{2}}  & \frac{n}{\norm{m^{t-1}_{\perp}}^{2}}  \end{array} \right],
\label{eq:Mt1_inverse0}
\ee
where we have used $\alpha^{t-1} = \frac{1}{n}\mathbf{M}_{t-1}^{-1} M_{t-1}^* m^{t-1}$ and $(M_{t-1}^* m^{t-1})^* \alpha^{t-1} = (m^{t-1})^* m^{t-1}_{\parallel}$.  Therefore,
\be
\mbf{M}_t^{-1}v= 
\begin{bmatrix}
\mbf{M}_{t-1}^{-1}v_{[t-1]} +  \alpha^{t-1}( (\alpha^{t-1})^*v_{[t-1]} - v_t )  \mathsf{a}_{t-1} \\
- ( (\alpha^{t-1})^*v_{[t-1]} - v_t)  \mathsf{a}_{t-1}
\end{bmatrix},
\ee
where  $\mathsf{a}_{r}:=n/\norm{m^r_{\perp}}^2$ for $r \in [t]$, and   $v_{[r]} \in \mathbb{R}^r$ denotes the vector consisting of the first $r$ elements of $v \in \mathbb{R}^t$.  Now, using the block inverse formula again to express $\mbf{M}_{t-1}^{-1}v_{[t-1]} $ and noting that $\alpha^{t-1}=(\alpha^{t-1}_0, \ldots, \alpha^{t-1}_{t-2})$, we obtain
  \ben
  \begin{split}
& \mbf{M}_t^{-1}v   \\
& =  \left[ 
\begin{array}{l}
 \mbf{M}_{t-2}^{-1}v_{[t-2]} + \alpha^{t-2} ( (\alpha^{t-2})^*v_{[t-2]} - v_{t-1} ) \mathsf{a}_{t-2}   \\
 \hspace{1in} +  \alpha^{t-1}_{[t-2]} ( (\alpha^{t-1})^*v_{[t-1]} - v_t )  \mathsf{a}_{t-1}   \\
 - ( (\alpha^{t-2})^*v_{[t-2]} - v_{t-1})  \mathsf{a}_{t-2}  \\
  \hspace{1in} +   \alpha^{t-1}_{t-2} ( (\alpha^{t-1})^*v_{[t-1]} - v_t )  \mathsf{a}_{t-1} \\
 - ( (\alpha^{t-1})^*v_{[t-1]} - v_t )  \mathsf{a}_{t-1}
 \end{array}
\right].
\end{split}
\een
Continuing in this fashion, we can express each element of $\mbf{M}_t^{-1}v$ as follows:
  \be
  \begin{split}
&[ \mbf{M}_t^{-1}v ]_k = \\
&\left\{ 
\begin{array}{ll}
v_1 \mathsf{a}_0 + \sum_{j=1}^{t-1} \alpha_0^j ( (\alpha^j)^*v_{[j]} - v_{j+1}) \mathsf{a}_j, & k=1, \\
-(   (\alpha^{k-1})^*v_{[k-1]} - v_k)  \mathsf{a}_{k-1}  &  \\
\hspace{0.3in} + \sum_{j=k}^{t-1} \alpha_{k-1}^j ( (\alpha^j)^*v_{[j]} - v_{j+1} ) \mathsf{a}_j,   & 2 \leq k < t, \\
-( (\alpha^{t-1})^*v_{[t-1]} - v_t )  \mathsf{a}_{t-1}, & k=t.
\end{array}
\right.
\label{eq:Mtv_k}
\end{split}
\ee
We will prove that each entry of  $\mbf{M}_t^{-1}v$ concentrates around $0$ by showing that each entry of $v$ concentrates around zero, and the entries of $\alpha^j, \mathsf{a}_j$ concentrate around constants for $j \in [t]$. 

For $k \in [t]$,   bound $\abs{v_k}$ as follows. Substituting $q^t_{\perp} = q^t - \sum_{j=0}^{t-1} \gamma^t_j q^j$ in the definition of $v$ and using the triangle inequality, we have
\be
\begin{split}
\abs{v_{k}} &\leq \Big \lvert \frac{(h^{k})^*q^t}{n} - \lambda_t \frac{(m^{k-1})^* m^{t-1}}{n}\Big \lvert +  \abs{\gamma_0^t} \Big \lvert \frac{(h^{k})^*q^0}{n}\Big \lvert \\
&+ \sum_{i=1}^{t-1} \abs{\gamma_{i}^t} \Big \lvert \frac{(h^{k})^*q^{i}}{n} - \lambda_{i} \frac{(m^{k-1})^*m^{i-1}}{n} \Big \lvert.
\label{eq:Bta1}
\end{split}
\ee
Therefore, 
\be
\begin{split}
& P( \abs{v_{k}} \geq \e )   \leq
P\Big( \Big \lvert \frac{1}{n}(h^{k})^*q^t - \lambda_t \frac{1}{n}(m^{k-1})^* m^{t-1}\Big \lvert \geq \e' \Big) \\
&+  P\Big(\abs{\gamma_0^t} \Big \lvert \frac{1}{n}(h^{k})^*q^0\Big \lvert \geq \e' \Big) \\
& + \sum_{i=1}^{t-1} P\Big(\abs{\gamma_{i}^t} \Big \lvert  \frac{1}{n}(h^{k})^*q^{i} - \lambda_{i} \frac{1}{n}(m^{k-1})^*m^{i-1} \Big \lvert \geq \e' \Big)
\end{split}
\label{eq:vk_conc1}
\ee
where $\e'=\frac{\e}{t+1}$. The first term in \eqref{eq:vk_conc1} can be bounded using Lemma \ref{products} and induction hypotheses $\mathcal{H}_t (f)$  and $\mathcal{B}_{t-1} (e)$   as follows.
\ben
\begin{split}
& P\Big( \Big \lvert \frac{(h^{k})^*q^t}{n} - \lambda_t \frac{(m^{k-1})^* m^{t-1}}{n}\Big \lvert \geq \e' \Big) \\
& \leq P\Big( \Big \lvert \frac{(h^{k})^*q^t}{n} - \hat{\lambda}_{t} \breve{E}_{k-1,t-1} \Big \lvert \geq \frac{\e'}{2} \Big) \\
&\qquad + P\Big( \Big \lvert \lambda_t \frac{(m^{k-1})^* m^{t-1} }{n} - \hat{\lambda}_t \breve{E}_{k-1,t-1} \Big \lvert \geq \frac{\e'}{2} \Big) \\
& \leq K_{t-1}e^{-\kappa \kappa_{t-1} n \e'^2} + 2K_{t-1}e^{-\frac{\kappa \kappa_{t-1} n \e'^2}{\max(1, \hat{\lambda}_t^2, \breve{E}_{k-1,t-1}^2)}}.
\end{split}
\een
For $k  \in [t]$, the second term in \eqref{eq:vk_conc1} can be bounded as 
\begin{align*}
& P\Big(\abs{\gamma_0^t} \Big \lvert \frac{1}{n}(h^{k})^*q^0\Big \lvert \geq \e' \Big) \\
&  \leq P\Big(( \abs{\gamma_0^t -  \hat{\gamma}_0^t} + \abs{\hat{\gamma}_0^t}) \abs{\frac{1}{n}(h^{k})^*q^0} \geq \e' \Big) \\ 
 & \leq P( \abs{ \gamma_0^t - \hat{\gamma}_0^t} \geq  \sqrt{\e'} ) \\
 & \hspace{0.5in} + P\Big( \abs{ \frac{1}{n}(h^{k})^*q^0} \geq \frac{\e'}{2} \min\{1, \abs{\hat{\gamma}_0^t}^{-1}\} \Big) \\
 & \leq K_{t-1}e^{-\kappa \kappa_{t-1} n\e'} + K_{t-1}e^{-\kappa \kappa_{t-1} n\e'^2},
\end{align*}
where the last inequality follows from induction hypotheses $\mathcal{H}_t (g)$ and $\mathcal{H}_t (c)$. Similarly,  for 
$k \in [t], \, i \in [t-1]$, the third term in  \eqref{eq:vk_conc1} can be bounded as 
\begin{align*}
& P\Big(\abs{\gamma_{i}^t} \Big \lvert  \frac{(h^{k})^*q^{i}}{n} - \lambda_{i} \frac{(m^{k-1})^*m^{i-1}}{n} \Big \lvert \geq \e' \Big) \\
& \leq  P\Big( (\abs{\gamma_{i}^t - \hat{\gamma}_{i}^t} + \abs{\hat{\gamma}_{i}^t})  \abs{\frac{(h^{k})^*q^{i}}{n} - \lambda_{i} \frac{(m^{k-1})^*m^{i-1}}{n}} \geq \e' \Big) \\
& \leq P( \abs{ \gamma_{i}^t - \hat{\gamma}_{i}^t} \geq \sqrt{\e'} ) \\
&\quad + P\Big(\Big \lvert \frac{(h^{k})^*q^{i}}{n} - \lambda_{i} \frac{(m^{k-1})^*m^{i-1}}{n} \Big \lvert \geq  \frac{\e'}{2} \min\{1, (\hat{\gamma}_i^t)^{-1}\} \Big) \\
&\leq K_{t-1} e^{-\kappa \kappa_{t-1}  n \e'} +2 K_{t-1}e^{-\kappa \kappa_{t-1}  n \e'^2}.
\end{align*}
Substituting $\e'=\frac{\e}{t+1}$ in each of the above bounds and using them in \eqref{eq:vk_conc1},
\be
 P( \abs{v_{k}} \geq \e )  \leq Kt K_{t-1} e^{ -\kappa \kappa_{t-1} \e^2/t^2}.
\label{eq:vk_conc2}
\ee
Furthermore, from induction hypotheses $\mc{B}_0(g) - \mc{B}_{t-1}(g)$, for $0 \leq i < j \leq (t-1)$:
\be
 P( \abs{ \alpha^j_i - \hat{\alpha}^j_i  } \geq \e )  \leq K_{t-1}e^{ - n \kappa_{t-1} \e^2}.
 \label{eq:alph_conc0}
\ee 
Also, using induction hypotheses $\mc{B}_0(h) - \mc{B}_{t-1}(h)$ and Lemma \ref{inverses}, for $0 \leq r  \leq (t-1)$:
\be
 P( \abs{ \mathsf{a}_r - (\tau_t^{\perp})^{-2}} \geq \e)  \leq K_{t-1}e^{ - n \kappa_{t-1} \e^2}.
 \label{eq:ar_conc0}
\ee 
Finally, from \eqref{eq:Mtv_k}, we have for $k \in [t]$,
\ben
\begin{split}
&P\Big( \abs{[ \mbf{M}_t^{-1}v]_k} \geq \e \Big) \\
& \stackrel{(a)}{\leq}    P\Big( \cup_{k \in [t]}\{  \abs{v_k} \geq \e\}  
\cup_{0\leq r<t}  \{ \abs{\mathsf{a}_r - (\tau_t^{\perp})^{-2}} \geq \kappa_1 \e /t \}  \Big. \\
&\qquad \qquad \Big. \cup_{0\leq i < j <t}  \{  \abs{ \alpha^j_i - \hat{\alpha}^j_i  } \geq \kappa_2 \e / t \}
\Big) \\ 
&\stackrel{(b)}{\leq}  \, K t^2 K_{t-1}e^{ -n \kappa \kappa_{t-1} \e^2/t^2 }.
\end{split}
\een
where in step $(a)$, $\kappa_1, \kappa_2$ are appropriately chosen positive constants, and step $(b)$ follows from the bounds in \eqref{eq:vk_conc2}, \eqref{eq:alph_conc0}, and \eqref{eq:ar_conc0}.
\end{IEEEproof}

\textbf{(a)}  Recall the definition of $\Delta_{t,t}$ from  \eqref{eq:Dtt}.  Then using Fact \ref{fact:gauss_p0}, it follows $\frac{1}{\sqrt{n}} \norm{q^t_{\perp}} \mathsf{P}^{\parallel}_{M_t} Z'_t \overset{d}{=} \frac{1}{n} \norm{q^t_{\perp}} \tilde{M}_t \bar{Z}'_t,$ where the columns of $\tilde{M}_{t} \in \mathbb{R}^{n \times t}$ form an orthogonal basis for the column space of $M_{t}$ with $\tilde{M}_{t}^* \tilde{M}_{t} = n\mathsf{I}_{t}$, and $\bar{Z}'_t \in \mathbb{R}^{t}$ is an independent random vector with i.i.d.\ $\mc{N}(0,1)$ entries.  Then,
\begin{align*}
\Delta_{t,t} = & \, \sum_{r=0}^{t-1} (\gamma^t_r - \hat{\gamma}^{t}_r) b^r + Z'_t \Big(\frac{1}{\sqrt{n}}\norm{q^t_{\perp}} - \sigma_{t}^{\perp}\Big) \\
&\quad - \frac{1}{n}\norm{q^t_{\perp}} \tilde{M}_t \bar{Z}'_t + M_t \mathbf{M}_t^{-1}v,
\end{align*}
where $ \mathbf{M}_t \in \mathbb{R}^{t \times t}$ and $v \in \mathbb{R}^t$ are defined in Lemma \ref{lem:Mv_conc}. Writing $M_t\textbf{M}_t^{-1}v = \sum_{j=0}^{t-1} m^j [\textbf{M}_t^{-1} v ]_{j+1}$ and using Lemma \ref{lem:squaredsums}, we have
\begin{align*}
\frac{\norm{\Delta_{t,t}}^2}{2(t+1)} & \leq \sum_{r=0}^{t-1} (\gamma^t_r - \hat{\gamma}^t_r)^2 \norm{b^r}^2 +\norm{Z'_t}^2 \Big(\frac{1}{\sqrt{n}}\norm{q^t_{\perp}} - \sigma_{t}^{\perp} \Big)^2 \\
&\qquad  + \frac{1}{n^2}\norm{q^t_{\perp}}^2 \norm{\tilde{M}_t \bar{Z}'_t}^2 + \sum_{j=0}^{t-1} \norm{m^j}^2[\textbf{M}_t^{-1} v ]_{j+1}^2, 
\end{align*}
Applying Lemma \ref{sums},
\begin{align}
P\Big(\frac{\norm{\Delta_{t,t}}^2}{n} \geq \epsilon \Big) &\leq \sum_{r=0}^{t-1} P\Big( \lvert\gamma^t_r - \hat{\gamma}^t_r \lvert \frac{\norm{b^r}}{\sqrt{n}} \geq  \sqrt{\tilde{\e}_t} \Big)\nonumber\\
& + P\Big(\frac{\norm{q^t_{\perp}}}{\sqrt{n}}  \frac{\norm{\tilde{M}_t \bar{Z}'_t}}{n} \geq  \sqrt{\tilde{\e}_t}\Big) \nonumber \\
& + P\Big(\Big \lvert \frac{\norm{q^t_{\perp}}}{\sqrt{n}} - \sigma_{t}^{\perp}\Big \lvert \frac{\norm{Z'_t}}{\sqrt{n}} \geq  \sqrt{\tilde{\e}_t} \Big) \nonumber \\
&+ \sum_{j=0}^{t-1} P\Big(\Big \lvert [\textbf{M}_t^{-1} v ]_{j+1} \Big \lvert \frac{\norm{m^j}}{\sqrt{n}} \geq  \sqrt{\tilde{\e}_t} \Big),
\label{eq:delttt_sq_conc}
\end{align}
where $\tilde{\e}_t := \frac{\e}{4(t+1)^2}$.  We now  bound each of the terms in \eqref{eq:delttt_sq_conc}.   

For  $0 \leq r \leq t-1$, the first term is bounded as 
\begin{align*}
& P\Big( \lvert\gamma^t_r - \hat{\gamma}^t_r  \lvert \frac{1}{\sqrt{n}}\norm{b^r} \geq  \sqrt{\tilde{\e}_t} \Big)  \\
&\leq P\Big( \lvert\gamma^t_r - \hat{\gamma}^t_r  \lvert \Big(  \Big\lvert\frac{1}{\sqrt{n}}\norm{b^r}  -\sigma_r \Big \lvert+\sigma_r\Big) \geq \sqrt{\tilde{\e}_t} \Big)  \\ 
& \leq P\Big( \lvert \gamma^t_r - \hat{\gamma}^t_r \lvert \geq \frac{\sqrt{\tilde{\e}_t}}{2}  \min\{1 , \frac{1}{\sigma_r} \} \Big)  \\
& \hspace{1in}+ P\Big(\Big \lvert \frac{\norm{b^r}}{\sqrt{n}} - \sigma_r \Big \lvert \geq \sqrt{\e}  \Big) \\
&\overset{(a)}{\leq} K_{t-1} e^{-\kappa \kappa_{t-1} n \tilde{\e}_t} + K_{t-1}e^{- \kappa \kappa_{t-1}  n \e} ,
\end{align*}
where step $(a)$ follows from induction hypotheses $\mathcal{H}_t (g)$, $\mathcal{B}_{0} (d)-\mathcal{B}_{t-1} (d)$, and Lemma \ref{sqroots}.  Next, the third term in \eqref{eq:delttt_sq_conc} is bounded as 
\ben
\begin{split}
 P&\Big(\Big \lvert \frac{1}{\sqrt{n}}\norm{q^t_{\perp}} - \sigma_{t}^{\perp}\Big \lvert  \frac{1}{\sqrt{n}}\norm{Z'_t} \geq \sqrt{\tilde{\e}_t} \Big) \\
 &\leq P\Big(\Big \lvert \frac{1}{\sqrt{n}}\norm{q^t_{\perp}} - \sigma_{t}^{\perp}\Big \lvert  \geq \frac{\sqrt{\tilde{\e}_t}}{\sqrt{2}} \Big)  +  P\Big( \frac{1}{\sqrt{n}}\norm{Z'_t} \geq \sqrt{2} \Big) \\
&\overset{(b)}{\leq}  K_{t-1}e^{-\kappa \kappa_{t-1} n \tilde{\e}_t} + e^{- n/8},
\end{split}
\een
where step $(b)$ is obtained using induction hypothesis $\mathcal{H}_t (h)$, Lemma \ref{sqroots}, and Lemma \ref{subexp}. Since $\frac{1}{\sqrt{n}}\norm{q^t_{\perp}}$ concentrates on $\sigma_{t}^{\perp}$ by $\mc{H}_t (h)$, the second term in \eqref{eq:delttt_sq_conc} can be bounded as 
\begin{equation}
 \label{eq:T3split}
\begin{split}
& P\Big(\frac{1}{\sqrt{n}}\norm{q^t_{\perp}} \cdot  \frac{1}{n}\norm{\tilde{M}_t \bar{Z}'_t} \geq  \sqrt{\tilde{\e}_t}  \Big) \\
 &\leq P\Big( \Big \lvert \frac{1}{\sqrt{n}}\norm{q^t_{\perp}} - \sigma_{t}^{\perp}  \Big \lvert \geq  \sqrt{\e} \Big) \\
 &\qquad + P\Big(\frac{1}{n}\norm{\tilde{M}_t\bar{Z}'_t} \geq 
 \frac{1}{2} \sqrt{\tilde{\e}_t}  \min\{ 1 , (\sigma_t^{\perp})^{-1}\} \Big) \\
 & \leq K_{t-1}e^{-\kappa \kappa_{t-1} n \tilde{\e}_t} + t K K_{t-1}e^{-\kappa \kappa_{t-1}n\tilde{\e}_t/t},
 \end{split}
\end{equation}
 where the last  inequality is obtained as follows. The concentration for $\norm{q^t_{\perp}}/\sqrt{n}$ has already been shown above. For the second term, denoting the  columns of $\tilde{M}_t$ by $\{\tilde{m}_0, \ldots, \tilde{m}_{t-1} \}$, we have
$ \| \tilde{M}_t \bar{Z}'_t\|^2 = \sum_{i=0}^{t-1} \norm{\tilde{m}_i}^2 (\bar{Z}'_{t_i})^2 =  n \sum_{i=0}^{t-1} (\bar{Z}'_{t_i})^2$
since the $\{\tilde{m}_i \}$ are orthogonal, and $\norm{\tilde{m}_i}^2 = n$ for $0 \leq i \leq t-1$.  Therefore,
\ben
\begin{split}
&P\Big(\frac{ 1}{n^2} \norm{\tilde{M}_t\bar{Z}'_t}^2 \geq \tilde{\e}_t \Big) = P\Big( \sum_{i=0}^{t-1}(\bar{Z}'_{t_i})^2 \geq n \tilde{\e}_t \Big) \\
&\overset{(c)}{\leq} \sum_{i=0}^{t-1}P\Big(|\bar{Z}'_{t_i}| \geq \sqrt{\frac{n\tilde{\e}_t}{t}} \Big)\overset{(d)}{\leq} 2t e^{-\frac{n\tilde{\e}_t}{2t}}.
\label{eq:MtZtconc}
\end{split}
\een
Step $(c)$ is obtained from Lemma \ref{sums}, and step $(d)$ from Lemma \ref{lem:normalconc}. This yields the second term in \eqref{eq:T3split}.

Finally, for $0 \leq j \leq (t-1)$, the last term in \eqref{eq:delttt_sq_conc} can be bounded by 
\ben
\begin{split}
& P\Big(\lvert [\textbf{M}_t^{-1} v ]_{j+1} \lvert \frac{\norm{m^j}}{\sqrt{n}} \geq \sqrt{\tilde{\e}_t} \Big) \\
&= P\Big( \lvert [\textbf{M}_t^{-1} v]_{j+1} \lvert \Big( \Big \lvert\frac{\norm{m^j}}{\sqrt{n}} - \tau_j \Big \lvert + \tau_j \Big) \geq \sqrt{\tilde{\e}_t} \Big)
\\  & \leq  P\Big(\Big \lvert \frac{\norm{m^j}}{\sqrt{n}} - \tau_j \Big \lvert \geq  \sqrt{\e} \Big)  \\
& \hspace{0.5in} + P\Big( \lvert [\textbf{M}_t^{-1} v]_{j+1} \lvert \geq \frac{ \sqrt{\tilde{\e}_t}}{2} \min\{1, \frac{1}{\tau_j}\} \Big) \\
   &\overset{(e)}\leq K_{t-1}e^{- \kappa \kappa_{t-1}  n \e} + Kt^2 K_{t-1} e^{- \kappa \kappa_{t-1}  n \tilde{\e}_t/ t^2},
\end{split}
\een
where step $(e)$ follows from induction hypothesis $\mc{B}_{t-1}(e)$, and Lemma \ref{lem:Mv_conc}. Substituting $\tilde{\e}_t = \frac{\e}{4(t+1)^2}$, we have bounded each term of \eqref{eq:delttt_sq_conc} as desired.


\textbf{(b).(iii)}  For brevity, let $\mathbb{E} \phi_b := \mathbb{E}[\phi_b(\sigma_0 \breve{Z}_0, ..., \sigma_t \breve{Z}_t, W)]$, and 
\be
a_i = (b^0_i, ..., b^{t}_i, w_i), \qquad 
c_i = ( {\bpure^0}_i, ..., {\bpure^t}_i, w_i).
\label{eq:Btfunc2a}
\ee
Using Lemma \ref{sums}, we have
\be
\begin{split}
& P\Big(\Big \lvert \frac{1}{n}\sum_{i=1}^n \phi_b(b^0_i, ..., b^{t}_i, w_{i})  - \mathbb{E} \phi_b \Big \lvert \geq \epsilon \Big) \\
&\leq P\Big(\Big \lvert \frac{1}{n}\sum_{i=1}^n \phi_b(c_i) - \mathbb{E} \phi_b \Big \lvert \geq \frac{\epsilon}{2} \Big) \\
&\quad + P\Big(\Big \lvert \frac{1}{n}\sum_{i=1}^n (\phi_b(a_i) -\phi_b(c_i)) \Big\lvert \geq \frac{\epsilon}{2} \Big).
\label{eq:Btfunc3a}
\end{split} 
\ee

Lemma \ref{lem:bhpure} (Eq.\ \eqref{eq:pure_bh_dist}) shows the joint distribution of $( {\bpure^0}_i, ..., {\bpure^t}_i)$ is jointly  Gaussian for $i \in [N]$. The first term in \eqref{eq:Btfunc3a} can therefore be bounded as 
\begin{align}
&P\Big(\Big \lvert \frac{1}{n}\sum_{i=1}^n \phi_b(c_i) - \mathbb{E} \phi_b \Big \lvert \geq \frac{\epsilon}{2} \Big)  
\nonumber \\
&=  P\Big(\Big \lvert \frac{1}{n}\sum_{i=1}^n \phi_b(\sigma_0 \breve{Z}_{0,i}, \ldots, \sigma_t \breve{Z}_{t,i} , w_i) - \mathbb{E} \phi_b \Big \lvert \geq \frac{\epsilon}{2} \Big) \nonumber \\
&\leq 2 e^{{-\kappa n \e^2}/{t^3}}, 
 \label{eq:phib_ci_conc}
\end{align}
where the last inequality is obtained from  Lemma \ref{lem:PLsubgaussconc}. Here $\kappa >0$ is a generic  absolute constant. 

We now bound the second term in \eqref{eq:Btfunc3a} using the pseudo-Lipschitz property of $\phi_b$. Denoting the pseudo-Lipschitz constant by $L$, we have
\begin{align}
&\Big \lvert \frac{1}{n}\sum_{i=1}^n (\phi_b(a_i) - \phi_b(c_i) ) \Big\lvert^2   \leq
\Big[\frac{1}{n}\sum_{i=1}^n \abs{ \phi_b(a_i) - \phi_b(c_i) } \Big]^2 \nonumber \\
&\leq \Big[\frac{L}{n}\sum_{i=1}^n( 1 + 2 \norm{c_i} + \norm{a_i - c_i}) \norm{a_i - c_i}\Big]^2 \nonumber \\ 
& \leq \frac{3L^2 }{n} \sum_{j=1}^n \norm{a_j - c_j}^2 \Big[1 + \frac{4}{n} \sum_{i=1}^n \norm{c_i}^2 + \frac{1}{n} \sum_{i=1}^n\norm{a_i - c_i}^2\Big] ,
\label{eq:phi_ac_diffv2}
\end{align}
where the last inequality is obtained by first applying Cauchy-Schwarz, and then using Lemma \ref{lem:squaredsums}.

For $j \in [N]$, note that
$  \expec \norm{c_j}^2 =  \sigma_1^2 + \ldots + \sigma_t^2 + \sigma^2$.  Now using \eqref{eq:phi_ac_diffv2} we  bound the second term in \eqref{eq:Btfunc3a} as follows.
\begin{align}
& P\Big( \Big \lvert \frac{1}{n}\sum_{i=1}^n (\phi_b(a_i) -\phi_b(c_i)) \Big\lvert \geq \frac{\epsilon}{2} \Big) \nonumber \\
&= P\Big( \Big \lvert \frac{1}{n}\sum_{i=1}^n (\phi_b(a_i) -\phi_b(c_i)) \Big\lvert^2 \geq \frac{\epsilon^2}{4} \Big)\nonumber \\
%
%
& \leq P\Big( \frac{1}{n} \sum_{i=1}^n \norm{a_i -c_i}^2 \geq \frac{\e^2\min\{1, \frac{1}{12L^2}\}}{ 2 +  8(\sigma_1^2 + \ldots + \sigma_t^2 + \sigma^2) }  \Big) \nonumber \\
& \quad +  P\Big( \frac{1}{n} \sum_{j=1}^n  \norm{c_j}^2 \geq 2(\sigma_1^2 + \ldots + \sigma_t^2 + \sigma^2)  \Big).
\label{eq:aici_and_cj2_v2}
\end{align}
Label the two terms above as $T_1$ and $T_2$. We bound $T_2$ as 
\begin{align}
 &P\Big( \frac{1}{n} \sum_{j=1}^n  \norm{c_j}^2 \geq 2(\sigma^2 + \sum_{r=1}^t \sigma_r^2)  \Big) \nonumber \\
& = P\Big( \frac{1}{n} \sum_{j=1}^n  \big( \norm{c_j}^2  - \expec \| c_j\|^2 \big) \, \geq (\sigma^2 + \sum_{r=1}^t \sigma_r^2)  \Big) \leq e^{ {-\kappa n }/{t^3}}
\label{eq:norm_cj_bnd}
\end{align}
for an absolute constant $\kappa >0$, where the last inequality is obtained by applying the concentration result in  Lemma \ref{lem:PLsubgaussconc} to the pseudo-Lipschitz function $\phi_b(c_j) = \| c_j \|^2$. 
\begin{align}
& \sum_{i=1}^n \norm{a_i -c_i}^2   =  \sum_{i=1}^n \sum_{k=0}^t (b^k_{pure_i} - b^k_i)^2    \nonumber \\
&=  \sum_{i=1}^n \sum_{k=0}^t \Big[\sum_{r=0}^k \sfc^k_r \, [\Delta_{r,r}]_i\Big]^2 \nonumber \\
 & \leq     \sum_{i=1}^n \sum_{k=0}^t \Big[\sum_{r'=0}^k (\sfc^k_{r'})^2 \sum_{r=0}^k ([\Delta_{r,r}]_i)^2 \Big] \nonumber \\
 &= \sum_{k=0}^t \Big[\sum_{r'=0}^k (\sfc^k_{r'})^2 \sum_{r=0}^k \norm{\Delta_{r,r}}^2 \Big]
 =  \sum_{r=0}^t \norm{\Delta_{r,r}}^2  \sum_{k=r}^t \sum_{r'=0}^k (\sfc^k_{r'})^2,
\label{eq:aici_diff_v2}
\end{align}
where the inequality is obtained by applying Cauchy-Schwarz. 

Comparing \eqref{eq:pure_bh_dist} and \eqref{eq:bhZZ} in   Lemma \ref{lem:bhpure}, we observe that for  $k \geq 0$ and $j \in  [n]$, 
\be 
\expec  ({\bpure^k}_{j})^2 =  \sigma_k^2 = \sum_{i=0}^t (\sigma^{\perp}_i)^2 (\sfc^k_i)^2. 
\ee
Therefore, 
\be 
\sum_{i=0}^k  (\sfc^k_i)^2 \leq \frac{ \sigma_t^2 }{ \min_{ 0 \leq i \leq k}  (\sigma^{\perp}_i)^2 } \leq \frac{ \sigma_k^2 }{ \varepsilon_2 },
\label{eq:sum_sfc_bnd}
\ee
where the last inequality follows from the stopping criterion in \eqref{eq:sigmat_taut_bounds}. Using \eqref{eq:sum_sfc_bnd} and \eqref{eq:aici_diff_v2} we have
\[ \frac{1}{n} \sum_{i=1}^n \norm{a_i -c_i}^2  \leq \frac{1}{n}  \sum_{r=0}^t \norm{\Delta_{r,r}}^2  \sum_{k=r}^t \frac{\sigma_k^2}{\varepsilon_2}.
\]
Therefore we can bound the first term $T_1$ in \eqref{eq:aici_and_cj2_v2}  as follows.
\begin{align}
& T_1 =  \nonumber \\
& P\Big(\frac{1}{n}  \sum_{r=0}^t \norm{\Delta_{r,r}}^2 \geq \frac{\varepsilon_2 (\sigma_1^2 + \ldots + \sigma_t^2)^{-1} \e^2\min\{1, \frac{1}{12L^2}\}}{ (2 +  8(\sigma_1^2 + \ldots + \sigma_t^2 + \sigma^2))}  \Big)  \nonumber \\
& \leq  \sum_{r=0}^t  P\Big( \frac{1}{n} \| \Delta_{r,r} \|^2  \leq   \frac{ \kappa \e^2}{t^3} \Big)  
 \stackrel{(a)}{\leq} K t^3 K_{t-1} e^{-{\kappa \kappa_{t-1} n \epsilon^2}/{ t^7}}, \label{eq:T1_sumdelta_bound_v2}
\end{align}
where $K, \kappa >0$ are some absolute constants. The inequality $(a)$ follows from steps $\mc{B}_0 (a) - \mc{B}_t (a)$.

Finally, substituting \eqref{eq:T1_sumdelta_bound_v2} and \eqref{eq:norm_cj_bnd} in \eqref{eq:aici_and_cj2_v2}, and then combining with \eqref{eq:phib_ci_conc} and \eqref{eq:Btfunc3a}, we obtain
\begin{align}
P&\Big(\Big \lvert \frac{1}{n}\sum_{i=1}^n \phi_b(b^0_i, ..., b^{t}_i, w_{i})  - \mathbb{E} \phi_b \Big \lvert \geq \epsilon \Big) 
\nonumber \\
&\quad \leq K t^3 K_{t-1} e^{-{\kappa \kappa_{t-1} n \epsilon^2}/{ t^7}}.
\end{align}

\textbf{(b).(iv)} For brevity, we write $\mathsf{b}_{t,i} : = \sum_{r=0}^{t-1} \hat{\gamma}^t_{r} b^{r}_i$.  Then using the conditional distribution of $b^{t}$ in \eqref{eq:Ba_dist} and Lemma \ref{sums}, we write
\begin{equation}
\begin{split}
& P\Big(\Big \lvert \frac{1}{n} \sum_{i=1}^n \psi_b(b^{t}_i, w_i) - \mathbb{E}[\psi_b( \sigma_{t} \breve{Z}_{t}, W)] \Big\lvert \geq \e \Big)  \\
&= P\Big(\Big \lvert \frac{1}{n} \sum_{i=1}^n \psi_b(\mathsf{b}_{t,i} + \sigma^{\perp}_t Z'_{t_i} + [\Delta_{t,t}]_i, w_i)  \\
& \hspace{1.3in} - \mathbb{E}[\psi_b(\sigma_{t} \breve{Z}_{t}, W)] \Big \lvert \geq \e \Big) \\
& \leq P \Big( \Big \lvert \frac{1}{n} \sum_{i=1}^n \Big[\psi_b(\mathsf{b}_{t,i} + \sigma^{\perp}_t Z'_{t_i} + [\Delta_{t,t}]_i, w_i)  \\
& \hspace{1in} - \psi_b(\mathsf{b}_{t,i} + \sigma^{\perp}_t Z'_{t_i}, w_i)\Big] \Big\lvert \geq \frac{\e}{3} \Big)  \\
& \ + P\Big ( \Big \lvert \frac{1}{n} \sum_{i=1}^n  \Big[ \psi_b(\mathsf{b}_{t,i} + \sigma^{\perp}_t Z'_{t_i}, w_i) \\
& \hspace{1in} - \mathbb{E}_{Z'_{t}}[ \psi_b(\mathsf{b}_{t,i} + \sigma^{\perp}_t Z'_{t_i}, w_i)]  \Big] \Big\lvert \geq \frac{\e}{3} \Big) \\
&  \ + P \Big( \Big \lvert \frac{1}{n} \sum_{i=1}^n \mathbb{E}_{Z'_{t}} [ \psi_b(\mathsf{b}_{t,i} + \sigma^{\perp}_t Z'_{t_i}, w_i)]  \\
& \hspace{1.3in} -  \mathbb{E}[\psi_b(\sigma_{t} \breve{Z}_{t}, W) ] \Big\lvert \geq \frac{\e}{3} \Big).
\end{split}
\label{eq:Bjt12}
\end{equation}
Label the terms of \eqref{eq:Bjt12} as $T_1 - T_3$. First consider $T_2$. Since $\psi_b$ is bounded, Hoeffding's inequality yields $T_2 \leq 2 e^{-\kappa n \e^2}$.

To bound $T_3$, first note that the $\mathbb{R}^2 \to \mathbb{R}$ function  $\expec_Z[\psi_b(x+Z, y)], \ Z \sim \mc{N}(0,1)$, is bounded and differentiable in the first argument (due to the smoothness of the Gaussian density).  Hence, using induction hypotheses $\mc{B}_{0} (b). \text{(iv)}-\mc{B}_{t-1} (b).$(iv), the probability of each of the following events is  bounded by $K_{t-1} \exp\{-\kappa_{t-1}  n \e^2 /t^2\}$:
\begin{equation}
\begin{split} 
 &\Big \lvert  \frac{1}{n} \sum_{i=1}^n \mathbb{E} \, \psi_b(\sum_{r=0}^{t-1} \hat{\gamma}^t_{r} b^{r}_i + \sigma^{\perp}_t Z'_{t_i}, w_i)
 \\
 &\qquad -  \mathbb{E}\, \psi_b(\sum_{r=0}^{t-2} \hat{\gamma}^t_{r} b^{r}_i + \hat{\gamma}^t_{t-1} \sigma_{t-1} \breve{Z}_{t-1} + \sigma^{\perp}_t Z'_{t_i}, W)  \Big \lvert  \geq \frac{\e}{t}, \\
& \Big \lvert  \frac{1}{n} \sum_{i=1}^n 
\mathbb{E}\, \psi_b(\sum_{r=0}^{t-2} \hat{\gamma}^t_{r} b^{r}_i + \hat{\gamma}^t_{t-1} \sigma_{t-1} \breve{Z}_{t-1} + \sigma^{\perp}_t Z'_{t_i}, W)   \\
&\qquad  - \mathbb{E} \, \psi_b(\sum_{r=0}^{t-3} \hat{\gamma}^t_{r} b^{r}_i + \hspace{-5pt} \sum_{r'=t-2}^{t-1} \hat{\gamma}^t_{r'} \sigma_{r'} \breve{Z}_{r'} + \sigma^{\perp}_t Z'_{t_i}, W)  \Big \lvert  \geq \frac{\e}{t},  \\
&\quad \vdots \\
&  \Big \lvert \frac{1}{n} \sum_{i=1}^n \mathbb{E}\, \psi_b(\hat{\gamma}^t_{0} b^{0}_i + \sum_{r'=1}^{t-1} \hat{\gamma}^t_{t-1} \sigma_{t-1} \breve{Z}_{t-1} + \sigma^{\perp}_t Z'_{t_i}, W)\} \\
& \qquad - \mathbb{E}\, \psi_b(\sum_{r'=0}^{t-1} \hat{\gamma}^t_{r'} \sigma_{r'} \breve{Z}_{r'} + \sigma^{\perp}_t Z'_{t_i}, W)  \Big \lvert  \geq \frac{\e}{t}.
\label{eq:seq_split_events_Bt}
\end{split}
\end{equation}
In the above, the expectation in each term is over  the random variables denoted in upper case. Recall from the proof of Lemma \ref{lem:bhpure} above that $\sum_{r'=1}^{t-1} \hat{\gamma}^t_{t-1} \sigma_{t-1} \breve{Z}_{t-1} + \sigma^{\perp}_t Z'_{t_i} \stackrel{d}{=} \sigma_t \breve{Z}_t$. Thus $T_3$, the third term in \eqref{eq:Bjt12}, can be bounded by the probability of the union of the events in \eqref{eq:seq_split_events_Bt}, which is no larger than  $t K_{t-1} \exp\{-\kappa_{t-1} n \e^2 /t^2\}$. 

Finally, consider $T_1$, the first term  of \eqref{eq:Bjt12}. From the definition of $\Delta_{t,t}$ in Lemma \ref{lem:hb_cond}, we have
$\mathsf{b}_{t,i} + \sigma^{\perp}_t Z'_{t_i} + [\Delta_{t,t}]_i =  \mathsf{b}_{t,i}+ \frac{1}{n}\norm{q^t_{\perp}} [ (\mathsf{I} -\mathsf{P}^{\parallel}_{M_{t}})Z'_{t}]_i + u_i,$
where $u=(u_1, \ldots, u_n)$ is defined  $u := \sum_{r=0}^{t-1} (\gamma^t_r - \hat{\gamma}^t_r) b^{r} + \sum_{j=0}^{t-1} m^j [\textbf{M}_t^{-1}v]_{j+1}$,
with $v$ and $\textbf{M}_t$ defined as in Lemma \ref{lem:Mv_conc}.  For $\e_0 > 0$ to be specified later, define the event $\mc{F}$ as 
\be
\begin{split}
\mc{F} := &\Big\{\Big \lvert \frac{\norm{q^t_{\perp}}}{\sqrt{n}} - \sigma_t^{\perp} \Big \lvert \geq \e_0 \Big\} \cup \Big\{ \frac{\norm{u}^2}{n} \geq \e_0\Big\}\\
& \qquad  \cup_{r=0}^{t-1} \Big\{\Big \lvert \frac{\norm{b^{r}}}{\sqrt{n}} - \sigma_r \Big \lvert \geq \e_0\Big\}.  \label{eq:BFtdef}
\end{split}
\ee
Denoting the event we are considering in $T_1$ by $\Pi_t$ and following steps analogous to \eqref{eq:PPi0}--\eqref{eq:prob2} in $\mc{H}_{1}(b)$.(ii), we obtain
\be
\begin{split} \label{eq:PPit}  
P(T_1) &\leq P(\mc{F}) + \expec[ P(\Pi_t \mid \mc{F}^c, \mscrs_{t,t}) \mid \mc{F}^c]  \\
&\leq  K t^2 K_{t-1} e^{-\kappa \kappa_{t-1} n \e_0^2/  t^4} + \expec [ P(\Pi_t \mid \mc{F}^c, \mscrs_{t,t}) \mid \mc{F}^c ], 
\end{split}
 \ee 
 where the bound on $P(\mc{F})$ is obtained by  the induction hypotheses $\mc{H}_t (h)$, $\mc{B}_0(d) -\mc{B}_{t-1}(d)$, Lemma \ref{sqroots}, and steps similar to the proof of $\mc{B}_t(a)$ for the concentration of $\norm{u}^2/n$ (cf. \eqref{eq:delttt_sq_conc}). 

For the second term in \eqref{eq:PPit}, we have

\begin{equation}
\begin{split}
& P(\Pi_t | \mc{F}^c, \mscrs_{t,t})  \\
 & = P\Big(\Big \lvert \frac{1}{n} \sum_{i=1}^n \Big[\psi_b(\mathsf{b}_{t,i}+ \frac{\norm{q^t_{\perp}}}{\sqrt{n}}[(\mathsf{I} -
 \mathsf{P}^{\parallel}_{M_t})Z'_{t}]_i + u_i,  w_i)  \\
 & \hspace{1.3in} - \psi_b(\mathsf{b}_{t,i} + \sigma^{\perp}_t Z'_{t_i},  w_i)\Big] \Big \lvert \geq \e  \Big) \\
&  \leq  P\Big(\Big \lvert \frac{1}{n} \sum_{i=1}^n \Big[\psi_b(\mathsf{b}_{t,i} + \frac{\norm{q^t_{\perp}}}{\sqrt{n}} Z'_{t_i} + u_i,  w_i)  \\
& \hspace{1.3in} - \psi_b(\mathsf{b}_{t,i}+ \sigma^{\perp}_t Z'_{t_i},  w_i)\Big] \Big \lvert \geq \frac{\e}{2}\Big) \\
& + \Big(\Big \lvert \frac{1}{n} \sum_{i=1}^n \Big[\psi_b(\mathsf{b}_{t,i}+ \frac{\norm{q^t_{\perp}}}{\sqrt{n}} [(\mathsf{I} - \mathsf{P}^{\parallel}_{M_t})Z'_{t}]_i + u_i,  w_i) \\
& \hspace{1in}  - \psi_b(\mathsf{b}_{t,i} + \frac{\norm{q^t_{\perp}}}{\sqrt{n}} Z'_{t_i} + u_i,  w_i)\Big] \Big \lvert \geq \frac{\e}{2}\Big),
\end{split}
\label{eq:Bprobt2}
\ee
where we have omitted the conditioning on the RHS to shorten notation. Label the two terms in \eqref{eq:Bprobt2} as $T_{1,a}$ and $T_{1,b}$.  To complete the proof we show that both terms are  bounded by $Ke^{-\kappa n  \e^2/t}$.   

First consider $T_{1,b}$.  We note that
\be
\mathsf{P}^{\parallel}_{M_t} Z'_t  = \sum_{r=0}^{t-1} \frac{\tilde{m}^r}{\sqrt{n}} \Big[ \frac{(\tilde{m}^r)^* Z'_{t}}{\sqrt{n}}  \Big] \overset{d}{=} \sum_{r=0}^{t-1} \frac{\tilde{m}^r}{\sqrt{n}} U_r,
\label{eq:PMt_rep}
\ee
where $\tilde{m}^r$, $0 \leq r \leq t-1$, are columns of $\tilde{M}_{t}$, which form an orthogonal basis for $M_{t}$ with $\tilde{M}^*_{t} \tilde{M}_{t} =n \mathsf{I}_t$, and $U_1, \ldots, U_t$ are i.i.d.\ $\sim \mc{N}(0,1)$.  Then,
\be
\begin{split}
& T_{1,b} \overset{(a)}{\leq} P\Big(\frac{C}{n} \sum_{i=1}^n\Big \lvert  \frac{\norm{q^t_{\perp}}}{\sqrt{n}} [\mathsf{P}^{\parallel}_{M_{t}} Z'_{t}]_i \Big \lvert \geq \frac{\e}{2} \Big) \\
& \overset{(b)}{\leq} P\Big( \frac{C}{n} \sum_{i=1}^n\Big \lvert  (\sigma_t^{\perp} + \e_0) [\mathsf{P}^{\parallel}_{M_{t}} Z'_{t}]_i \Big \lvert \geq \frac{\e}{2} \Big)  \\
& = P\Big( \Big \lvert \frac{C}{n}  \sum_{i=1}^n  \sum_{r=0}^{t-1}  \frac{\tilde{m}^r_i  U_r}{\sqrt{n}} \Big \lvert \geq \frac{\e}{2 \abs{\sigma_t^{\perp} + \e_0}} \Big) \\
& \stackrel{(c)}{=} P\Big( \Big \lvert \frac{C}{n}  \sum_{i=1}^n  \Big( \sum_{r=0}^{t-1} (\tilde{m}^r_i)^2  \Big)^{1/2} \frac{Z}{\sqrt{n}} \Big \lvert  \geq \frac{\e}{2 \abs{\sigma_t^{\perp} + \e_0}} \Big) \\
& \overset{(d)}{\leq} P \Big ( \sqrt{\frac{t}{n}} \abs{Z}  \geq \frac{\e}{2C \abs{\sigma_t^{\perp} + \e_0}} \Big )  \leq 2 e^{-\kappa n \e^2/t}. 
\end{split}
\label{eq:Bprobt5_1}
\ee
In the above, $(a)$ follows from Fact \ref{fact:lip_deriv} for a suitable constant $C>0$.  Step $(b)$ holds since we are conditioning on event $\mc{F}^c$ defined in \eqref{eq:BFtdef}. In step $(c)$, $Z \sim \mc{N}(0,1)$ since $ \sum_{r}  \tilde{m}^r_i  U_r$ is a zero-mean Gaussian with variance $\sum_{r} (\tilde{m}^r_i)^2$. Step $(d)$ uses the Cauchy-Schwarz inequality and the fact that $\norm{\tilde{m}^r} = \sqrt{n}$ for $0 \leq r < t$.

Finally $T_{1,a}$, the first term in \eqref{eq:Bprobt2}, can be bounded using Hoeffding's inequality. Noting that all quantities except $Z'_t$ are in $\mscrs_{t, t}$, define the shorthand  $\textsf{diff}(Z'_{t_i}) :=  \psi_b(\sum_{r=0}^{t-1} \hat{\gamma}^t_r b^{r}_i+ \frac{1}{\sqrt{n}}\norm{q^t_{\perp}}  Z'_{t_i} + u_i, w_i ) - \psi_b(\sum_{r=0}^{t-1} \hat{\gamma}^t_r b^{r}_i + \sigma^{\perp}_t Z'_{t_i}, w_i )$. Then the  upper tail of $T_{1,a}$  can be written as 
 \be
 P\Big( \frac{1}{n} \sum_{i=1}^n\textsf{diff}(Z'_{t_i}) - \mathbb{E}[\textsf{diff}(Z'_{t_i})]  \geq \frac{\e}{2} -  \frac{1}{n} \sum_{i=1}^n\mathbb{E}[\textsf{diff}(Z'_{t_i})] \Big).
 \label{eq:diffi_boundBt}
 \ee
Using the conditioning on $\mc{F}^c$ and steps similar to those in $\mc{B}_0(b)$.(iv), we can show that $\frac{1}{n} \sum_{i} \mathbb{E}[\textsf{diff}(Z'_{t_i})]]$  $\leq \frac{1}{4} \e$ for $\e_0 \leq C (\sigma_t^{\perp}) \e$, where the constant $C > 0$ can be explicitly computed. For such $\e_0$, using Hoeffding's inequality the probability in 
\eqref{eq:diffi_boundBt} can be bounded by $e^{-n\e^2/(32B^2)}$, where $B$ is the upper bound on $\abs{\textsf{diff}(\cdot)}$. A similar bound holds for the lower tail of $T_{1,a}$. Thus both terms of \eqref{eq:Bprobt2} are bounded by $K\exp\{-\kappa n  \e^2/t\}$.

The proof is completed by collecting the above bounds for each of the terms in \eqref{eq:Bjt12}, and observing that the overall  bound is dominated by $P(T_1)$ in \eqref{eq:PPit}. Hence the final bound is of the form $K t^2 K_{t-1} \exp\{-\kappa \kappa_{t-1} n \e^2/  t^4\}$.   


\textbf{(c)} The function $\phi_b(b^t_i, w_{i}) := b^t_i w_i  \in PL(2)$ by Lemma \ref{lem:Lprods}.  Then by $\mathcal{B}_t (b).\text{(iii)}$,
$ \frac{1}{n}(b^t)^*w \doteq \sigma_t  \mathbb{E} [\breve{Z}_t W ] =0$.  

\textbf{(d)} The function $\phi_b(b^r_i, b^t_i, w_{i}) := b^r_i b^t_i  \in PL(2)$ by Lemma \ref{lem:Lprods}. The result then follows from $\mathcal{B}_t (b).\text{(iii)}$.

\textbf{(e)} The function $\phi_b(b^r_i,b^t_i, w_{i})$  $:= g_r(b^r_i, w_i) g_t(b^t_i, w_i) \in PL(2)$  since $g_t$ is Lipschitz continuous (by Lemma \ref{lem:Lprods}).  Then by $\mathcal{B}_t (b).\text{(iii)}$,
\begin{align*}
 \frac{1}{n}(m^r)^* m^t \doteq \mathbb{E}[g_r(\sigma_r \breve{Z}_r, W ) g_t( \sigma_t \breve{Z}_t, W )]= \breve{E}_{r,t}.
\end{align*}
where the last equality is due to the definition in \eqref{eq:Edef}.  

\textbf{(f)} The concentration of $\xi_t$ around $\hat{\xi_t}$ follows from $\mathcal{B}_t (b).$(iv) applied to the function $\psi_b(b^t_i,w_i):=g_t'(b^t_i,w_i)$. 
Next, for $r \leq t$,  $\phi_b(b^0_i, \ldots, b^t_i, w_{i}) := b^r_i  g_t(b^t_i, w_i)= b^r_i m_i \in PL(2)$, by Lemma \ref{lem:Lprods}. Thus by $\mathcal{B}_t (b).\text{(iii)}$,
\ben
\begin{split}   
\frac{1}{n}(b^r)^* m^t &\doteq \mathbb{E}[\sigma_r \breve{Z}_r \, g_t( \sigma_t \breve{Z}_t, W )]
 \end{split}
 \een
 and
 \ben
\begin{split}   
\mathbb{E}[\sigma_r \breve{Z}_r \, g_t( \sigma_t \breve{Z}_t, W )] &\stackrel{(a)}{=} 
\sigma_r \sigma_t  \expec[\breve{Z}_r \breve{Z}_t] \expec[ g'_t( \sigma_t \breve{Z}_t, W )] \\
&= \tilde{E}_{r,t} \expec[ g'_t( \sigma_t \breve{Z}_t, W )] = \tilde{E}_{r,t} \hat{\xi}_t,
 \end{split}
 \een
where $(a)$ holds due to Stein's lemma (Fact \ref{fact:stein}).

\textbf{(g)}    For 
$1 \leq r,s \leq t$, note that $[ \mathbf{M}_t]_{r,s}= \frac{1}{n}(m^{r-1})^*m^{s-1}$.  Hence by $\mc{B}_{t-1}(e)$,  $[ \mathbf{M}_t ]_{r,s}$  concentrates on $[\breve{C}^{t}]_{r,s} = \breve{E}_{r-1,s-1}$.  We first show \eqref{eq:Msing}. By Fact \ref{fact:eig_proj}, if $\frac{1}{n}\norm{m^r_{\perp}}^2 \geq c >0$ for all $0 \leq r \leq t-1$, then $\mathbf{M}_t$ is invertible.  Note from  $\mc{B}_{t-1}(h)$ that $\frac{1}{n}\norm{m^r_{\perp}}^2$ concentrates on $(\tau_r^{\perp})^2$, and $(\tau_r^{\perp})^2 > \varepsilon_3$ by the stopping criterion assumption. Choosing $c = \frac{1}{2}\varepsilon_3$, we therefore have
\be
\begin{split}
&P(\mathbf{M}_t \text{ singular})  \leq \sum_{r=0}^{t-1} P\Big(\Big \lvert \frac{1}{n}\norm{m^r_{\perp}}^2 - (\tau_r^{\perp})^2 \Big \lvert \geq   \frac{1}{2}\varepsilon_3 \Big) \\
& \leq \sum_{r=0}^{t-1} K_{r-1} e^{-  \kappa_{r-1} n (\varepsilon_3)^2/4}
\leq t K_{t-1} e^{- \kappa \kappa_{t-1} n (\varepsilon_3)^2},
  \end{split}
 \label{eq:mr_perp_conc}
\ee
where the second inequality follows from  $\mc{B}_0 (h) -\mc{B}_{t-1} (h)$.

Next, we show \eqref{eq:Bg}. Recall the expression for $\mbf{M}_t^{-1}$ from \eqref{eq:Mt1_inverse0}:
\be
 \mathbf{M}_{t}^{-1} = \left[ \begin{array}{cc}
\mathbf{M}_{t-1}^{-1} +  \frac{n\alpha^{t-1}(\alpha^{t-1})^*}{\norm{m^{t-1}_{\perp}}^{2}}  & - \frac{n \alpha^{t-1}}{\norm{m^{t-1}_{\perp}}^{2}}  \\
- \frac{n(\alpha^{t-1})^*}{\norm{m^{t-1}_{\perp}}^{2}}   & \frac{n}{\norm{m^{t-1}_{\perp}}^{2}}  \end{array} \right],
\label{eq:Mt1_inverse}
\ee
Block inversion can be similarly used to decompose $\breve{C}^t$ in terms of 
$\breve{C}^{t-1}$, which gives the concentrating values of the elements in \eqref{eq:Mt1_inverse}.

Let $\mc{F}_{r}$ denote the event that $\mathbf{M}_{r}^{-1}$  is invertible, for $r \in [t]$. Then, for $i,j \in [t]$, we have
\be
\begin{split}
& P\Big( \Big\lvert [\mathbf{M}_{t}^{-1} ]_{i,j} - [\breve{C}_{t}^{-1}]_{i,j}\Big \lvert  \geq \e \mid \mc{F}_t \Big)  \\
& \leq P(\mc{F}_{t-1}^c) + P\Big(\Big \lvert [\mathbf{M}_{t}^{-1} ]_{i,j} - [\breve{C}_{t}^{-1}]_{i,j} \Big\lvert \geq \e \mid \mc{F}_t, \mc{F}_{t-1} \Big) \\
& \leq  (t-1)K_{t-2} e^{- \kappa \kappa_{t-2} n } \\
&\qquad +  P\Big(\Big \lvert [\mathbf{M}_{t}^{-1} ]_{i,j} - [\breve{C}_{t}^{-1}]_{i,j} \Big \lvert \geq \e \mid \mc{F}_t, \mc{F}_{t-1} \Big),
\end{split} 
\label{eq:Mt_split}
\ee
where the final inequality follows from the inductive hypothesis $\mathcal{B}_{t-1}(g)$.  Using the representation in \eqref{eq:Mt1_inverse}, we bound the second  term in \eqref{eq:Mt_split} for  $i,j \in [t]$. In what follows, we drop the conditioning on $\mc{F}_t, \mc{F}_{t-1}$ for brevity.

First, consider the entry at $i=j=t$. By $\mc{B}_{t-1}(h)$ and Lemma \ref{inverses}, 
\[ P( \abs{n \| m_{t-1}^{\perp} \|^{-2} -  (\tau_{t-1}^{\perp})^{-2} } \geq \e ) \leq K_{t-1} e^{- \kappa \kappa_{t-1} n \e^2}. \]
Next, consider the $i^{th}$ element of $-n \norm{m^{t-1}_{\perp}}^{-2} \alpha^{t-1}$. For $i \in [t-1]$,
\begin{align}
& P (\lvert n \norm{m^{t-1}_{\perp}}^{-2} \alpha^{t-1}_{i-1} - (\tau_{t-1}^{\perp})^{-2} \hat{\alpha}^{t-1}_{i-1}  \lvert \geq \e) \nonumber  \\
& \leq  2K_{t-1} e^{-\kappa \kappa_{t-1} n \e^2}, 
\label{eq:matlem_bound1}
\end{align}
which follows from $\mc{B}_{t-1} (g)$, the concentration bound obtained above for $n \norm{m^{t-1}_{\perp}}^{-2}$, and combining these via Lemma \ref{products}.  

Finally consider element $(i,j)$ of $\mathbf{M}_{t-1}^{-1} + n \norm{m^{t-1}_{\perp}}^{-2} \alpha^{t-1}(\alpha^{t-1})^*$  for $i,j \in [t-1]$. We have
\begin{align*}
& P \Big(\Big \lvert [\mathbf{M}_{t-1}^{-1}]_{i,j} + \frac{n \alpha^{t-1}_{i-1} \alpha^{t-1}_{j-1}}{\norm{m^{t-1}_{\perp}}^{2}} - [\breve{C}_{t}^{-1}]_{i,j} -\frac{ \hat{\alpha}^{t-1}_{i-1} \hat{\alpha}^{t-1}_{j-1}}{(\tau_{t-1}^{\perp})^{2}} \Big \lvert \geq \e\Big) \\
&\overset{(a)}{\leq} P \Big( \Big \lvert [\mathbf{M}_{t-1}^{-1}]_{i,j}  - [\breve{C}_{t}^{-1}]_{i,j} \Big \lvert \geq \frac{\e}{2}\Big)   \\
& \quad+ P \Big( \lvert \alpha^{t-1}_{j-1} -  \hat{\alpha}^{t-1}_{j-1} \lvert \geq \frac{\e'}{2}\Big) \\
&\quad  + P \Big(\lvert n \norm{m^{t-1}_{\perp}}^{-2} \alpha^{t-1}_{i-1} - (\tau_{t-1}^{\perp})^{-2} \hat{\alpha}^{t-1}_{i-1}  \lvert \geq \frac{\e'}{2}\Big)  \\
&\overset{(b)}{\leq} K_{t-1} e^{-\frac{\kappa_{t-1}n \e^2}{4}} + 2 K_{t-1} e^{-\frac{\kappa \kappa_{t-1} n \e'^2}{4}} + K_{t-1} e^{-\frac{\kappa_{t-1} n \e'^2}{4}} \\
&\leq 4 K_{t-1} e^{- \kappa \kappa_{t-1}n \e^2}.
\end{align*}
Step $(a)$ follows from Lemma \ref{sums} and Lemma \ref{products} with
$\e' := \min \Big(\sqrt{\frac{\e}{3}}, \frac{\e (\tau_{t-1}^{\perp})^{2}}{3 \hat{\alpha}^{t-1}_{i-1}}, \frac{\e}{3  \hat{\alpha}^{t-1}_{j-1}}\Big)$.
Step $(b)$ follows from the inductive hypothesis, $\mc{H}_t(g)$, and \eqref{eq:matlem_bound1}.

Next, we prove the concentration of $\alpha^t$ around $\hat{\alpha}^t$. Recall from Section \ref{subsec:defs} that $\alpha^t = \frac{1}{n} \textbf{M}_t^{-1} M_t^* m^t $ where $\mathbf{M}_t := \frac{1}{n} M_{t}^* M_{t}$. Thus 
 for $1 \leq i \leq t$, $\alpha^t_{i-1} = \frac{1}{n} \sum_{j=1}^{t} [\textbf{M}_t^{-1}]_{i,j} (m^{j-1})^* m^t$.
Then from the definition of $\hat{\alpha}^{t}$ in \eqref{eq:hatalph_hatgam_def}, for  $1 \leq i \leq t$,
\begin{align*}
& P( \lvert \alpha^t_{i-1} - \hat{\alpha}^t_{i-1} \lvert \geq \epsilon ) =  \\
&  P\Big(\Big \lvert \sum_{j=1}^{t} \Big[\frac{1}{n}[\textbf{M}_t^{-1}]_{i,j} (m^{j-1})^* m^t- [(\breve{C}^t)^{-1}]_{i,j} \breve{E}_{j-1,t} \Big] \Big \lvert \geq \e \Big) \\
&\overset{(a)}{\leq} \sum_{j=1}^{t}P\Big(\Big \lvert \frac{1}{n} (m^{j-1})^* m^t - \breve{E}_{j-1,t} \Big \lvert \geq \tilde{\e}_{j} \Big) \\
&\qquad +\sum_{j=1}^{t}P( \lvert [\textbf{M}_t^{-1} ]_{i,j} - [(\breve{C}^t)^{-1}]_{i,j} \lvert \geq \tilde{\e}_j )  \\
&\overset{(b)}{\leq}  K t^4 K_{t-1} e^{-\kappa \kappa_{t-1}n\e^2/ t^{9}} + 4 t K_{t-1} e^{-\kappa \kappa_{t-1} t^{-2} n \e^2} .
\end{align*}
Step $(a)$ uses
$\tilde{\e}_j := \min \Big \{\sqrt{\frac{\epsilon}{3t}} , \frac{\epsilon}{3t \breve{E}_{j-1,t}} , \frac{\epsilon}{3t [(\breve{C}^t)^{-1}]_{k,j}}  \Big\}$ and follows from Lemma \ref{sums} and Lemma \ref{products}.
Step $(b)$ uses $\mathcal{B}_{t} (e)$ and the work above. 

\textbf{(h)} First, note that 
$\norm{m^t_{\perp}}^2 =\| m^t\| ^2 -\| m^t_{\parallel}\| ^2 =\| m^t\|^2 -\| M_t \alpha^t\|^2$. 
Using the definition of $\tau_t^{\perp}$ in \eqref{eq:sigperp_defs}, 
\begin{equation}
\begin{split}
& P\Big(\Big \lvert \frac{1}{n}\norm{m^t_{\perp}}^2 - (\tau_{t}^{\perp})^2 \Big\lvert \geq \epsilon \Big) \\
&= P\Big(\Big \lvert \frac{1}{n}\norm{m^t}^2 - \frac{1}{n}\norm{M_t \alpha^t}^2 - \tau_{t}^2 + (\hat{\alpha}^{t})^* \breve{E}_t \Big \lvert \geq \epsilon\Big)  \\
&\leq P\Big(\Big \lvert \frac{\norm{m^t}^2}{n} - \tau_t^2 \Big \lvert \geq \frac{\epsilon}{2} \Big)  \\
& \quad + P\Big(\Big \lvert \frac{\norm{M_t \alpha^t}^2}{n} - (\hat{\alpha}^{t})^* \breve{E}_t \Big \lvert \geq \frac{\epsilon}{2}\Big).
\label{eq:Bgt1}
\end{split}
\ee
The bound for the first term in \eqref{eq:Bgt1} follows by $\mathcal{B}_{t} (e)$. 
For the second term,
\ben
\begin{split}
\norm{M_t \alpha^t}^2 = n (\alpha^t)^* \textbf{M}_t \alpha^t &\stackrel{(a)}{=} (\alpha^t)^*\textbf{M}_t \textbf{M}_t^{-1} {M_t^* m^t} \\
&= (\alpha^t)^* {M_t^* m^t} = \sum_{i=0}^{t-1} \alpha^t_i (m^i)^* m^t, 
\end{split}
\een 
where $(a)$ holds because $\alpha^t = \textbf{M}_t^{-1} {M_t^* m^t}/n$. Hence 
\begin{align*}
& P\Big(\Big \lvert \frac{1}{n}\norm{M_t \alpha^t}^2 - (\hat{\alpha}^{t})^* \breve{E}_t \Big \lvert \geq \frac{\epsilon }{2} \Big) \\
& = P\Big(\Big\lvert \sum_{i=0}^{t-1} \Big[\frac{1}{n}\alpha^t_i (m^i)^* m^t - \hat{\alpha}^{t}_i \breve{E}_{i,t} \Big] \Big \lvert \geq \frac{\epsilon}{2} \Big) \nonumber \\
&\overset{(a)}{\leq} \sum_{i=0}^{t-1} P( \lvert \alpha^t_i - \hat{\alpha}^{t}_i  \lvert \geq \tilde{\e}_i ) + \sum_{i=0}^{t-1} P\Big(\Big \lvert \frac{1}{n}(m^i)^* m^t - \breve{E}_{i,t} \Big \lvert \geq \tilde{\e}_i \Big) \\
&\overset{(b)}{\leq}  K t^5 K_{t-1} e^{-\kappa \kappa_{t-1}n\e^2/ t^{11} }  + K t^4 K_{t-1} e^{-\kappa \kappa_{t-1}n\e^2/ t^{9}}.
\end{align*}
Step $(a)$ follows Lemma \ref{sums} and Lemma \ref{products}, using
$\tilde{\e}_i := \min \Big \{\sqrt{\frac{\epsilon}{6t}} , \frac{\epsilon}{6t \breve{E}_{i,t}} , \frac{\epsilon}{6t \hat{\alpha}^{t}_i}  \Big\}$,
and step $(b)$ using $\mathcal{B}_{t} (e)$ and the proof of $\mathcal{B}_{t} (g)$ above.


\subsection{Step 4: Showing $\mc{H}_{t+1}$ holds}

The statements in $\mc{H}_{t+1}$ are proved assuming that $\mc{B}_{t}, \mc{H}_t$ hold due to the induction hypothesis. 

\textbf{(a)} The proof of $\mc{H}_{t+1} (a)$ is similar to that of  $\mc{B}_t (a)$, and uses the following lemma, which is analogous to Lemma \ref{lem:Mv_conc}.
\begin{lem}
\label{lem:Qv_conc}
Let $v := \frac{1}{n}B^*_{t+1} m_t^{\perp} - \frac{1}{n}Q_{t+1}^*(\xi_t q^t - \sum_{i=0}^{t-1} \alpha^t_i \xi_i q^i)$ and $\mathbf{Q}_{t+1} := \frac{1}{n}Q_{t+1}^* Q_{t+1}$.  Then for $j \in [t+1]$,
\ben
P(\lvert [\mathbf{Q}_{t+1}^{-1} v]_{j} \lvert \geq \e) \leq Kt^2 K'_{t-1} \exp\{-\kappa'_{t-1} n\e^2/t^2\}.
\een
\end{lem}


\textbf{(b)--(h)} The proofs of the results in $\mc{H}_{t+1} (b) - \mc{H}_{t+1} (h)$ are along the same lines as $\mc{B}_t (b) - \mc{B}_t  (h)$. By the end of step  $\mc{H}_{t+1} (h)$, we will similarly pick up a $t^5 K$ term in the pre-factor in front of the exponent, and a $\kappa t^{-11}$ term in the exponent.  It then follows that the $K_t, \kappa_t$ are as given in \eqref{eq:Kkappa_def}.

\appendices
\renewcommand{\theequation}{A.\arabic{equation}}
\section{Concentration Lemmas}  \label{app:conc_lemma}

In the following, $\e >0$ is assumed to be a generic constant, with additional conditions specified whenever needed.


\begin{applem}[Hoeffding's inequality]
\label{lem:hoeff_lem}
If $X_1, \ldots, X_n$ are bounded random variables such that $a_i \leq X_i \leq b_i$, then for $\nu = 2[\sum_{i} (b_i -a_i)^2]^{-1}$
\begin{align*}
P\Big( \frac{1}{n}\sum_{i=1}^n (X_i - \expec X_i) \geq \e \Big) &\leq e^{ -\nu n^2 \e^2}, \\
P\Big( \Big \lvert \frac{1}{n}\sum_{i=1}^n (X_i - \expec X_i) \Big \lvert \geq \e \Big) &\leq 2e^{ -\nu n^2 \e^2}.
\end{align*}
\end{applem}

\begin{applem}[Concentration of Sums]
\label{sums}
If random variables $X_1, \ldots, X_M$ satisfy $P(\abs{X_i} \geq \e) \leq e^{-n\kappa_i \e^2}$ for $1 \leq i \leq M$, then 
\ben
P\Big(  \lvert \sum_{i=1}^M X_i  \lvert \geq \e\Big) \leq \sum_{i=1}^M P\left(|X_i| \geq \frac{\e}{M}\right) \leq M e^{-n (\min_i \kappa_i) \e^2/M^2}.
\een
\end{applem}

\begin{applem}[Concentration of Products]
\label{products} 
For random  variables $X,Y$ and non-zero constants $c_X, c_Y$, if
\ben
P\left(  | X- c_X |  \geq \e \right) \leq K e^{-\kappa n \e^2},
\een
and 
\ben
P\left(  | Y- c_Y  |  \geq \e \right) \leq K e^{-\kappa n \e^2},
\een
then the probability  $P\left( \left | XY - c_Xc_Y \right |  \geq \e \right)$ is bounded by 
\begin{align*}  
&  P\Big(  | X- c_X  |  \geq \min\Big( \sqrt{\frac{\e}{3}}, \frac{\e}{3 c_Y} \Big) \Big) \\
&\qquad  +  
P\Big(| Y- c_Y |  \geq \min\Big( \sqrt{\frac{\e}{3}}, \frac{\e}{3 c_X} \Big) \Big) \\
&\leq 2K e^{-\frac{\kappa n \e^2}{9\max(1, c_X^2, c_Y^2)}}.
\end{align*}
\end{applem}

\begin{IEEEproof}
The probability of interest, $P\left( \left | XY - c_Xc_Y \right |  \geq \e \right) $, equals
\[P\left( \left | (X -c_X)(Y-c_Y) +  (X-c_X)c_Y  + (Y-c_Y)c_X \right |  \geq \e \right). \]
The result follows by noting that if $ \left | X- c_X \right |  \leq \min( \sqrt{\frac{\e}{3}}, \frac{\e}{3 c_Y})$ and $ \left | Y- c_Y \right |  \leq \min ( \sqrt{\frac{\e}{3}}, \frac{\e}{3 c_X} )$, then the following terms are all bounded by $\frac{\e}{3}$:
\ben
\abs{(X-c_X)c_Y}, \abs{(Y-c_X)c_Y}, \text{ and } \abs{(X -c_X)(Y-c_Y)}. \qedhere
\een
\end{IEEEproof}

\begin{applem}[Concentration of Square Roots]
\label{sqroots}
Let $c \neq 0$.
\ben
\text{If } P( \lvert X_n^2 - c^2 \lvert \geq \epsilon ) \leq e^{-\kappa n \epsilon^2},
\een
then
\ben
P (\lvert \abs{X_n} - \abs{c} \lvert \geq \epsilon) \leq e^{-\kappa n \abs{c}^2 \epsilon^2}.
\een
\end{applem}

\begin{IEEEproof}
If $\e \leq c^2$, then the event  $c^2 - \epsilon \leq X_n^2 \leq c^2 + \epsilon$ implies that $\sqrt{c^2 - \e} \leq \abs{X_n} \leq \sqrt{c^2 + \e}$.  On the other hand, if $\e \geq c^2$, then $c^2 - \epsilon \leq X_n^2 \leq c^2 + \epsilon$ implies that $0 \leq \abs{X_n} \leq \sqrt{c^2 + \e}$.  Therefore, 
$\lvert X_n^2 - c^2 \lvert \leq \epsilon$ implies
\ben
\left \lvert \abs{X_n} - \abs{c} \right \lvert \leq \abs{c} \max (1 - \sqrt{(1 - (\e/c^2))_+}, \sqrt{1 + (\e/c^2)} - 1),
\een
where $x_+ := \max\{x,0 \}$. Note, $(1 + x)^{1/2} \leq 1 + \frac{1}{2}x$ for $x \geq 0$,  and $(1 - x)^{1/2} \geq 1-x$ for $x \in (0,1)$.  Using these, we conclude that $\lvert X_n^2 - c^2 \lvert \leq \epsilon$ implies
\ben
\begin{split}
\left \lvert \abs{X_n} - \abs{c} \right \lvert &  \leq \abs{c} \max\Big(1 - \sqrt{\Big(1 - \frac{\epsilon}{c^2}\Big)_+}, \sqrt{1 + \frac{\epsilon}{c^2}} - 1\Big) \\
& \leq \abs{c} \max\Big(\frac{\e}{c^2}, \frac{\epsilon}{2c^2}\Big) = \frac{\e}{\abs{c}}. \qedhere
\end{split} 
\een
\end{IEEEproof}

\begin{applem}[Concentration of Powers]
\label{powers}
Assume $c \neq 0$ and $0 < \e \leq 1$.  Then for any integer $k \geq 2$,
\ben
\text{if } P( \lvert X_n - c \lvert \geq \epsilon ) \leq e^{-\kappa n \epsilon^2},
\een
then
\ben
P(\lvert X_n^k - c^k  \lvert \geq \epsilon) \leq e^{ {-\kappa n \e^2}/[{(1+\abs{c})^k -\abs{c}^k}]^2}.
\een
\end{applem}
\begin{IEEEproof}
  Without loss of generality, assume that $c>0$. First consider the case where $\e < c$. Then  $c - \epsilon \leq X_n \leq c + \epsilon$ implies
  \[ (c-\e)^k -c^k \leq X_n^k -c^k \leq  (c+\e)^k -c^k = \sum_{i=1}^k {k \choose i} c^{k-i} \e^{i}.  \]
  Hence, $\abs{X_n -c} \leq \e$ implies $\abs{X_n ^k-c^k} \leq \e c_0$, where
  \[  c_0 =  \sum_{i=1}^k {k \choose i} c^{k-i} \e^{i-1} <  \sum_{i=1}^k {k \choose i} c^{k-i} =(1+c)^k - c^k.   \]
  Therefore,
  \be  P(\lvert X_n^k - c^k \lvert \geq \epsilon) \leq P\left(\left \lvert X_n - c \right \lvert \geq {\epsilon}/{c_0}   \right)  
  \leq e^{{-\kappa n \e^2}/{[(1+c)^k - c^k}]^2}.  \label{eq:Xkck}  \ee
  
  For the case where $0 < c < \e <1$, $X_n \in [c - \epsilon, c + \epsilon]$ implies $(c-\e)^k -c^k \leq X^k -c^k \leq  (c+\e)^k -c^k$.  Using $\e <1$, we note that the absolute values of 
  \[ (c-\e)^k -c^k =\sum_{i=1}^k {k \choose i} c^{k-i} (-\e)^{i},\]
and
\[(c+\e)^k -c^k = \sum_{i=1}^k {k \choose i} c^{k-i} \e^{i},\]
  are bounded  by  $c_1:= (1+c)^k - c^k$.  Thus  $\abs{X_n -c} \leq \e$ implies $\abs{X_n ^k-c^k} \leq \e c_1$. 
Therefore the same bound as in \eqref{eq:Xkck} holds when $0<c<\e<1$ (though a tighter bound could be obtained in this case).
\end{IEEEproof}

\begin{applem}[Concentration of Scalar Inverses]
\label{inverses} Assume $c \neq 0$ and $0<\e <1$. If
\ben
P( \lvert X_n - c \lvert \geq \epsilon ) \leq e^{-\kappa n \epsilon^2},
\een
then
\ben
P( \lvert X_n^{-1} - c^{-1} \lvert \geq \epsilon ) \leq 2 e^{-n \kappa \e^2 c^2 \min\{c^2, 1\}/4}.
\een
\end{applem}

\begin{IEEEproof}
Without loss of generality, we can assume that $c >0$. We have 
\ben
\begin{split}
&P(\lvert X_n^{-1} - c^{-1}  \lvert \leq \epsilon) \\
&\qquad =  P( c^{-1}  - \epsilon \leq X_n^{-1}  \leq c^{-1} + \epsilon ).
\end{split}
\een
First consider the case $0 < \e < c^{-1}$. Then, $X_n$ is strictly positive in the interval of interest, and therefore
\begin{align}
\label{eq:ce_lt1}
&P( c^{-1}  - \epsilon \leq X_n^{-1} \leq c^{-1}  + \epsilon) \\
&= P\Big(  \frac{-\e c}{c^{-1} + \e} \leq X_n -c \leq \frac{\e c}{c^{-1} - \e} \Big) \nonumber \\
&\geq 1- e^{-n \kappa \e^2 c^2/(\e + c^{-1})^2}  \geq 1-  e^{-n \kappa \e^2 c^4/4}.
\end{align}
Next consider $0 < c^{-1} < \e <1$. The probability to be bounded can be written as 
\be
\label{eq:ce_gt1}
\begin{split}
&P( X_n^{-1}  \geq c^{-1}  + \epsilon ) + P( - ( \e - c^{-1} ) \leq X_n^{-1}  < 0   )  \\
& = P\Big( X_n - c  \leq \frac{-\e c}{\e + c^{-1}} \Big) + P\Big( \frac{-\e c}{\e - c^{-1}} \leq X_n - c \leq - c \Big) \\
 & \leq  e^{-\frac{n \kappa \e^2 c^2}{(\e + c^{-1})^2}} +  e^{-n \kappa  c^2}  \leq e^{-n \kappa  c^2/4}  + e^{-n \kappa  c^2}  \leq 2 e^{-n \kappa  c^2/4},
 \end{split}
 \ee
where the last two inequalities are obtained using $\e > c^{-1}$ and $\e <1$, respectively. The bounds \eqref{eq:ce_lt1} and \eqref{eq:ce_gt1} together give the result of the lemma.
\end{IEEEproof}

\section{Gaussian and Sub-Gaussian Concentration}  \label{app:gauss_lemmas}

\renewcommand{\theequation}{B.\arabic{equation}}

\begin{applem}
\label{lem:normalconc}
For a random variable $Z \sim \mc{N}(0,1)$ and  $\e > 0$,
$P\Big( \abs{Z} \geq \e \Big) \leq 2e^{-\frac{1}{2}\e^2}$.
\end{applem}

\begin{applem}[$\chi^2$-concentration]
For  $Z_i$, $i \in [n]$ that are i.i.d.\ $\sim \mc{N}(0,1)$, and  $0 \leq \e \leq 1$,
\[P\Big(\Big \lvert \frac{1}{n}\sum_{i=1}^n Z_i^2 - 1\Big \lvert \geq \e \Big) \leq 2e^{-n \e^2/8}.\]
\label{subexp}
\end{applem}

\begin{applem}\cite{BLMConc} Let $X$ be a centered sub-Gaussian random variable with variance factor $\nu$, i.e., $\ln \expec[e^{tX}] \leq \frac{t^2 \nu}{2}$, for all $ t \in \mathbb{R}$. Then $X$ satisfies:
\begin{enumerate}
\item For all $x> 0$, $P(X > x)  \vee P(X <-x) \leq e^{-\frac{x^2}{2\nu}}$, for all $x >0$.
\item For every integer $k \geq 1$,
\be
\expec[X^{2k}] \leq 2 (k !)(2 \nu)^k \leq (k !)(4 \nu)^k.
\label{eq:subgauss_moments}
\ee
\end{enumerate}
\end{applem}

\begin{applem}
\label{lem:PLsubgaussconc}
Let $Z_1, \ldots, Z_t \in \reals^N$ be random vectors such that $(Z_{1,i}, \ldots, Z_{t,i})$ are i.i.d.\ across $i \in [n]$, with  $(Z_{1,i}, \ldots, Z_{t,i})$ being jointly Gaussian with zero mean, unit variance and covariance matrix $K \in \reals^{t \times t}$. Let $G \in  \mathbb{R}^N$ be a random vector with entries  $G_1, \ldots, G_N$  i.i.d.\   $\sim p_{G}$, where $p_G$ is sub-Gaussian with variance factor $\nu$.   Then for any pseudo-Lipschitz function $f: \reals^{t+1} \to \reals$, non-negative constants $\sigma_1, \ldots, \sigma_t$, and  $0< \e \leq 1$, we have
\begin{align*}
& P\Big(\Big\lvert \frac{1}{N}\sum_{i=1}^N f(\sigma_1 Z_{1,i}, \ldots, \sigma_tZ_{t,i}, G_i) \\
&\qquad \qquad \qquad  - \mathbb{E}[f(Z_{1,1}, \ldots, Z_{t,1}, G)] \Big \lvert \geq \e \Big) \\
& \leq 2 \exp \Big\{ \frac{ - N \e^2}{ 128 L^2 (t+1)^2(  \nu + 4 \nu^2 + \sum_{m=1}^t  (\sigma_m^{2} +  4  \sigma_m^{4})) }  \Big\},
\end{align*}
where $L >0$ is an absolute constant. ($L$ can be bounded above  by three times the pseudo-Lipschitz constant of $f$.)  
\end{applem}
\begin{IEEEproof}
Without loss of generality, assume $\mathbb{E}[ f(\sigma_1 Z_{1,i}, \ldots, \sigma_tZ_{t,i}, G_i) ]  = 0$ for $i \in [N]$.  In what follows we demonstrate the upper-tail bound:
\be
P \Big(\frac{1}{N}\sum_{i=1}^N f(\sigma_1 Z_{1,i}, \ldots, \sigma_tZ_{t,i}, G_i) \geq \e \Big) \leq  \exp \Big\{ \frac{-N \e^2}{4 \tilde{\kappa}_t} \Big\},
\label{eq:PLsubgauss1}
\ee
where
 \be 
\tilde{\kappa}_t = 32 L^2(t+1)^2 (  \nu + 4 \nu^2 + \sum_{m=1}^t  (\sigma_m^{2} +  4  \sigma_m^{4})).
\label{eq:tkt}
\ee 
 The lower-tail bound follows similarly.
 
  Using the Cram{\'e}r-Chernoff method, for any $s>0$ we can write
\begin{align}
& P \Big(\frac{1}{N}\sum_{i=1}^N f(\sigma_1 Z_{1,i}, \ldots, \sigma_tZ_{t,i}, G_i) \geq \e \Big)   \\
  & \qquad \leq \mathbb{E}\Big[e^{s\sum_{i=1}^N f(\sigma_1 Z_{1,i}, \ldots, \sigma_tZ_{t,i}, G_i)} \Big] e^{-sN\e}.
\label{eq:Cram_Cher}
\end{align}
To prove \eqref{eq:PLsubgauss1},  we will show that for $0 <  s < \sqrt{\frac{1}{\tilde{\kappa}_t}}$,
\be
\mathbb{E} \Big[\exp\{s\sum_{i=1}^N f(\sigma_1 Z_{1,i}, \ldots, \sigma_tZ_{t,i}, G_i)\}\Big] \leq  \exp\{ N \tilde{\kappa}_t  s^2\}.
\label{eq:PLsubgauss2}
\ee
Then, using \eqref{eq:PLsubgauss2} in \eqref{eq:Cram_Cher} and taking $s = \e/{2 \tilde{\kappa}_t}$ yields the upper tail bound in \eqref{eq:PLsubgauss1}.

We now prove \eqref{eq:PLsubgauss2}.  For $i \in [N]$, let $(\tZ_{1,i}, \ldots, \tZ_{t,i}, \tG_i)$ be an independent copy of $({Z}_{1,i}, \ldots, {Z}_{t,i}, G_i)$. Since $\mathbb{E}[ f(\sigma_1 \tZ_{1,i}, \ldots, \sigma_t \tZ_{t,i}, \tG_i)] = 0$, using Jensen's inequality we have 
\ben
\begin{split}
&\mathbb{E}[\exp(-s  f(\sigma_1 \tZ_{1,i}, \ldots, \sigma_t \tZ_{t,i}, \tG_i) )] \\
&\qquad \geq \exp(-s \mathbb{E} [f(\sigma_1 \tZ_{1,i}, \ldots, \sigma_t \tZ_{t,i}, \tG_i)]) = 1.
\end{split}
\een
Therefore, using the independence of  $\tilde{Z}$ and $Z$ we write
\begin{align}
&\mathbb{E}[e^{s f(\sigma_1 Z_{1,i}, \ldots, \sigma_tZ_{t,i}, G_i)  }] \\
&\leq \mathbb{E}[e^{s f(\sigma_1 Z_{1,i}, \ldots, \sigma_tZ_{t,i}, G_i) }] \cdot \mathbb{E}[e^{-s  f(\sigma_1 \tZ_{1,i}, \ldots, \sigma_t \tZ_{t,i}, \tG_i)}] \nonumber \\
& =  \mathbb{E}[e^{s (f(\sigma_1 Z_{1,i}, \ldots, \sigma_tZ_{t,i}, G_i) - f(\sigma_1 \tZ_{1,i}, \ldots, \sigma_t \tZ_{t,i}, \tG_i) )}].
\label{eq:PLsubgauss5}
\end{align}
Using \eqref{eq:PLsubgauss5} we prove \eqref{eq:PLsubgauss2} by demonstrating that for each $i \in [N]$,
\be
\mathbb{E}[e^{s (f(\sigma_1 Z_{1,i}, \ldots, \sigma_tZ_{t,i}, G_i) - f(\sigma_1 \tZ_{1,i}, \ldots, \sigma_t \tZ_{t,i}, \tG_i) )}] \leq  \exp\{\tilde{\kappa}_t  s^2\},
\label{eq:PLsubgauss6}
\ee
for $0 <  s < \sqrt{\frac{1}{\tilde{\kappa}_t}}$.
For $i \in [N]$ we have
\begin{align}
& \mathbb{E} [e^{s(f(\sigma_1 Z_{1,i}, \ldots, \sigma_tZ_{t,i}, G_i) - f(\sigma_1 \tZ_{1,i}, \ldots, \sigma_t \tZ_{t,i}, \tG_i)  )}]  \nonumber  \\
&= \sum_{q=0}^{\infty} \frac{s^q}{q!}  \mathbb{E} \Big[ f(\sigma_1 Z_{1,i}, \ldots, \sigma_tZ_{t,i}, G_i) \\
&\qquad \qquad  \qquad \qquad - f(\sigma_1 \tZ_{1,i}, \ldots, \sigma_t \tZ_{t,i}, \tG_i)  \Big]^q
\nonumber \\
&  \overset{(a)}{=} \sum_{k=0}^{\infty} \frac{s^{2k}}{(2k)!}  \mathbb{E} \Big[f(\sigma_1 Z_{1,i}, \ldots, \sigma_tZ_{t,i}, G_i) \\
&\qquad \qquad  \qquad \qquad  - f(\sigma_1 \tZ_{1,i}, \ldots, \sigma_t \tZ_{t,i}, \tG_i) \Big]^{2k},
\label{eq:EFZftildZ}
\end{align}
where step $(a)$ holds because the odd moments of the difference  equal $0$.  Next, using the pseudo-Lipschitz property of $f$,  for an absolute constant $L >0$, we have for $k \geq 1$:
\begin{align}
& \Big[f(\sigma_1 Z_{1,i}, \ldots, \sigma_tZ_{t,i}, G_i) - f(\sigma_1 \tZ_{1,i}, \ldots, \sigma_t \tZ_{t,i}, \tG_i) \Big]^{2k}  \nonumber \\
& \leq  L^{2k} \Big[1 +  \sum_{m=1}^t \sigma_m^2 ( Z_{m,i}^2 + \tZ_{m,i}^2 ) + G_i^2  +  \tG_i^2 \Big]^{k} \times \nonumber
\\
&\qquad \qquad  \qquad
\Big[  \sum_{m=1}^t \sigma_m^2 ( Z_{m,i}- \tZ_{m,i})^2  + (G_i - \tG_i)^2 \Big]^{k}  \nonumber   \\
&\stackrel{(a)}{\leq}  L^{2k} \Big[1 +  \sum_{m=1}^t \sigma_m^2 ( Z_{m,i}^2 + \tZ_{m,i}^2 ) + G_i^2  +  \tG_i^2 \Big]^{k}\times \nonumber
\\
&\qquad \qquad  \qquad 
2^k \Big[ \sum_{m=1}^t \sigma_m^2 ( Z_{m,i}^2 + \tZ_{m,i}^2 ) + G_i^2  +  \tG_i^2 \Big]^{k}  \nonumber  \\
& \stackrel{(b)}{\leq} (2L^{2})^{k}  
\Big[ \sum_{m=1}^t \sigma_m^2 ( Z_{m,i}^2 + \tZ_{m,i}^2 ) + G_i^2  +  \tG_i^2  \nonumber  \\
&\hspace{0.5in}  + (2t+2) \Big( \sum_{m=1}^t \sigma_m^4 ( Z_{m,i}^4 + \tZ_{m,i}^4 ) + G_i^4  +  \tG_i^4 \Big)  \Big]^k,  \nonumber  \\
 & \stackrel{(c)}{\leq}  \frac{(2 L^{2} (4t +4))^{k}}{4t +4} \Big[ \sum_{m=1}^t \sigma_m^{2k} ( Z_{m,i}^{2k} + \tZ_{m,i}^{2k} ) + G_i^{2k}  +  \tG_i^{2k} \Big]\nonumber
\\
& \  +    \frac{(2 L^{2} (4t +4) (2t+2))^{k}}{4t +4} \Big[\Big( \sum_{m=1}^t \sigma_m^{4k} ( Z_{m,i}^{4k} + \tZ_{m,i}^{4k} ) \nonumber   \\
&  \hspace{2.2in} + G_i^{4k}  +  \tG_i^{4k} \Big)  \Big]  \nonumber  \\ 
 & \leq \frac{( 2 L (2t+2))^{2k}}{4t+4} \Big[ \sum_{m=1}^t \sigma_m^{2k} ( Z_{m,i}^{2k} + \tZ_{m,i}^{2k} ) + G_i^{2k}  +  \tG_i^{2k}  \Big] \nonumber
\\
& \ + \frac{( 2 L (2t+2))^{2k}}{4t+4} \Big[  \sum_{m=1}^t \sigma_m^{4k} ( Z_{m,i}^{4k} + \tZ_{m,i}^{4k} ) + G_i^{4k}  +  \tG_i^{4k}   \Big],  \label{eq:diff_2k}
\end{align}
where inequalities $(a), (b), (c)$ are all obtained using using Lemma \ref{lem:squaredsums}. 
Using \eqref{eq:diff_2k} in \eqref{eq:EFZftildZ} and recalling that $\{  (Z_{m,i})_{1 \leq k \leq t}, G_i \}$  are identically distributed as  $\{  (\tZ_{m,i})_{1 \leq k \leq t}, \tG_i \}$, we get
\begin{align}
& \mathbb{E} [e^{s(f(\sigma_1 Z_{1,i}, \ldots, \sigma_tZ_{t,i}, G_i) - f(\sigma_1 \tZ_{1,i}, \ldots, \sigma_t \tZ_{t,i}, \tG_i)  )}]  \nonumber  \\
& \leq  1 +  \sum_{k=1}^{\infty} \frac{(s 2 L (2t+2))^{2k}}{(2k)! (4t+4)}  2  \Big[ \sum_{m=1}^t \sigma_m^{2k} \, \expec Z_{m,i}^{2k}  + \expec G_i^{2k}    \nonumber \\
& \hspace{1.6in} + \sum_{m=1}^t \sigma_m^{4k} \, \expec Z_{m,i}^{4k}  + \expec G_i^{4k}    \Big]  \nonumber \\
 & \stackrel{(a)}{\leq}  1 +  \sum_{k=1}^{\infty} \frac{(s 2 L (2t+2))^{2k}}{(2k)! (2t+2)} \Big[ \sum_{m=1}^t \sigma_m^{2k} 2 (k!) 2^{k}    \nonumber
\\
&\hspace{0.5in} 
+ 2 (k!) (2\nu)^{k}   + \sum_{m=1}^t \sigma_m^{4k} \, 2 (2k!) 2^{2k}  +  2 (2k!) (2\nu)^{2k}     \Big]  \nonumber  \\
 &\stackrel{(b)}{\leq}   1 +  \sum_{k=1}^{\infty} \frac{(s 2 L (2t+2))^{2k} }{t+1}   \Big[   \sum_{m=1}^t  \frac{\sigma_m^{2k}}{k!}  +  \frac{\nu^{k}}{k!} \nonumber \\
& \hspace{1.5in} + \sum_{m=1}^t (4 \sigma_m^4)^{k} +  (4\nu^2)^{k}     \Big]  \nonumber  \\
 & \leq  1 +  \sum_{k=1}^{\infty} (s 2 L (2t+2))^{2k} \Big[ \nu + 4 \nu^2 +  \sum_{m=1}^t  (\sigma_m^{2} +  4  \sigma_m^{4})   \Big]^k  \nonumber \\ 
 & \stackrel{(c)}{=} \Big(1 - s^2 16 L^2 (t+1)^2 [  \nu + 4 \nu^2 +  \sum_{m=1}^t  (\sigma_m^{2} +  4  \sigma_m^{4})] \Big)^{-1} \nonumber  \\
 & \stackrel{(d)}{\leq} e^{ s^2 32 L^2(t+1)^2 [  \nu + 4 \nu^2 + \sum_{m=1}^t  (\sigma_m^{2} +  4  \sigma_m^{4})]}.
\end{align}
In the chain of inequalities above, $(a)$ is obtained using the sub-Gaussian moment bound \eqref{eq:subgauss_moments}; step $(b)$ using the inequality  $\frac{(2k)!}{k!} \geq 2^k k!$, which can be seen as follows.
\[   \frac{(2k)!}{k!} = \prod_{j=1}^{k}(k+j) = k!  \  \prod_{j=1}^{k} \Big(\frac{k}{j}+ 1\Big)  \geq (k!) 2^k.  \] 
The equality $(c)$ holds because $s$ lies in the range specified by \eqref{eq:PLsubgauss2}, and $(d)$ holds because $\frac{1}{1-x} \leq e^{2x}$ for $x \in[0,\frac{1}{2}]$. This completes the proof of \eqref{eq:PLsubgauss6}, and hence  the result.
\end{IEEEproof}

\section{Other Useful Lemmas}  \label{app:lip_lemmas}

\begin{applem}[Product of Lipschitz Functions is PL(2)] \label{lem:Lprods}
Let $f: \mathbb{R}^p \rightarrow \mathbb{R}$ and $g: \mathbb{R}^p \rightarrow \mathbb{R}$ be Lipschitz continuous.  
Then the product function $h: \mathbb{R}^p \rightarrow \mathbb{R}$ defined as  $h(x): =f(x) g(x)$ is pseudo-Lipschitz of order 2.
\end{applem}

\begin{applem} \label{lem:PLexamples}
Let $\phi: \mathbb{R}^{t+2} \rightarrow \mathbb{R}$ be $PL(2)$. For $(c_1, \ldots, c_{t+1})$ constants and $Z \sim \mc{N}(0,1)$, the function $\tilde{\phi}: \mathbb{R}^{t+1} \rightarrow \mathbb{R}$ defined as
$
\tilde{\phi}(v_1, \ldots, v_t,  w) = \mathbb{E}_{Z}[\phi(v_1, \ldots, v_t, \sum_{r=1}^{t} c_{r} v_{r} + c_{t+1} Z, w)]
$ is then also PL(2).

\end{applem}

\begin{applem}
For any scalars $a_1, ..., a_t$ and positive integer $m$, we have  $\left(\abs{a _1} + \ldots + \abs{a_t} \right)^m \leq t^{m-1} \sum_{i=1}^t \abs{a_i}^m$.
Consequently, for any vectors $\un{u}_1, \ldots, \un{u}_t \in \mathbb{R}^N$, $\norm{\sum_{k=1}^t \un{u}_k}^2 \leq t \sum_{k=1}^t \norm{\un{u}_k}^2$.
\label{lem:squaredsums}
\end{applem}
\begin{IEEEproof} The first result follows from applying H{\"o}lder's inequality to the length-$t$ vectors $(\abs{a_1}, \ldots, \abs{a_t})$ and $(1, \ldots, 1)$. The second statement is obtained by applying the result with $m=2$.
\end{IEEEproof}

\section{Supplementary Material: Proof of Lemma \ref{lem:main_lem} parts (b).(ii) and (b).(iv)}   \label{supA}
The supplement available at \url{http://bit.ly/2iWMgbr} contains the proof of Lemma \ref{lem:main_lem} parts $(b)$.(ii) and $(b)$.(iv) for the case where the denoising functions $\{\eta_t(\cdot)\}_{t > 0}$ are differentiable in the first argument  except at a finite number of points.  The proof in Sec. \ref{sec:main_lem_proof}  covers the case where the denoising functions $\{\eta_t(\cdot)\}_{t > 0}$ are differentiable everywhere.  The proof of the general case is longer and somewhat tedious, so we include it in the supplement.

\section*{Acknowledgment}
We thank Andrew Barron for helpful discussions regarding certain technical aspects of the proof.

\IEEEtriggeratref{13}


\end{document}